\documentclass[12pt]{article}

\RequirePackage[OT1]{fontenc}
\RequirePackage{amsthm,amssymb,amsfonts,graphicx,subfigure,caption,bm,fullpage,wrapfig,setspace}
\RequirePackage[colorlinks,citecolor=blue,urlcolor=blue]{hyperref}
\usepackage{amsmath,amssymb,enumerate}
 \usepackage{float}
\usepackage{algpseudocode,algorithm,algorithmicx}
\usepackage{mathrsfs}
\usepackage{multirow}
\usepackage{forest, adjustbox}
\usepackage{bbm}
\usepackage{lscape}
\usetikzlibrary{calc}
\usepackage{relsize}
\usepackage{float}

\usepackage{epstopdf}
\usepackage{tikz}
\usetikzlibrary{positioning,decorations.pathreplacing,calc}
 \usetikzlibrary{bayesnet}
\tikzset{fontscale/.style = {font=\relsize{#1}}
    }
\numberwithin{equation}{section}
\theoremstyle{plain}
                          \newtheorem{lem}{Lemma}
                          
\theoremstyle{remark}         
\theoremstyle{definition} 

\newcommand\T{\mathcal{T}}

\newcommand\mI{\mathcal{I}}
\newcommand\mL{\mathcal{L}}

\newcommand\bpsi{\bm{\psi}}

\newcommand\bc{\bm{c}}

\newcommand\bp{\bm{p}}
\newcommand\bx{\bm{x}}
\newcommand\by{\bm{y}}

\newcommand\bmu{\bm{\mu}}
\newcommand\bnu{\bm{\nu}}
\newcommand\bSigma{\bm{\Sigma}}

\newcommand\bY{\bm{Y}}

\newcommand\balp{\bm\alpha}
\newcommand\bbet{\bm\beta}

\newcommand\bthe{\bm\theta}

\newcommand\btau{\bm\tau}

\newcommand\bgam{\bm\gamma}

\newcommand\ddfrac[2]{\frac{\displaystyle #1}{\displaystyle #2}}

\DeclareMathOperator*{\argmin}{arg\,min}

\newcommand\bpi{\boldsymbol{\pi}}
\newcommand{\RN}[1]{%
  \textup{\uppercase\expandafter{\romannumeral#1}}%
}

\newcommand\bi{\begin{itemize}}
\newcommand\ei{\end{itemize}}

\newcommand{\comment}[1]{}

 \allowdisplaybreaks

\newcommand{\beginsupplement}{%
        \setcounter{table}{0}
        \renewcommand{\thetable}{S\arabic{table}}%
        \setcounter{figure}{0}
        \renewcommand{\thefigure}{S\arabic{figure}}%
}

\algdef{SE}[SUBALG]{Indent}{EndIndent}{}{\algorithmicend\ }%
\algtext*{Indent}
\algtext*{EndIndent}

\def\woMR#1{\w@MR#1MR#1MR\relax}%
\def\w@MR#1MR#2MR#3\relax{#2}

\def\@MR#1 #2\relax#3{%
 \href{http://www.ams.org/mathscinet-getitem?mr=#1}%
 {\MRfixed{#3}}}%

\def\MRfixed{MR\woMR}%

\usepackage{natbib}

\usepackage{varioref}		
\labelformat{section}{Section~#1}
\labelformat{figure}{Figure~#1}
\labelformat{table}{Table~#1}

\usepackage{booktabs}
\usepackage{multirow}
\usepackage{calc}
\usepackage{tikz}
\usepackage{pgfplots}
\usepackage{makecell}

\usepackage{mathtools}

\DeclarePairedDelimiterX{\infdivx}[2]{(}{)}{%
  #1\;\delimsize\|\;#2%
}
\newcommand{\infdiv}{D_{\text{KL}}\infdivx}

\usepackage{lineno}

\title{Dirichlet-tree multinomial mixtures for clustering microbiome compositions}
\author{Jialiang Mao \and Li Ma}
\date{\textit{Duke University, Durham, NC} \\ \hspace{1mm} \\ \today}
\begin{document}
\maketitle

\begin{abstract}

Studying the human microbiome has gained substantial interest in recent years, and a common task in the analysis of these data is to cluster microbiome compositions into subtypes. This subdivision of samples into subgroups serves as an intermediary step in achieving personalized diagnosis and treatment. In applying existing clustering methods to modern microbiome studies including the American Gut Project (AGP) data, we found that this seemingly standard task, however, is very challenging in the microbiome composition context due to several key features of such data. Standard distance-based clustering algorithms generally do not produce reliable results as they do not take into account the heterogeneity of the cross-sample variability among the bacterial taxa, while existing model-based approaches do not allow sufficient flexibility for the identification of complex within-cluster variation from cross-cluster variation. Direct applications of such methods generally lead to overly dispersed clusters in the AGP data and such phenomenon is common for other microbiome data. To overcome these challenges, we introduce Dirichlet-tree multinomial mixtures (DTMM) as a Bayesian generative model for clustering amplicon sequencing data in microbiome studies. DTMM models the microbiome population with a mixture of Dirichlet-tree kernels that utilizes the phylogenetic tree to offer a more flexible covariance structure in characterizing within-cluster variation, and it provides a means for identifying a subset of signature taxa that distinguish the clusters. We perform extensive simulation studies to evaluate the performance of DTMM and compare it to state-of-the-art model-based and distance-based clustering methods in the microbiome context, and carry out a validation study on a publicly available longitudinal data set to confirm the biological relevance of the clusters. Finally, we report a case study on the fecal data from the AGP to identify compositional clusters among individuals with inflammatory bowel disease and diabetes. Among our most interesting findings is that enterotypes (i.e., gut microbiome clusters) are not always defined by the most dominant species as previous analyses had assumed, but can involve a number of less abundant OTUs, which cannot be identified with existing distance-based and method-based approaches.

\end{abstract}


\clearpage

\section{Introduction}


The human microbiome is the collective genomes of all microbes that inhabit the human body. It has been associated with various aspects of our physiology \citep{turnbaugh2009core, qin2012metagenome, karlsson2013gut} and is suggested as a way towards precision medicine \citep{kuntz2017introducing}. The development of next-generation sequencing strategies enables us to profile the microbiome fast and economically, through either amplicon sequencing on target genes (usually the 16S ribosomal RNA gene) or shotgun sequencing on the entire microbial genome. In this work, we focus on datasets obtained from amplicon sequencing studies. Traditionally, the sequencing reads are sent to preprocessing pipelines such as QIIME \citep{caporaso2010qiime} to construct clusters named operational taxonomic units (OTUs) based on certain predefined similarity threshold (typically $97\%$). In contrast, more recently developed pipelines such as DADA2 \citep{callahan2016dada2} directly resolve amplicon sequence variants (ASVs), which is shown to outperform OTUs in terms of accuracy and interpretability \citep{callahan2017exact}. OTUs and ASVs serve as the unit for downstream statistical analyses and provide the same interface: each sample is a vector of counts on a list of units (OTUs or ASVs), representing the composition of the underlying community. The methodology developed in this work applies to both OTUs and ASVs, we thus use the customary OTU to refer to the unit.

 
Given the heterogeneous nature of microbiome samples, a useful idea in microbiome analysis is to first group individual samples into clusters and seek to understand the relation of these clusters with the host environment and other health outcomes. For example, in the context of the human gut microbiome, these clusters are referred to as ``enterotypes'' \citep{arumugam2011enterotypes, costea2018enterotypes}, which are shown to be associated with long-term dietary habits and risks for obesity and Crohn's disease \citep{wu2011linking,holmes2012dirichlet,quince2013impact}. 

In applying this strategy to analyze several modern microbiome data sets including the American Gut Project (AGP) \citep{mcdonald2015context,mcdonald2018american}, however, we found that reliably clustering microbiome samples is in fact very challenging. Off-the-shelf clustering algorithms such as $k$-means, Partitioning Around Medoids (PAM) and hierarchical clustering that are distance-based are not satisfactory when applied to microbiome data. 
This is true even when using popular distance metrics specifically tailored for microbiome compositions such as the Bray-Curtis dissimilarity and the Unifrac distances \citep{lozupone2005unifrac}. Other authors including \cite{koren2013guide} also showed that in clustering microbiome compositions, different methods for selecting the number of clusters in distance-based methods can yield inconsistent results and that these algorithms are highly sensitive to the distance metrics chosen. We believe a main reason for such inconsistency is that different distance metrics induce different weighting on the OTUs that are inconsistent with the actual patterns of heterogeneous cross-ample variability in the data.

In addition, we have found that existing non-distance-based methods, such as those based on probabilistic models like mixtures  also suffer similar issues for microbiome compositional data. In particular, the arguably most widely used model-based approach for clustering microbiome data, called the Dirichlet multinomial mixture (DMM) model \citep{holmes2012dirichlet}, which adopts a multinomial sampling scheme and generates the sample-specific multinomial parameters from a finite mixture of Dirichlet components, often results in very large and overly dispersed clusters.
A key reason for the poor performance of this model, we believe, is its lack of flexibility in characterizing the cross-sample variability, which in turn hampers the ability to differentiate within-cluster variability from between-cluster variability.
In particular, the single dispersion (also called concentration) parameter of the Dirichlet distribution is insufficient for characterizing the often complex variation among samples within each cluster. Moreover, the Dirichlet distribution implies independence among OTU compositions up to the sum to one constraint \citep{aitchison1982statistical}, which is restrictive in the microbiome context \citep{wang2017dirichlet}.

To overcome these difficults and achieve more reliable cluster analysis on the AGP study and other microbiome data, we introduce a new probabilitistic model for clustering microbiome compositions called Dirichlet-tree multinomial mixtures (DTMM). Similar to the DMM, our method uses mixture modeling to achieve clustering under the Bayesian inference framework. 
The difference is two-fold. First, by utilizing a natural hierarchical relationship among the OTUs in terms of the phylogenetic tree, DTMM adopts the Dirichlet-tree distribution (DT) \citep{dennis1991hyper,wang2017dirichlet} as the mixture component, 
in contrast to the Dirichlet component used by DMM. The DT mixing component incorporates multiple dispersion parameters, one for each node in the phylogenetic tree, 
thereby allowing much more flexible and realistic cross-sample variation among the OTUs. 
In addition, motivated by the fact that microbiome clusters are often determined by a subset of the taxa, we incorporate a model selection feature into the DTMM framework that allows (i) the signature taxa that distinguish the clusters to be identified and (ii) the common features across clusters (e.g., groups of ``house-keeping'' taxa) to be more accurately characterized through borrowing information among clusters.



With the proposed method, we report a case study of the AGP data to find and explore enterotypes of samples that are diagnosed with Inflammatory Bowel Disease (IBD) or diabetes. Enterotypes have been established and compared among samples from different geographical locations \citep{arumugam2011enterotypes, costea2018enterotypes} or with different host dietary patterns \citep{wu2011linking}. Our analysis provides another important facet to this thread of work. Both IBD and diabetes were shown to be related to the human gut microbiome \citep{kostic2014microbiome, qin2012metagenome}. It is thus natural to expect different enterotype patterns in samples with these diseases. Interesting, contrary to traditional wisdom, our analysis shows that enterotypes are not always characterized by a small number of highly abundant dominant taxa, but can arise from combinations of several taxa.


The rest of the paper is organized as follows. In \ref{sec:method} we introduce a phylogenetic tree-based decomposition of the multinomial counts and proposes the DTMM model for clustering OTU counts based on this decomposition. In \ref{sec:num_exam} we conduct a series of representative numerical experiments to evaluate the performance of DTMM. We also validate the clusters found by DTMM with a longitudinal microbiome dataset for which the actual clustering pattern is roughly known due to the experiment design. In \ref{sec:app} we report our case study of the AGP data. \ref{sec:discussion} concludes with a few remarks.


\section{Method}
\label{sec:method}

\subsection{DM and DMM}

Consider a microbiome dataset with OTU counts of $n$ samples $\by_1, \by_2,\ldots,\by_n$. Each sample is a vector of counts of the $M$ OTUs in the study denoted by $\Omega = \{\text{OTU}_2, \text{OTU}_2, \ldots, \text{OTU}_M \} = \{\omega_1,\omega_2, \ldots, \omega_M \}$. Let the $i$-th sample and the counts in that sample be $\by_i = (y_{i1}, y_{i2},\ldots, y_{iM})$ and $N_i=\sum^{M}_{j=1}y_{ij}$, where $y_{ij}$ is the count of OTU $j$. The samples can be stacked into an OTU table denoted by $\bY$, as shown in \ref{tab:otuexample}. In this work, we treat the total counts $N_i$'s as given since they are artificial quantities that depend on the sequencing depth. 
We consider the following Dirichlet-multinomial model (DM) \citep{knights2011bayesian,la2012hypothesis}:
\begin{equation}
\begin{aligned}
\by_i \mid N_i, \bp_i &  \overset{\rm{ind}}{\sim}  \text{Multi}(N_i, \bp_i) \qquad \text{and} \qquad \bp_i \mid \balp \overset{\rm{iid}}{\sim} \rm{Dir}(\balp),
\end{aligned}
\label{eq:dm}
\end{equation}
where $\bp_i = (p_{i1},p_{i2}, \ldots, p_{iM})$, $p_{ij}$ is the probability that a count in sample $i$ belongs to OTU $j$, $\balp = (\alpha_1, \alpha_2, \ldots, \alpha_M)$ with $\alpha_j > 0$ for $j=1,\ldots,M$.

Viewing each sample as randomly drawn from an underlying community characterized by its multinomial parameter \citep{holmes2012dirichlet}, DM models all the communities as realizations of a single metacommunity governed by $\balp$. \cite{holmes2012dirichlet} extend DM to Dirichlet multinomial mixtures (DMM) by replacing the single Dirichlet prior in DM by a finite mixture of $K$ Dirichlets:  
\begin{equation}
\begin{aligned}
\bp_i \mid \bpi, \balp_1,\ldots,\balp_K  &  \overset{\rm{iid}}{\sim} \sum\limits^{K}_{k=1} \pi_k\text{Dir}(\balp_k) \quad \text{and} \quad \bpi = (\pi_1,\ldots,\pi_K) \sim \text{Dir}(\bm{b}_0),
\end{aligned}
\label{eq:dmm}
\end{equation}
where $\balp_k=(\alpha_{k1},\alpha_{k2}, \ldots, \alpha_{kM})$, $\bpi$ the weights of the metacommunities with $\sum^{K}_{k=1}\pi_k=1, \pi_k \geq 0$ for $k=1,\ldots,K$. In DMM, each sample is viewed as a draw from a unique community that is itself drawn from one of the $K$ metacommunities. 
 \begin{center}\resizebox {\textwidth} {!}{  
 \begin{minipage}{\textwidth}   
  \begin{minipage}[b]{0.45\textwidth}
    \centering
    \begin{tabular}{@{}c | llll | c@{}}
\toprule
Sample   &  $\omega_1$ & $\omega_2$& $\cdots$ & $\omega_M$ & Sum \\ \midrule
1 & $y_{11}$ & $y_{12}$ & $\cdots$ & $y_{1M}$  & $N_{1}$\\
2 &  $y_{21}$ & $y_{22}$ & $\cdots$ & $y_{2M}$ & $N_{2}$\\
$\vdots$  & $\vdots$ & $\vdots$ & $\ddots$ & $\vdots$ & $\vdots$ \\
$n$ &  $y_{n1}$ & $y_{n2}$ & $\cdots$ & $y_{nM}$ & $N_{n}$ \\ \bottomrule
\end{tabular}
\captionof{table}{An $n\times M$ OTU table.}
        \label{tab:otuexample}
  \end{minipage}
  \hspace{0.5cm} 
   \begin{minipage}[b]{0.45\textwidth}
    \centering
    \includegraphics[width = 6.6cm ]{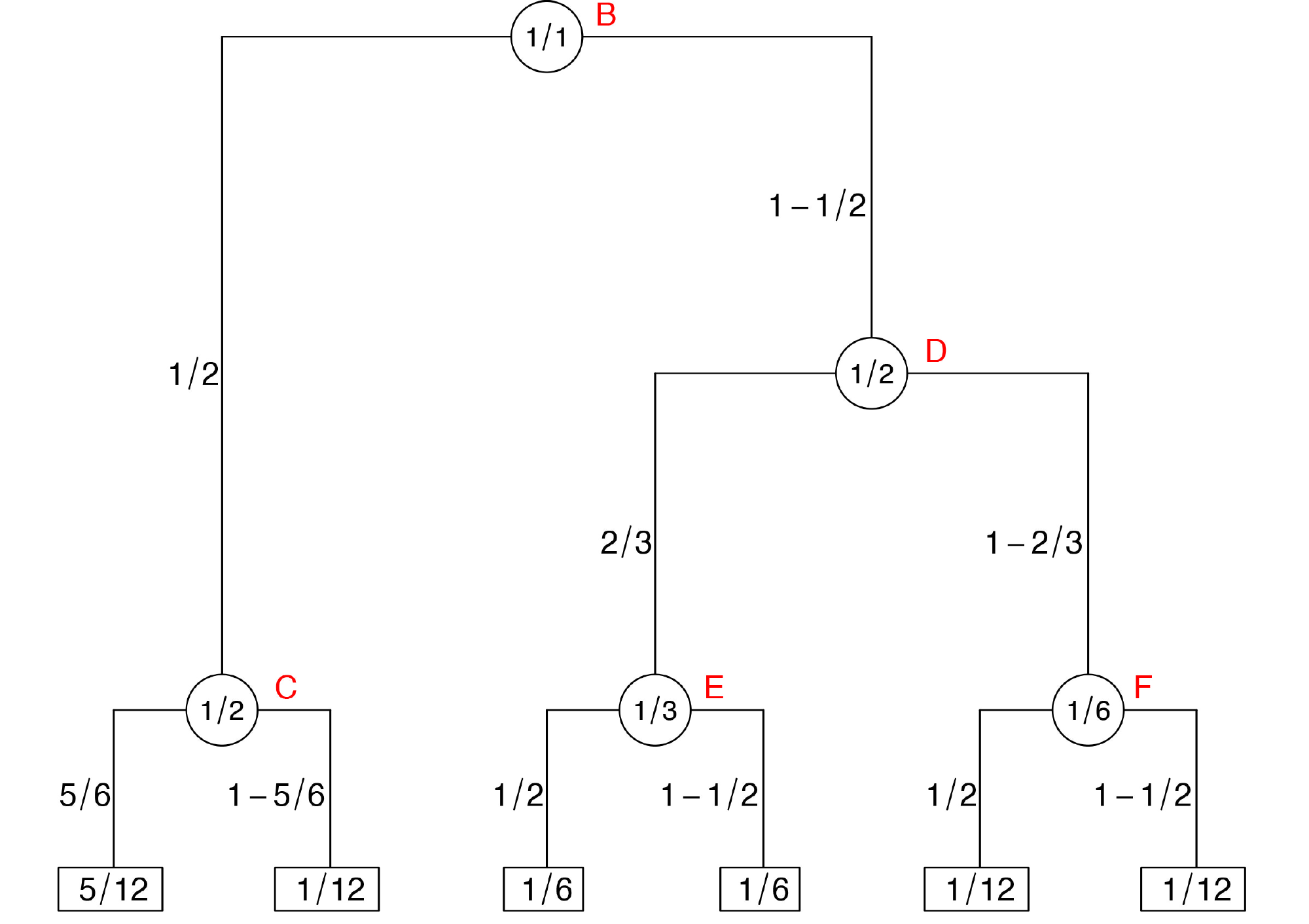}
    \captionof{figure}{A tree based generation of $\bp_{\text{ex}}$.}
        \label{fig:sec_2_tree}
  \end{minipage}   
       \hfill
  \end{minipage} 
  }
  \end{center}

As a clustering method, DMM has several limitations. Most importantly, it could not adequately model the within cluster variation of the microbial composition. This can be seen by writing the cluster-specific Dirichlet parameter $\balp_k$ as $\balp_k = \alpha_{k0} \cdot\bar\balp_k$, where $\bar\balp_k$ lying in the $(M-1)$-dimensional simplex represents the prior mean of the multinomial probabilities in cluster $k$, $\alpha_{k0} = \sum^{M}_{j=1}\alpha_j$ determines the within cluster variation of all these probabilities around $\bar\balp_k$ simultaneously. 
Secondly, the multinomial parameters in DM are modeled independently up to the sum to one constraint \citep{mosimann1962compound}, which is not suitable in the microbiome context since the OTUs are functionally and evolutionarily related. 
Although DMM is specified within the Bayesian framework, the posterior inference is performed by optimization through an EM algorithm with Laplace approximations of the marginal likelihoods. When the number of OTUs is moderate, which is typical in microbiome studies, these techniques are numerically unstable and cannot provide reliable uncertainty quantifications.


\subsection{Dirichlet-tree multinomial mixtures}


OTUs in a microbiome study are functionally and evolutionarily related. Typically, this relationship can be summarized into a rooted phylogenetic tree, where each internal node can be viewed as a ``taxa" that represents the most recent common ancestor of its descendant OTUs. Let $\T = \T(\mI, \mathcal{U}; \mathcal{E})$ be a rooted full binary phylogenetic tree over the $M$ OTUs in the study, where $\mI$, $\mathcal{U}$ and $\mathcal{E}$ denote the set of internal nodes, leaves and edges of $\T$, respectively. We denote each node $A\in \mI\cup\mathcal{U}$ by the set of its descendant OTUs. In particular, $A=\Omega$ denotes the root of $\T$; $A=\{\omega_j\}$ represents the leaf that contains OTU $j$ for $j=1,\ldots, M$. With our notation, $\mathcal{U} = \{ \{\omega\}: \omega\in \Omega \}$. For $A\in\mI$, let $A_l$ and $A_r$ be the left and right children of $A$, respectively. For $A\in\mI\cup\mathcal{U}\setminus\{\Omega\}$, let $A_p$ be its parent and $A_s$ be its sibling (i.e., the node in $\T$ that has the same parent as $A$).

Given $\T$, it can be shown that the multinomial likelihood of $\by_i$ factorizes into a series of binomial likelihoods at the internal nodes of $\T$:
\begin{equation}
\begin{aligned}
\mL_M(\by_i\mid \bp_i) \propto  \prod\limits_{\{ A: A\in\mI \}} \mL_{B}(y_i(A_l)\mid y_i(A), \theta_i(A)),
\end{aligned}
\label{eq: like_fac}
\end{equation}
where 
\begin{equation}
\begin{aligned}
&y_i(A)=\sum_{\{j:\omega_j\in A\} }y_{ij}, \quad \theta_i(A) = \frac{\sum_{\{j:\omega_j\in A_l\}} p_{ij}}{\sum_{\{j:\omega_j\in A\}} p_{ij}},\\
&y_i(A_l) \mid y_i(A), \theta_i(A)  \overset{\rm{ind}}{\sim} \text{Binom}(y_i(A),\theta_i(A)).
\end{aligned}
\label{eq: trans1}
\end{equation}
Note that for $j = 1,\ldots, M$, there is a unique path $\mathscr{P}^j = A_0^j = \Omega\rightarrow A_1^j \rightarrow\cdots \rightarrow A^j_{l_j} \rightarrow \omega_j$ in $\T$ connecting the root with $\omega_j$ such that 
\begin{equation}
\begin{aligned}
p_{ij} = \prod_{l = 0}^{l_j} \theta_i(A^j_l).
\end{aligned}
\label{eq: trans2}
\end{equation}
We denote $\bthe_i = \{\theta_i(A):A\in\mI\}$. Let $\bthe_i = tr(\bp_i)$ and $\bp_i = tr^{-1}(\bthe_i)$ be the ``tree-based ratio transform'' and the ``inverse tree-based ratio transform'' defined in (\ref{eq: trans1}) and (\ref{eq: trans2}). $\bp_i$ and $\bthe_i$ give two equivalent parameterizations of the distribution of $\by_i$. \ref{fig:sec_2_tree} gives an example of how a specific multinomial parameter $\bp_{\text{ex}}=(\frac{5}{12},\frac{5}{12},\frac{1}{6}, \frac{1}{6}, \frac{1}{12},\frac{1}{12})$ on $6$ OTUs can be generated sequentially along a given tree.

The likelihood factorization in (\ref{eq: like_fac}) provides an orthogonal decomposition of the empirical evidence about $\bp_i$ into pieces of evidence about $\theta_i(A)$ at $A\in\mI$, which suggests a divide-and-conquer strategy of doing inference on $\bp_i$ through learning the ``probability assignment parameters'' $\bthe_i$. To this end, we take the Bayesian strategy and put independent beta priors on the binomial parameters:
\begin{equation}
\begin{aligned}
\theta_i(A)\mid \theta(A), \tau(A) & \overset{\rm{ind}}{\sim} \text{Beta}(\theta(A)\tau(A), (1-\theta(A))\tau(A)),
\end{aligned}
\label{eq:}
\end{equation}
where $\theta(A)\in (0, 1)$ is the prior mean of $\theta_i(A)$, $\tau(A) > 0$ is a dispersion parameter that controls the variability of $\theta_i(A)$ around its mean. Note that the independent priors on $\bthe_i$ together with the relation in (\ref{eq: trans2}) induce a joint prior on $\bp_i$, which falls into the family of Dirichlet-tree distributions (DT) \citep{dennis1991hyper}. Let $\bthe = \{\theta(A): A\in\mI \}$ and $\btau = \{\tau(A):A\in\mI \}$, we shall denote the Dirichlet-tree prior on $\bp_i$ as 
\begin{equation}
\begin{aligned}
\bp_i  = tr^{-1}(\bthe_i), \quad\bthe_i  \sim \text{DT}_{\T}(\bthe, \btau).
\end{aligned}
\label{eq:treebasedtrans}
\end{equation}
When $\tau(A) = \tau(A_l) + \tau(A_r)$ for every $A\in\mI$ that has non-leaf children, DT degenerates to the Dirichlet distribution. Without this constraint, DT offers a more flexible way to model the variability of $\bp_i$ around its cluster centroid. 

DT also induces a more flexible covariance structure than the Dirichlet distribution. This can be seen by considering the covariance between any two different categories $j_1$ and $j_2$. Suppose that the first $(L+1)$ nodes in $\mathscr{P}^{j_1}$ and $\mathscr{P}^{j_2}$ are shared and let the shared path be $\Omega= A_0\rightarrow A_1\rightarrow \cdots\rightarrow A_L$. It can be shown that \citep{dennis1991hyper} 
\begin{equation}
\begin{aligned}
{\rm{Cov}}(p_{ij_1}, p_{ij_2}) = \left[\frac{\tau(A_L)}{\tau(A_L) + 1} \prod\limits_{1\leq t \leq L} \frac{ [a(A_t) +1] \tau(A_{t-1})}{a(A_t) [\tau(A_{t-1})+1]} - 1\right]\mathbb{E}(p_{ij_1}) \mathbb{E}(p_{ij_2}),
\end{aligned}
\label{eq:}
\end{equation}
where $a(A_t) = \theta(A_{t-1})\tau(A_{t-1})$ if $A_t$ is the left child of $A_{t-1}$ and $a(A_t) = (1-\theta(A_{t-1}))\tau(A_{t-1})$ otherwise. In DT, the covariance between categories depends not only on their means and the sum of the pseudo count as in the Dirichlet distribution, but also on the tree structure. 
This offers a more flexible covariance structure among OTU counts governed by the phylogenetic information. For example, since $a(A_t) < \tau(A_{t-1})$, $[a(A_t) +1] \tau(A_{t-1})/ a(A_t) [\tau(A_{t-1})+1]> 1$, $p_{ij_1}$ and $p_{ij_2}$ can be positively correlated if $j_1$ and $j_2$ share a series of common ancestors in the phylogenetic tree. On the other hand, if $j_1$ and $j_2$ are far away in the phylogenetic tree such that their only common ancestor is $\Omega$, ${\rm{Cov}}(p_{ij_1}, p_{ij_2})  = -\mathbb{E}(p_{ij_1}) \mathbb{E}(p_{ij_2})/(\tau(\Omega) + 1)$ as in the Dirichlet distribution. When the phylogenetic tree gives decent summaries of the functional relationship among OTUs, this introduces suitable covariance structure among the OTU counts and can improve the inference substantially. 

DT has been used for microbiome modelings in various context for different purposes. For example, \cite{wang2017dirichlet} apply the DT multinomial model to study the association between OTU counts and a set of covariates; \cite{tang2018phylogenetic} and \cite{mao2020bayesian} use the tree decomposition to motivate a divide-and-conquer strategy to increase the statistical power when comparing the OTU composition of groups of samples. 

In this work, we replace the Dirichlet component in DMM with DT mixing components to give a more suitable clustering model for microbiome data. 
In DMM, if the counts of one or a small number of OTUs are highly variable, the single dispersion parameter would be estimated large in adjustment of this variation. As a result, less variable cluster signatures contained in other OTUs would be buried and the samples would be modeled as drawn from a single or otherwise small number of highly heterogeneous clusters. In contrast, the set of dispersion parameters in DT are able to account for different levels of variation across OTUs and thus prevent the signals from being contaminated by the noises.

Specifically, we take the multinomial sampling scheme as in DMM. Following (\ref{eq:treebasedtrans}), let
\begin{equation}
\begin{aligned}
\bthe_i\mid \bpi, \{ (\bthe^*_{k}, \btau^*_k)\}^K_{k=1} \overset{\rm{iid}}{\sim}  \sum\limits^{K}_{k=1}\pi_k \text{DT}_{\T}(\bthe^*_{k}, \btau^*_k) \quad \text{and} \quad 
\bpi = (\pi_1,\ldots,\pi_K) \sim \text{Dir}(\bm{b}_0),
\end{aligned}
\end{equation}
where $(\bthe^*_{k}, \btau^*_k)  \overset{\rm{iid}}{\sim} \text{DT}_{\T}(\bthe_0, \bnu_0) \times F(\btau)$, $\text{DT}_{\T}(\bthe_0, \bnu_0)$ the prior for the cluster centroid, $F(\btau) = \prod_{A\in\mI} F^A(\tau(A))$ the prior for the within-cluster dispersion. Note that $(\bthe_k^*, \btau_k^*)$ determines the $k$-th meta-community. We shall for now refer to this model as the Dirichlet-tree multinomial mixtures (DTMM).


\subsection{Discriminative taxa selection}
\label{subsec:node}

In DMM, all OTUs are treated equally in the clustering procedure. In many applications, however, it is expected that only a (possibly small) subset of OTUs determine the underlying clusters. 
When this is the case, not only can identifying these signature taxa enhance the sensitivity for separating the clusters but it will improve the interpretability of the resulting inference. 

In this section, we incorporate model selection into the model to allow automatic taxa selection. At the same time, in specifying the DTMM model we adopt a standard nonparametric modeling approach in dealing with the difficulty in setting the number of clusters beforehand by replacing the finite mixture with a Dirichlet process (DP) mixture. 
Formally, for $A\in\mI$, let $\gamma(A) \in \{0,1\}$ be an indicator of whether node $A$ can be contributive to the latent clustering: $\gamma(A) = 1$ if $A$ can play a role in defining clusters, and $0$ otherwise. If $\gamma(A) = 1$, $A$ is ``active'' in clustering and we allow different clusters to have cluster-specific branching probabilities at $A$; otherwise, $A$ in ``inactive'' and we force all the clusters at $A$ to share the same branching probability. For this reason, we shall refer to $\gamma(A)$ and $\lambda(A)$ as the activation indicator and the prior activation probability on $A$. Let $\bgam = \{\gamma(A):A\in\mI \}$ be the collection of activation indicators of all the internal nodes.

Let $F(\cdot)$ be a probability measure on $(0,\infty)$ and $\delta_{x}(\cdot)$ the Dirac measure. The model can be written in the following hierarchical form:

\begin{itemize}
\item
sampling model on $\by_i$:
\begin{equation}
\begin{aligned}
\by_i \mid N_i, \bp_i & \overset{\rm{ind}}{\sim} \text{Multi}(N_i, \bp_i);
\end{aligned}
\end{equation}

\item
priors for the sample-specific probability assignment vector $\bp_i$:
\begin{equation}
\begin{aligned}
\bp_i  & = tr^{-1}(\bthe_i) \quad \text{and} \quad
\bthe_i \mid \bthe_i^\prime, \btau_i^\prime & \overset{\rm{ind}}{\sim}   \text{DT}_{\T}(\bthe_i^\prime,  \btau_i^\prime) ;
\end{aligned}
\label{eq:dt}
\end{equation}

\item
priors for the cluster-specific probability assignment vector:
\begin{equation}
\begin{aligned}
(\bthe_i^\prime,  \btau_i^\prime) \mid G & \overset{\rm{iid}}{\sim} G \quad \text{and} \quad
G & \sim {\rm{DP}}( G_0(\bthe, \btau \mid \bgam, \tilde\bthe, \tilde\btau) ; \beta) ;
\end{aligned}
\label{eq:dp}
\end{equation}

\item
the base measure in DP:
\begin{equation}
\begin{aligned}
G_0(\bthe, \btau \mid \bgam, \tilde\bthe, \tilde\btau) & = \prod\limits_{A\in\mI}G_0^A(\theta(A), \tau(A) \mid \gamma(A),\tilde\theta(A), \tilde\tau(A))\\
G_0^A(\theta(A), \tau(A) \mid \gamma(A) = 1,\tilde\theta(A), \tilde\tau(A)) & = \text{Beta}(\theta_0(A)\nu_0(A),(1-\theta_0(A))\nu_0(A) ) \times F^A(\tau)\\
G_0^A(\theta(A), \tau(A) \mid \gamma(A) = 0, \tilde\theta(A), \tilde\tau(A)) & =  \delta_{(\tilde\theta(A), \tilde\tau(A))};
\end{aligned}
\label{eq:model_base}
\end{equation}

\item
priors for the hyperparameters in the base measure: for $A\in\mI$,
\begin{equation}
\begin{aligned}
\gamma(A) &\overset{\rm{ind}}{\sim} \text{Binom}(\lambda(A)) \\
(\tilde\theta(A), \tilde\tau(A)) & \overset{\rm{ind}}{\sim}  \text{Beta}(\theta_0(A)\nu_0(A),(1-\theta_0(A))\nu_0(A) ) \times F^A(\tau)\\
\lambda(A) & \overset{\rm{ind}}{\sim} \text{Beta}(a_0(A), b_0(A)).
\end{aligned}
\label{eq:model_hyper}
\end{equation}
\end{itemize}

 \begin{minipage}{\linewidth}
 \begin{minipage}[b]{0.4\linewidth}
 \begin{figure}[H]
      \centering
       \scalebox{.85}{
  \tikz{ %
    \node[latent] (g) {$G$} ; %
    \node[latent, left=of g] (beta) {$\beta$} ; %
    \node[latent, below=of g, xshift = -20pt] (theta_i) {$\bthe^\prime_i$} ; %
    \node[latent, below=of g, xshift = 20pt] (tau_i) {$\btau^\prime_i$} ; %
   
    \node[latent, below=of theta_i, xshift = 20pt] (p) {$\bp_i$} ; %
    \node[obs, right=of p] (y) {$\by_i$} ; %
     
    \node[latent, above=of g, xshift = -35pt, yshift = 5pt] (theta_t) {\tiny$\tilde\theta(A)$} ; %
    \node[latent, above=of g, xshift = 0pt, yshift = 5pt] (tau_t) {\tiny$\tilde\tau(A)$} ; %
    \node[latent, above=of g, xshift = 35pt, yshift = 5pt] (gamma) {\tiny$\gamma(A)$} ; %
     
    \node[latent, right=of gamma] (lambda) {\tiny$\lambda(A)$} ; %
    

    \plate[inner sep=0.25cm, xshift=0.0cm, yshift=0.12cm] {plate1} {(theta_i)(tau_i)(p) (y)} {$n$}; %
     {\tikzset{plate caption/.append style={below right=7pt and -20pt of #1.south east}}
    \plate[inner sep=0.3cm, xshift=-0.1cm, yshift=0.14cm] {plate2} {(gamma)(theta_t)(tau_t)(lambda)} {$A\in\mI$}; %
    }
    \edge{beta}{g};%

    \edge{p}{y};%
    \edge{theta_i}{p};
    \edge{tau_i}{p};
 
    \edge{g}{theta_i};
    \edge{g}{tau_i};
    
    \edge{gamma}{g};
    \edge{tau_t}{g};
    \edge{theta_t}{g};
    
    \edge{lambda}{gamma};
    
  } }
  \caption{A graphical model representation of DTMM.}
  \label{fig:graph3}
\end{figure}
\end{minipage}
   \hspace{0.05\linewidth}
\begin{minipage}[b]{0.45\linewidth}
 \begin{figure}[H]
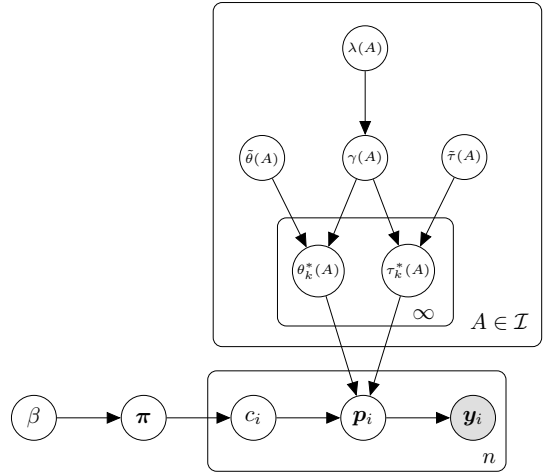

       \scalebox{.85}{
  \tikz{ %

    \node[latent] (pi) {$\bpi$} ; %
    \node[latent, left=of pi] (beta) {$\beta$} ; %
    \node[latent, right=of pi] (c) {$c_i$} ; %
    \node[latent, right=of c] (p) {$\bp_i$} ; %
    \node[obs, right=of p] (y) {$\by_i$} ; %
    
    \node[latent, above = of p, xshift = -20pt, yshift = 15pt] (theta_d) {\tiny$\theta^*_k(A)$} ; %
    \node[latent, above = of p, xshift = 20pt, yshift = 15pt] (tau_d) {\tiny$\tau^*_k(A)$} ; %

    \node[latent, above= of tau_d, xshift = 25pt] (F_A) {\tiny$\tilde\tau(A)$} ; %
    \node[latent, above= of theta_d, xshift = -25pt] (theta_t) {\tiny$\tilde\theta(A)$} ; %

    \node[latent, right= of theta_t, xshift = -3pt] (gamma) {\tiny$\gamma(A)$} ; %
    \node[latent, above= of gamma, xshift = 0pt] (lambda) {\tiny$\lambda(A)$} ; %
      
    
    \plate[inner sep=0.25cm, xshift=-0.1cm, yshift=0.12cm] {plate1} {(c)(p) (y)} {$n$}; %
    \plate[inner sep=0.25cm, xshift=0.03cm, yshift=0.15cm] {plate2} {(tau_d)(theta_d)} {$\infty$}; %
    
     {\tikzset{plate caption/.append style={below right=7pt and -8pt of #1.south east}}
    \plate[inner sep=0.3cm, xshift=-0.1cm, yshift=0.05cm] {plate2} {(tau_d)(theta_d)(F_A)(theta_t)(lambda)} {$A\in\mI$}; %
     }
    \edge{beta}{pi};%
    \edge{pi}{c};%
    \edge{c}{p};%
    \edge{p}{y};%
    
    \edge{theta_d}{p};
    \edge{tau_d}{p};
    \edge{F_A}{tau_d};
    
    \edge{theta_t}{theta_d};
    \edge{gamma}{theta_d};
    \edge{gamma}{tau_d};
    
    
    
    \edge{lambda}{gamma};

  } }
  \caption{An alternative graphical model representation of DTMM.}
  \label{fig:graph4}
\end{figure}
\end{minipage}

\end{minipage}

\begin{minipage}{\linewidth}
 \begin{minipage}[b]{0.4\linewidth}
\begin{figure}[H]
\scalebox{.8}{
\tikz{ %
    \node[latent] (pi) {$\bpi$} ; %
    \node[latent, left=of pi] (beta) {$\bm{b}_0$} ; %
    \node[latent, right=of pi] (c) {$c_i$} ; %
    \node[latent, right=of c] (p) {$\bp_i$} ; %
    \node[obs, right=of p] (y) {$\by_i$} ; %
    \node[latent, above=of p, yshift=0.5cm] (alpha) {$\balp_k$} ; %
    
    \plate[inner sep=0.25cm, xshift=-0.1cm, yshift=0.12cm] {plate1} {(c)(p) (y)} {$n$}; %
    \plate[inner sep=0.22cm, xshift=0.0cm, yshift=0.12cm] {plate2} {(alpha)} {$K$}; %
    
 
    \edge{pi}{c};%
    \edge{c}{p};%
    \edge{p}{y};%
    \edge{beta}{pi}
    
    \edge{alpha}{p};%
  } }
 \caption{A graphical model representation of DMM.}
 \label{fig:graph1}
\end{figure}
\end{minipage}
   \hspace{0.05\linewidth}
\begin{minipage}[b]{0.45\linewidth}
      \begin{figure}[H]
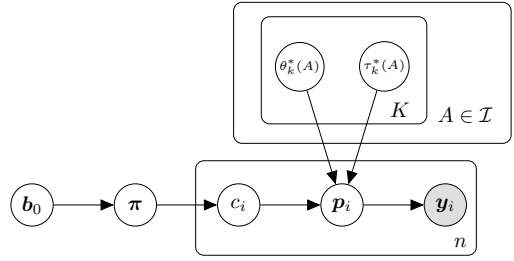

       \scalebox{.8}{
  \tikz{ %

    \node[latent] (pi) {$\bpi$} ; %
    \node[latent, left=of pi] (beta) {$\bm{b}_0$} ; %
    \node[latent, right=of pi] (c) {$c_i$} ; %
    \node[latent, right=of c] (p) {$\bp_i$} ; %
    \node[obs, right=of p] (y) {$\by_i$} ; %
    
    \node[latent, above = of p, xshift = -20pt, yshift = 15pt] (theta_d) {\tiny$\theta^*_k(A)$} ; %
    \node[latent, above = of p, xshift = 20pt, yshift = 15pt] (tau_d) {\tiny$\tau^*_k(A)$} ; %


    \plate[inner sep=0.25cm, xshift=-0.1cm, yshift=0.12cm] {plate1} {(c)(p) (y)} {$n$}; %
    \plate[inner sep=0.25cm, xshift=0.03cm, yshift=0.15cm] {plate2} {(tau_d)(theta_d)} {$K$}; %
    
     {\tikzset{plate caption/.append style={below right=7pt and 13pt of #1.south east}}
    \plate[inner sep=0.49cm, xshift=-0.19cm, yshift=0.15cm] {plate2} {(tau_d)(theta_d)} {$A\in\mI$}; %
     }
    \edge{beta}{pi};%
    \edge{pi}{c};%
    \edge{c}{p};%
    \edge{p}{y};%
    
    \edge{theta_d}{p};
    \edge{tau_d}{p};
     
  } }
  \caption{A graphical model representation of DTMM without taxa selection.}
  \label{fig:graph2}
\end{figure}
\end{minipage}
\end{minipage}

For simplicity, we shall refer to this model with taxa selection and infinite mixture still simply as the Dirichlet-tree multinomial mixtures (DTMM). The graphical model representation of DTMM is shown in \ref{fig:graph3}. Note that $G$ in (\ref{eq:dp}) is supported on a countable number of values since samples from a Dirichlet process are discrete, implying ties in the iid samples $(\bthe_i^\prime,  \btau_i^\prime) $'s and thus a clustering on $i$. This becomes clear with the stick-breaking construction of the Dirichlet process \citep{sethuraman1994constructive}, from which we can rewrite (\ref{eq:dt}) and (\ref{eq:dp}) as
\begin{equation}
\begin{aligned}
\bp_i\mid \bpi, \{ (\bthe^*_{k}, \btau^*_k)\}^\infty_{k=1} & \overset{\rm{iid}}{\sim}  \sum\limits^{\infty}_{k=1}\pi_k \text{DT}_{\T}(\bthe^*_{k}, \btau^*_k) \\
\pi_k  & =  v_k\prod\limits^{k-1}_{j=1}(1-v_j), \text{ where } v_1,v_2,\ldots\mid\beta \overset{\rm{iid}}{\sim} \text{Beta}(1, \beta)\\
(\bthe^*_{k}, \btau^*_k) & \overset{\rm{iid}}{\sim}  G_0(\bthe,\btau\mid\bgam,\tilde\bthe,\tilde\btau).
\end{aligned}
\end{equation}
For $i=1,\ldots,n$, let $c_i \in\mathbb{N}^+$ be the cluster label for the $i$-th sample such that $\bp_i\mid c_i, \{ (\bthe^*_{k}, \btau^*_k)\}^\infty_{k=1}\sim \text{DT}_{\T}(\bthe^*_{c_i}, \btau^*_{c_i})$. 
We can equivalently illustrate DTMM as in \ref{fig:graph4}. For comparison, we can introduce the latent cluster labels to DMM and DTMM without taxa selection 
and write their graphical model representations as in~\ref{fig:graph1} and~\ref{fig:graph2}. ~\ref{fig:graph2} and ~\ref{fig:graph4} illustrate how DTMM is generalized in this section.


\textit{Prior specification.} To complete the model specification, we need to choose $a_0(A)$, $b_0(A)$, $\theta_0(A)$, $\nu_0(A)$ and $F^A(\tau)$ for each $A\in\mI$. Ideally, informative prior knowledge shall be incorporated in choosing these parameters. If instead no prior knowledge is available, we treat these parameters (priors) as global such that they do not dependent on $A$ and remove the ``$(A)$'s" from the notations. 

For 
$\lambda$, we could set $a_0 = b_0 = 1$ such that $\lambda$ has a uniform distribution \textit{a priori}, which yields the following prior probability on the $\gamma(A)$'s \citep{scott2010bayes}: 
\begin{equation}
\begin{aligned}
\Pr(\bgam) = \frac{1}{M}{{M-1}\choose{\sum_{A\in\mI}\gamma(A)}}^{-1}.
\end{aligned}
\end{equation}
This prior allows multiplicity adjustment in the taxa selection. A default choice for $(\theta_0, \nu_0)$ is $(0.5, 1)$, which yields the Jeffrey's prior on $\theta^*_k(A)$ and $\theta^*(A)$. For $F(\gamma)$, any prior with a reasonably large support that covers a wide range of dispersion levels can be chosen. For example, we let $F(\tau)$ have density $f(\tau)=(\tau \times5\log10)^{-1}\mathbbm{1}_{(0.1\leq\tau\leq 10^4)}$, which is equivalent to putting the $\text{Unif}(-1, 4)$ prior on $\log_{10}\tau$. 
In our software, we use a discrete approximation of this prior induced by drawing $\log_{10}\tau$ uniformly from $\{-1, -0.5, 0, 0.5, 1, \ldots, 4 \}$.

\textit{Model behavior.} In our formulation, $\bthe$ and $\btau$ with a superscript ``$*$'' are cluster-specific parameters that govern the centroid and the within cluster variance of each cluster. $\bthe$ and $\btau$ with a ``$\sim$'' on top are parameters that determine the centroid and the variability of the shared base distribution. For $i=1,\ldots,n$, recall that $c_i \in\mathbb{N}^+$ is the cluster label for the $i$-th sample. Moreover, let $\bc = (c_1,c_2,\ldots, c_n)$, $\bc^*$ the set of distinct values in $\bc$ and $k^* = |\bc^*|$ the number of distinct clusters.  We note that the actual values of $c_i$ bear no significance and thus assume that the $c_i$'s take integer values between $1$ and $|\bc^*|$. At each node $A\in\mI$, $\gamma(A)$ serves as a selector: $\tilde\theta(A)$ and $\tilde\tau(A)$ become relevant only if $\gamma(A) = 0$. If $\gamma(A) = 1$, they are masked and not used by the model. We note that this masking happens at the level of the base distribution of the Dirichlet process mixture model. If $\gamma(A) = 0$, the base distribution $G^A_0$ is a point mass. Thus $(\theta^\prime_i(A), \tau^\prime_i(A))$'s must share the same value although $(\bthe_i^\prime,  \btau_i^\prime)$'s may not be the same. In a special case when $\gamma(A) = 0$ for all $A\in\mI$, the entire base distribution is a point mass and $(\bthe_i^\prime,  \btau_i^\prime)$'s are all the same. In this case, the cluster labels $\bc$ are only nominal---the samples are from a single cluster although it is possible that $|\bc^*| > 1$. Similarly, $\gamma(A)$ as an OTU selector is also nominal---$A$ is not necessarily relevant to clustering even if $\gamma(A) = 1$. In real applications, what we care are not these ``nominal'' parameters $\bc$ and $\bgam$ \textit{per se}, but their ``actual'' counterparts. Specifically, let $g_i \in \mathbb{N}^+$ be the ``actual'' cluster label of sample $i$, 
and let $s(A)\in\{0,1\}$ be the ``actual'' indicator of whether $A$ is relevant to clustering. 
Moreover, let $\bm{g} = (g_1,\ldots,g_n)$ and $\bm{s} = \{ s(A): A\in\mI\}$. We have 
\begin{equation}
\begin{aligned}
\bm{g} = \begin{cases}
\bc, & \text{ if }  \bgam\not = \bm{0}_{M-1} \text{ and } \bc\not=\bm{1}_n\\
\bm{1}_n, & \text{ if } \bgam = \bm{0}_{M-1} \text{ or } \bc=\bm{1}_n,
\end{cases} \text{, $\quad$  } \bm{s} = \begin{cases}
\bgam, & \text{ if } \bc\not=\bm{1}_n, \text{ and } \bgam\not = \bm{0}_{M-1}\\
\bm{0}_{M-1}, & \text{ if } \bc=\bm{1}_n \text{ or } \bgam = \bm{0}_{M-1}.
\end{cases}
\end{aligned}
\label{eq:g_trans}
\end{equation}
Unlike $\bc$ and $\bgam$, $\bm{g}$ and $\bm{s}$ are directly interpretable. For example, $A\in\mI$ is relevant to clustering if and only if $s(A) = 1$. In microbiome applications, it is typically expected that the samples have a latent clustering pattern. Therefore, it is common that $\bm{g}=\bc$ and $\bm{s} = \bgam$.
 

\subsection{Inference strategy}
\label{subsec:inference}

Under DTMM, we are interested in inferring the nominal cluster labels $\bc$ and the nominal activation indicator $\bgam$ from which the actual cluster labels $\bm{g}$ and the actual activation indicators $\bm{s}$ can be obtained. Let $\by_{-i}$ denote all observations other than $\by_i$. Bayesian inference for DTMM can be achieved by constructing a Markov chain that converges to the joint posterior of $(\bc, \bgam)$. Techniques for Dirichlet process mixture models, such as those described in \cite{neal2000markov} or \cite{ishwaran2001gibbs}, can be applied here. 

For $c\in\bc^*$, let $\bpsi_c^* = (\bthe^*_c, \btau^*_c)$ be the parameters that define the cluster indicated by $c$ (we also let $\bpsi_c^*(A) = (\theta^*_c(A), \tau^*_c(A))$ for $A\in\mI$). Similarly, let $\tilde\bpsi=(\tilde\bthe, \tilde\btau)$ be the shared parameters at the coupled nodes and $\tilde\bpsi(A)=(\tilde\theta(A), \tilde\tau(A))$ for $A\in\mI$. The set of unknown parameters in DTMM is $\{ \{\bthe_i, c_i\}_{i=1}^{n}, \{ \bpsi^*_c\}_{c=1}^{k^*}, \bgam, \tilde\bpsi, \beta, \lambda \}$. In this work, we construct a collapsed Gibbs sampler that iteratively samples from the joint posterior of $(\bc, \bgam, \beta, \lambda)$. The key to our inference strategy is to compute the marginal likelihoods of samples from a given cluster, integrating out both the sample-specific parameter $\bthe_i$ and the cluster-specific parameter $\bpsi^*_c$. 
This can be achieved numerically due to two facts. Firstly, the beta-binomial conjugacy makes it easy to integrate out the sample-specific compositional probabilities $\bthe_i$. Secondly, the tree-based decomposition of the Dirichlet distribution and the multinomial likelihood provides a divide-and-conquer strategy to marginalize out the high-dimensional cluster-specific parameters $\bpsi^*_c$ through performing a series of low-dimensional integrals at the internal nodes of the tree.

Specifically, for any $c\in\bc^*$, let $\bY_c^I = \{\by_i: c_i = c, i \in I \}$ be a set of samples in cluster $c$ where $I\subset [n]:=\{1,\ldots,n\}$. We also let $\bY_c = \bY_c^{[n]}$ be the set of all samples in cluster $c$ and $\bY_c^{-i} = \bY_c^{[n]\setminus \{i\} }$ be the set of samples in cluster $c$ excluding sample $i$. For $A\in\mI$, let $ \mL^A(\bY^I_c\mid \bpsi^*_c(A), \gamma(A), \tilde\bpsi(A) )$ be the marginal likelihood of $\bY^I_c$ at node $A$ by marginalizing out the sample-specific parameters. The beta-binomial conjugacy yields
\begin{equation}
\begin{aligned}
&\hspace{5mm} \mL^A(\bY^I_c\mid \bpsi^*_c(A), \gamma(A), \tilde\bpsi(A) ) \\
& =   \prod\limits_{\{i\in I:c_i=c\}}  {y_i(A) \choose y_i(A_l)} \frac{B(\theta^*_c(A)\tau^*_c(A) + y_i(A_l), (1 - \theta^*_c(A))\tau^*_c(A) + y_i(A_r))}{B(\theta^*_c(A)\tau^*_c(A), (1 - \theta^*_c(A))\tau^*_c(A))}.
\end{aligned}
\label{eq:}
\end{equation}
We then further integrate out $\bpsi^*_c(A)$ to obtain the marginal likelihood of $\bY^I_c$ at node $A$ given only the activation indicators and the base parameters:
{\small\begin{equation}
\begin{aligned}
 \mL_1^A(\bY^I_c) & :=   \iint   \mL^A(\bY^I_c\mid \bpsi^*_c(A), \gamma(A) = 1, \tilde\bpsi(A) ) d\Pi(\bpsi^*_c(A) \mid \gamma(A)=1, \tilde\bpsi(A)) \\
&  =   \iint    \prod\limits_{\{i\in I:c_i=c\}}  {y_i(A) \choose y_i(A_l)} \frac{B(\theta(A)\tau(A) + y_i(A_l), (1 - \theta(A))\tau(A) + y_i(A_r))}{B(\theta(A)\tau(A), (1 - \theta(A))\tau(A))}  \\
& \hspace{1.3cm} \times  \frac{\theta(A)^{\theta_0(A)\nu_0(A) - 1} (1 - \theta(A))^{(1 - \theta_0(A))\nu_0(A) - 1}}{B(\theta_0(A)\nu_0(A), (1 - \theta_0(A))\nu_0(A))}d\theta(A) dF^A(\tau)  ,
\end{aligned}
\label{eq:m1now}
\end{equation}  
\begin{equation}
\begin{aligned}
 \mL_0^A(\bY^I_c \mid \tilde\bpsi(A)) & :=    \iint   \mL^A(\bY^I_c\mid \bpsi^*_c(A), \gamma(A) = 0, \tilde\bpsi(A) ) d\Pi(\bpsi^*_c(A) \mid \gamma(A)=0, \tilde\bpsi(A)) \\
 & =\prod\limits_{\{i\in I:c_i=c\}}  {y_i(A) \choose y_i(A_l)} \frac{B(\tilde\theta(A)\tilde\tau(A) + y_i(A_l), (1 - \tilde\theta(A))\tilde\tau(A) + y_i(A_r))}{B(\tilde\theta(A)\tilde\tau(A), (1 - \tilde\theta(A))\tilde\tau(A))} .
\end{aligned}
\end{equation}}
 
Integrals in (\ref{eq:m1now}) are two-dimensional integrals that are easy to evaluate numerically. In comparison, to perform a fully Bayesian inference for DMM, the high dimensional cluster centroids $\balp_k$'s in (\ref{eq:dmm}) has to either be integrated out directly or be sampled in the MCMC procedure. With these marginal likelihoods, we can construct our Gibbs sampler for posterior inference. Details on deriving and implementing the Gibbs sampler are given in Section 1.1 and 1.2 in the supplementary material. 


 After running the chain for $T$ iterations, we discard the first $B$ samples as burn-in and obtain $(T-B)$ posterior samples denoted as {\small $\big[\{\bc^{(B+1)},\bgam^{(B+1)},\beta^{(B+1)},\lambda^{(B+1)}\}, \ldots, \{\bc^{(T)},\bgam^{(T)},$ $\beta^{(T)},\lambda^{(T)}\}\big]$}. Based on these posterior samples, we can compute the posterior samples for $\bm{g}$ and $\bm{s}$ based on (\ref{eq:g_trans}). We denote these posterior samples as $[\{\bm{g}^{(B+1)}, \bm{s}^{(B+1)} \}, \ldots, \{ \bm{g}^{(T)}, \bm{s}^{(T)} \}]$

For each sample $\bm{g}^{(t)}$, let $\Gamma^{(t)}$ be the corresponding $n\times n$ association matrix whose $(i_1,i_2)$ element is $1$ if $g^{(t)}_{i_1} = {g}^{(t)}_{i_2}$ and $0$ otherwise. Element-wise average of $\Gamma^{(B+1)},\ldots,\Gamma^{(T)}$ provides an estimation $\hat\Pi$ of the pairwise clustering probability matrix $\Pi$ whose $(i_1,i_2)$ element is $\Pr(\by_{i_1} \text{ and } \by_{i_2} \text{ in the same cluster}).$ To yield a representative clustering, we can report the least-squares model-based clustering \citep{dahl2006model}, defined as
\begin{equation}
\begin{aligned}
\bm{C}_{LS} = \argmin\limits_{\{ \bm{g}^{(t)}:B < t\leq T\} } \sum\limits_{1\leq i_1\leq n}\sum\limits_{1\leq i_2\leq n} (\Gamma^{(t)}_{i_1i_2} - \hat\Pi_{i_1i_2})^2.
\end{aligned}
\label{eq:cls}
\end{equation}
$\bm{C}_{LS}$ has the advantage that it incorporates information from all posterior samples while output one of the observed clustering in the Markov Chain \citep{dahl2006model}. Other representative clusterings such as the MAP clustering or the clustering given by the last iteration are also frequently used.

Given any representative clustering and the corresponding activation indicators, we can portray the cluster centroids by computing the posterior means of the cluster-specific parameters. Details are provided in Section 1.3 in the supplementary material. We also note that the DTMM framework can also be used in the supervised setting to achieve sample classification based on a training microbiome dataset. Details of classification under the DTMM framework can be found in Section 1.4 in the online supplementary material.


\section{Numerical experiments}

\label{sec:num_exam}

\subsection{Simulation studies}

We first carry out a series of simulation studies to evaluate the performance of DTMM and compare it to several other methods for clustering microbiome count data---namely, the Dirichlet multinomial mixtures (DMM) \citep{holmes2012dirichlet}, the $k$-means algorithm (K-ms) \citep{lloyd1982least}, the partitioning around medoids algorithm (PAM) \citep{kaufman2009finding}, hierarchical clustering (Hclust) \citep{kaufman2009finding}, and spectral clustering (Spec) \citep{ng2002spectral}.

\subsubsection{Simulation setup}

In the numerical examples, we simulate datasets with $n$ samples and six OTUs. 
In each dataset, the $n$ samples are denoted as $\by_i = (y_{i1},\ldots, y_{i6})$, $i=1,\ldots, n$, which are generated from the following model:
\begin{equation}
\begin{aligned}
\by_i \mid N_i, \bp_i  &\overset{\rm{ind}}{\sim}  \text{Multi}(N_i, \bp_i) \quad \text{and} \quad
\bp_i \overset{\rm{ind}}{\sim} \sum\limits^{K}_{k=1}\pi_k \cdot H_k(\bp_i \mid \bbet_k),
\end{aligned}
\label{eq:sec_3_dg}
\end{equation}
where the mixture kernel $H_k(\bp_i \mid \bbet_k)$ is a distribution on the 5-simplex with parameter $\bbet_k$, $N_i  \overset{\rm{iid}}{\sim} \text{Neg-Binom}(m, s)$. We take the tree in \ref{fig:sec_2_tree} as the ``phylogenetic tree'' over the six OTUs and 
consider 5 different simulation scenarios by choosing different mixture kernels $H_k(\bp_i \mid \bbet_k)$ in (\ref{eq:sec_3_dg}). In each scenario, we let $n = 90$ or $180$, $K=3$ and $(\pi_1, \pi_2, \pi_3) = (\frac{4}{9}, \frac{3}{9},\frac{2}{9})$. Parameters for the negative-binomial distribution are chosen as $m = 15000$, $s = 20$ such that the generated total counts has mean $15000$ and standard deviation $3346$, with $95\%$ of them fall into the range $(9158, 22258)$. In the 5 simulation scenarios, the mixture kernels are chosen as shown in \ref{tab:data_generate}. Details for the simulation setups can be found in Section 2.1 in the supplementary material.
 
\begin{table}[!ht]
\caption{Mixture kernels for generating the simulated datasets.}
\label{tab:data_generate}
\centering \resizebox {\textwidth} {!}{
\begin{tabular}{@{}clccclc@{}}
\toprule
                   & \multirow{2}{*}{Kernel} & \multicolumn{2}{c}{Signal} &  &  \multicolumn{2}{c}{\multirow{2}{*}{$\bbet_k$}} \\ \cmidrule(lr){3-4}
                   &                         & Level      & Parameter  &    & \multicolumn{2}{c}{}                  \\ \midrule
\multirow{3}{*}{$\RN{1}$} & \multirow{3}{*}{DT}     & W          &    $\alpha = 1$    &   & \multirow{3}{*}{ \hspace{2mm}\resizebox {0.19\textwidth} {!}{
\begin{tikzpicture}[sibling distance=6em, level distance = 3.7em,
  every node/.style = {shape=rectangle, rounded corners,
     align=left, 
    top color=white, bottom color=white}]]
  \node {$(12\alpha, 12\alpha)$}
    child { node[draw] { $\bnu_1 = (10\alpha, 2\alpha)$ \\ $\bnu_2 = (6\alpha, 6\alpha)$ \\ $\bnu_3 = (2\alpha, 10\alpha)$} }
    child { node {$(8\gamma, 4\gamma)$}
      child { node {$(4\gamma, 4\gamma)$}}
      child { node {$(2\gamma, 2\gamma)$} } };
\end{tikzpicture} 
}
} &  \multirow{3}{*}{ \hspace{-15mm}$\gamma = 0.1$}     \\ \cmidrule(lr){3-4}
                   &                         & M          &     $\alpha =3$     &      &              &       \\ \cmidrule(lr){3-4}
                   &                         & S          &     $\alpha =6$      &     &              &                \\  \midrule
\multirow{3}{*}{$\RN{2}$} & \multirow{3}{*}{Dir}    & W          &       $\alpha_0 = 1$    &      & \multicolumn{2}{c}{$\balp_1 = (2,2,5,2,3,1)\cdot \alpha_0$}                  \\ \cmidrule(lr){3-4}
                   &                         & M          &    $\alpha_0 = 3$      &       & \multicolumn{2}{c}{$\balp_2 = (2,4,3,2,1,3)\cdot \alpha_0$}                  \\ \cmidrule(lr){3-4}
                   &                         & S          &     $\alpha_0 = 6$     &        & \multicolumn{2}{c}{$\balp_3 = (2,6,1,2,2,2)\cdot \alpha_0$}                  \\ \midrule
\multirow{3}{*}{$\RN{3}$} & \multirow{3}{*}{LN}   & W          &   $\alpha = 3$      &        &    \multirow{3}{*}{ \hspace{5mm} \resizebox {0.17\textwidth} {!}{\thead{ $q_k = \text{DT}_{\T_6}(\bnu_k;\alpha; 0.5)$ \\ $\bnu_1=(10\alpha,2\alpha)$ \\$\bnu_2=(6\alpha, 6\alpha)$ \\ $\bnu_3=(2\alpha, 10\alpha)$ } }   }               &    \multirow{3}{*}{ \hspace{-10mm} \resizebox {0.18\textwidth} {!}{ \thead{$\bmu_k = \mathbb{E}_{q_k}\left[\log\left(\frac{\bx_{-6}}{x_6} \right)\right]$ \\ \\$\bSigma_k = \mathbb{V}_{q_k}\left[\log\left(\frac{\bx_{-6}}{x_6} \right)\right]$} }   }            \\ \cmidrule(lr){3-4}
                   &                         & M          &   $\alpha = 6$            &         &         &                   \\ \cmidrule(lr){3-4}
                   &                         & S          &     $\alpha = 9$          &         &       &                   \\ \midrule
\multirow{3}{*}{$\RN{4}$} & \multirow{3}{*}{LN}   & W          &     $a=5,b=3$      &   &      $\bmu_1 = (3,1,a,b,0) $             &   \multirow{3}{*}{{\resizebox {0.2\textwidth} {!}{$\Sigma_{1,2,3} = {\small\begin{pmatrix}
   0.05 & & & &  \\
   & 0.05 & & &\\
   & & 1 & &  \\
   & &  & 1 & \\
   & & & & 1
\end{pmatrix}} $ }}  }            \\ \cmidrule(lr){3-4}
                   &                         & M          &   $a=2,b=2$      &       &           $\bmu_2 = (2.43,2.43,a,b,0)$ ,           &                   \\ \cmidrule(lr){3-4}
                   &                         & S          &    $a=1,b=1$      &      &            $\bmu_3 = (1, 3 ,a,b,0) $         &                   \\ \midrule
\multirow{3}{*}{$\RN{5}$} & \multirow{3}{*}{LN}   & W          &     $c = 6, d = 6$   &       &   $\bmu_1 = (c, d, 3.5, 3, 2.5)$                  &         \multirow{3}{*}{{\resizebox {0.2\textwidth} {!}{$\Sigma_{1,2,3} = {\small\begin{pmatrix}
   1 & & & &  \\
   & 1 & & &\\
   & & 0.05 & &  \\
   & &  & 0.05 & \\
   & & & & 0.05
\end{pmatrix}} $ }}  }             \\ \cmidrule(lr){3-4}
                   &                         & M          &  $c = 3, d = 3$      &         &   $\bmu_2 = (c, d, 2.5, 3.5, 3)$ ,                 &                   \\ \cmidrule(lr){3-4}
                   &                         & S          &   $c = 1, d = 1$      &        &     $\bmu_3 = (c, d, 3, 2.5, 3.5)$              &                   \\ \bottomrule
\end{tabular} }
\end{table}


In each scenario, a ``null'' case is also considered by setting $K = 1$ in the case with the medium signal level. For each $(\text{kernel}, \text{signal level})$ combination, we conduct 100 rounds of simulations. For each simulated dataset with $K=3$, we calculate the following $R^2$ as a measure of the strength of the signal \citep{anderson2001new}:
\[
R^2 =\frac{\text{SSW}}{\text{SST}} =  \frac{\sum\limits^{3}_{k=1}\sum\limits^{N-1}_{i = 1}\sum\limits^{N}_{j = i + 1}d_{\text{BC}}(\by_i, \by_j)^2\epsilon^k_{ij}/n_k}{\sum\limits^{N-1}_{i = 1}\sum\limits^{N}_{j = i + 1}d_{\text{BC}}(\by_i, \by_j)^2/n},
\]
where $d_{\text{BC}}(\cdot, \cdot)$ is the Bray-Curtis dissimilarity, $n_k$ the number of samples in cluster $k$, $\epsilon^k_{ij} = 1$ if the samples $i$ and $j$ are both in cluster $k$ and $0$ otherwise. For example, the average $R^2$'s of the 100 simulated datasets in each experiment are reported in \ref{tab:rmse_90} for $n=90$.

In each simulation round, we run the Gibbs sampler for DTMM for 2000 iterations and discard the first half of the chain as burn-in. The priors and hyper-parameters for DTMM are set to the recommended choice in Section~\ref{subsec:node}. The initial values for the clustering labels in the Markov chain are set to the labels of running the $k$-means algorithm with $k=5$. For DTMM, we output $\bm{C}_{LS}$ as a representative clustering. For PAM and Hclust, we use the Bray-Curtis dissimilarity on the relative abundance as the underlying distance measure between samples. For all competitors other than DMM, the number of clusters is required as a tuning parameter, we set this parameter to the truth 3 when running these methods.


\subsubsection{Analyses}

To compare the performance of different methods, we compute the Jaccard index \citep{jaccard1912distribution} between the clusters obtained by each method and the truth. For a specific clustering $\bc$ and the true clustering $\bc_0$, the Jaccard index between $\bc$ and $\bc_0$ is defined as $J(\bc, \bc_0) = \mathcal{N}_{\bc\cap\bc_0}/\mathcal{N}_{\bc\cup\bc_0}$, where $\mathcal{N}_{\bc\cap\bc_0}$ is the number of pairs of samples that are in the same cluster under both $\bc$ and $\bc_0$, $\mathcal{N}_{\bc\cup\bc_0}$ the number of pairs of samples that are in the same cluster under at least one of $\bc$ and $\bc_0$. When $\bc$ gives the same clustering as $\bc_0$, $J(\bc, \bc_0) = 1$. In each simulation scenario, we compare the root mean squared error of each method $m$: $\text{RMSE}^{(m)} = \sqrt{\sum^{100}_{r = 1}[J(\bc_r^{(m)}, \bc_0) - 1]^2/100}$, where $\bc_r^{(m)}$ is the clustering obtained by method $m$ in simulation round $r$. As some references, let $\bc_0 = (1\cdot\bm{1}_{40}^\top, 2\cdot\bm{1}_{30}^\top,3\cdot\bm{1}_{20}^\top)$, $\bc_1 = (1\cdot\bm{1}_{90}^\top)$, $\bc_2 = (1\cdot\bm{1}_{30}^\top, 2\cdot\bm{1}_{30},  3\cdot\bm{1}_{30})$ and $\bc_3 = (1\cdot\bm{1}_{40}^\top, 2\cdot\bm{1}_{50})$, where $\bm{1}_n$ is the $n$-dimensional vector with all element equal to 1. We have $\sqrt{[J(\bc_1, \bc_0) - 1]^2} = 0.65$, $\sqrt{[J(\bc_2, \bc_0) - 1]^2} = 0.50$ and $\sqrt{[J(\bc_3, \bc_0) - 1]^2} = 0.30$. The RMSE of DTMM and the competitors under all simulation scenarios when $n=90$ is shown in \ref{tab:rmse_90}. The RMSE table for $n=180$ as well as boxplots of the Jaccard index reported by each method in different simulation scenarios can be found in Table S1, Figure S3 and Figure S4 in Section 2.3 of the supplementary material.


\begin{table}[!ht]
\caption{RMSE of the Jaccard index (small sample size). Cells with the lowest RMSE in each row are highlighted.}
\label{tab:rmse_90}
\centering \resizebox {\textwidth} {!}{
\begin{tabular}{@{}llccllp{30pt}p{30pt}p{30pt}p{30pt}p{30pt}c@{}}
\toprule
 \multicolumn{11}{c}{$n = 90$}  \\\toprule
\multicolumn{2}{c}{}                        & \multicolumn{2}{c}{Signal} &  & \multicolumn{6}{c}{Method}       \\ \cmidrule(lr){3-5}\cmidrule(lr){6-11} 
\multicolumn{2}{c}{\multirow{-2}{*}{Expt}}  & Level                & $R^2$ &  & {\color[HTML]{111111} DTMM} & DMM & K-ms & PAM & Hclust & Spec \\ \midrule
&                       &   --   & --    &    & {\bf 0.43}   &  0.51     &   \multicolumn{1}{c}{--}     &     \multicolumn{1}{c}{--}      &    \multicolumn{1}{c}{--}   &    \multicolumn{1}{c}{--}                  \\
&                       & W                    &  0.30    &  &  \bf{0.56}     &   0.64   &    0.67     &   0.71   &  0.65      &       0.71       \\
&                       & M                     &  0.35   &  & {\bf 0.33}     &  0.65   &    0.69     &  0.69  &   0.64     &     0.71          \\
\multirow{-4}{*}{\RN{1}}&\multirow{-4}{*}{DT}  & S                    &  0.37  &  & \bf{0.17}     &    0.65  & 0.69     & 0.71    & 0.65       &    0.70          \\ \midrule
&                       & --    & --    &     &   0.35   & {\bf 0.00}  &   \multicolumn{1}{c}{--}        &   \multicolumn{1}{c}{--}    &      \multicolumn{1}{c}{--}    &    \multicolumn{1}{c}{--}     	   \\
&                       & W                    &  0.35   &  & \bf{0.53}     & {\bf 0.53}    &  0.55       &  0.59   &   0.58     &    0.57          \\
&                       & M                    &  0.52   &  & {\bf  0.18}     &  0.30   &  0.37       &  0.33   &    0.37    &   0.33            \\
\multirow{-4}{*}{\RN{2}}&\multirow{-4}{*}{Dir} & S                    &  0.60  &   & {\bf 0.04}     & 0.09   &  0.32       &  0.19   &   0.38     &     0.22          \\ \midrule
&                       & --   & --   &       &  0.46                 &  {\bf 0.06}   &   \multicolumn{1}{c}{--}     &   \multicolumn{1}{c}{--}    &     \multicolumn{1}{c}{--}     &       \multicolumn{1}{c}{--}          \\
&                       & W                    &   0.37  &  & {\bf 0.49}     & 0.64    &   0.53      & 0.53    &    0.54   &   0.54           \\
&                       & M                     &   0.38   &  & {\bf 0.23 }     &   0.64  &     0.50    &  0.46   &    0.55    & 0.47              \\
\multirow{-4}{*}{\RN{3}}&\multirow{-4}{*}{LN-A} & S                 &  0.39   &  & {\bf 0.10 }   &   0.64  & 0.48        &  0.44   &  0.53      &   0.46            \\ \midrule
&                       & --  &  --   &       &  0.60                  & {\bf 0.54}    &    \multicolumn{1}{c}{--}       &    \multicolumn{1}{c}{--}   &    \multicolumn{1}{c}{--}      &           \multicolumn{1}{c}{--}      \\
&                       & W                    &  0.10  &  & {\bf 0.35}     & 0.72   &    0.77     &  0.78   &   0.73    &   0.74        \\
&                      & M                     &  0.41    &  & {\bf 0.21}     & 0.54     &  0.59       &  0.54   &   0.60     &      0.53         \\
\multirow{-4}{*}{\RN{4}}&\multirow{-4}{*}{LN-S} & S                  &   0.60  &  &  {\bf 0.17}      &    0.37  &  0.36       &  0.24   &   0.41  &     0.27           \\ \midrule
&                       & --  & --   &        & {\bf 0.41}                          &   0.61  &    \multicolumn{1}{c}{--}     &   \multicolumn{1}{c}{--}  &  \multicolumn{1}{c}{--}      &       \multicolumn{1}{c}{--}        \\
&                       & W                     &   0.04  &  & {\bf 0.20}     &  0.78   &   0.78      &  0.79   &     0.78   &  0.76              \\
&                       & M                     &   0.23  &  & {\bf 0.14 }    &   0.65  &   0.76      &   0.70    &    0.74    &   0.68            \\
\multirow{-4}{*}{\RN{5}} &\multirow{-4}{*}{LN-M} & S                 &  0.53   &  & {\bf  0.17}     &  0.49   &  0.22       & 0.20   &   0.39     &    0.22           \\ \bottomrule
\end{tabular} }
\end{table}

When $K = 3$, DTMM is always one of the top two methods under comparison. When it is not the best method, its performance is close to the best. Without utilizing the information provided by the phylogenetic tree, all competitors of DTMM suffer when the signal is weak or medium. Moreover, these competitors rely on global distance measures between samples and treat the six OTUs equivalently. As a result, in scenario like $\RN{1}$ and $\RN{4}$ where the signal is local to a single internal node of the phylogenetic tree, these methods have poor performance. Even in scenario $\RN{5}$ where half of the OTUs are relevant for clustering, these methods still suffer unless the signal is very strong. In scenario $\RN{2}$ where the signal is global, all methods perform reasonably well. In this scenario, DTMM can outperform DMM when $n=90$ even the latter is the true model. This is because DMM relies on a Laplace approximation to a six dimensional integral when computing the marginal likelihoods to choose the number of clusters. When the sample size is small, DMM tends to choose less than three clusters due to the poor approximation. When $n=180$, DMM is more likely to choose the right number of clusters even with the inaccurate approximation. Thus the performance of DMM improves significantly with more samples. Our experience suggests that DMM tends to underestimate the number of clusters in most cases. For example, in scenario $\RN{1}$ and $\RN{3}$, DMM simply puts all samples in a same cluster when $n=90$. 

In our simulation settings, there are two factors that determine the effect of the increase of sample size on the performance of the two model-based clustering methods. On the one hand, since more samples are available per cluster, the models have a better chance to capture the cluster centroids well once they identify the correct number of clusters. On the other hand, more samples makes it harder to get the number of clusters right. These two fighting forces together determine the overall performance shift of the two model-based methods, yet which force prevails is unclear. For DTMM, when the model is mis-specified (as in scenarios $\RN{3}$, $\RN{4}$ and $\RN{5}$), the model tends to identify too many small clusters, resulting in a worse overall performance. For the distance-based clustering methods, these two factors play no role since we assume that the number of clusters is known. In general, our observations suggest that these methods benefit a little from more samples when the signal is strong. Among the distance-based methods, PAM and Spec have a better overall performance. We thus recommend using these two methods to help choose the initial values of DTMM.

\begin{figure}[!ht]
\begin{center}
\includegraphics[width = 0.95\textwidth]{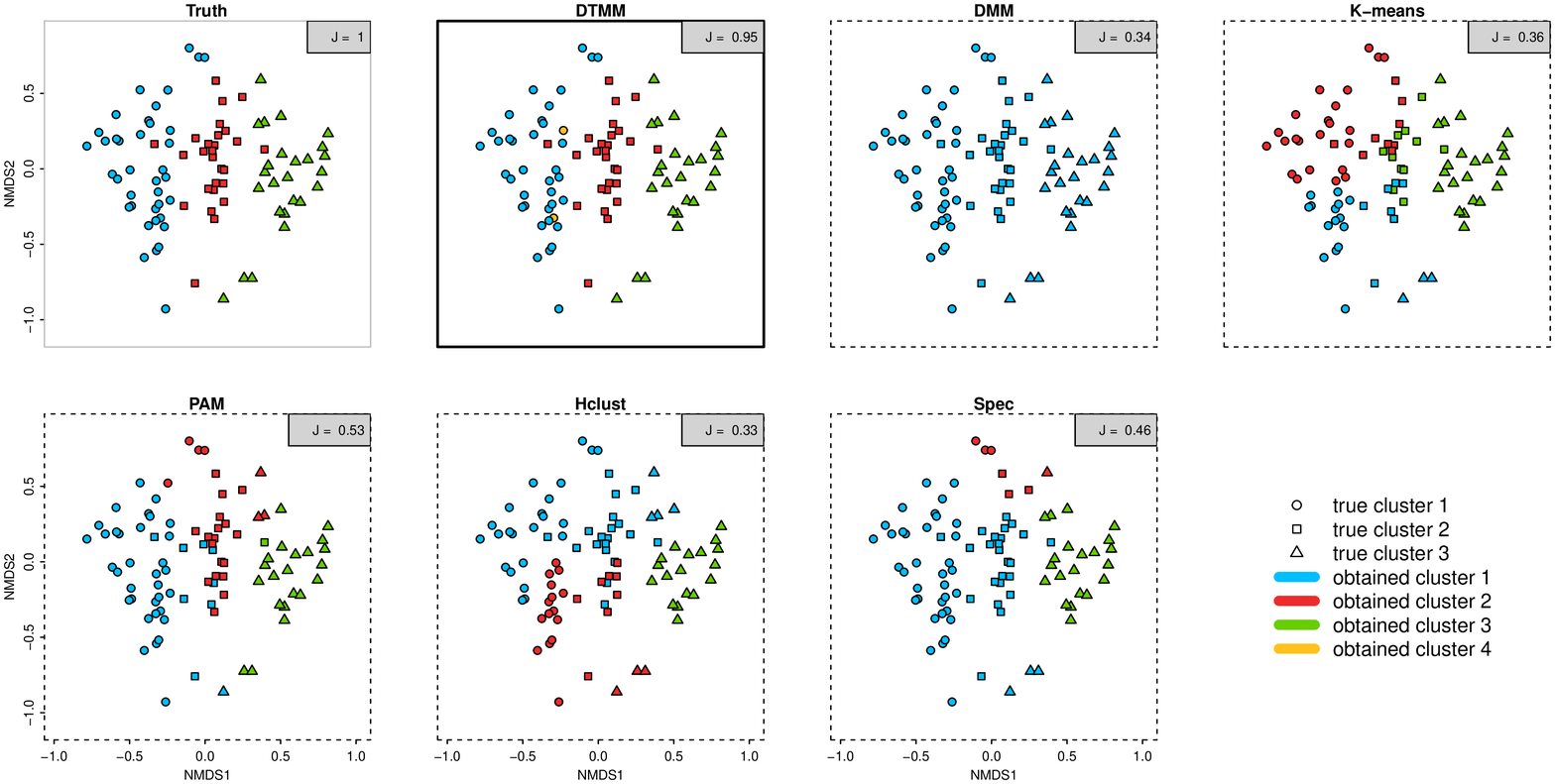}
\caption{2D NMDS plot of samples in a simulation round in scenario $\RN{4}$ ($n = 90$, medium noise level). In each sub-plot, the true clustering is indicated by the shape of the points while the clustering obtained is indicated by the color.}
\label{fig:sec_3_nmds_single}
\end{center}
\end{figure}

 \begin{figure}[h!]
 \centering
\resizebox {\textwidth} {!}{
  \begin{minipage}[b]{0.52\textwidth}
    \centering
   \includegraphics[width = 1\textwidth]{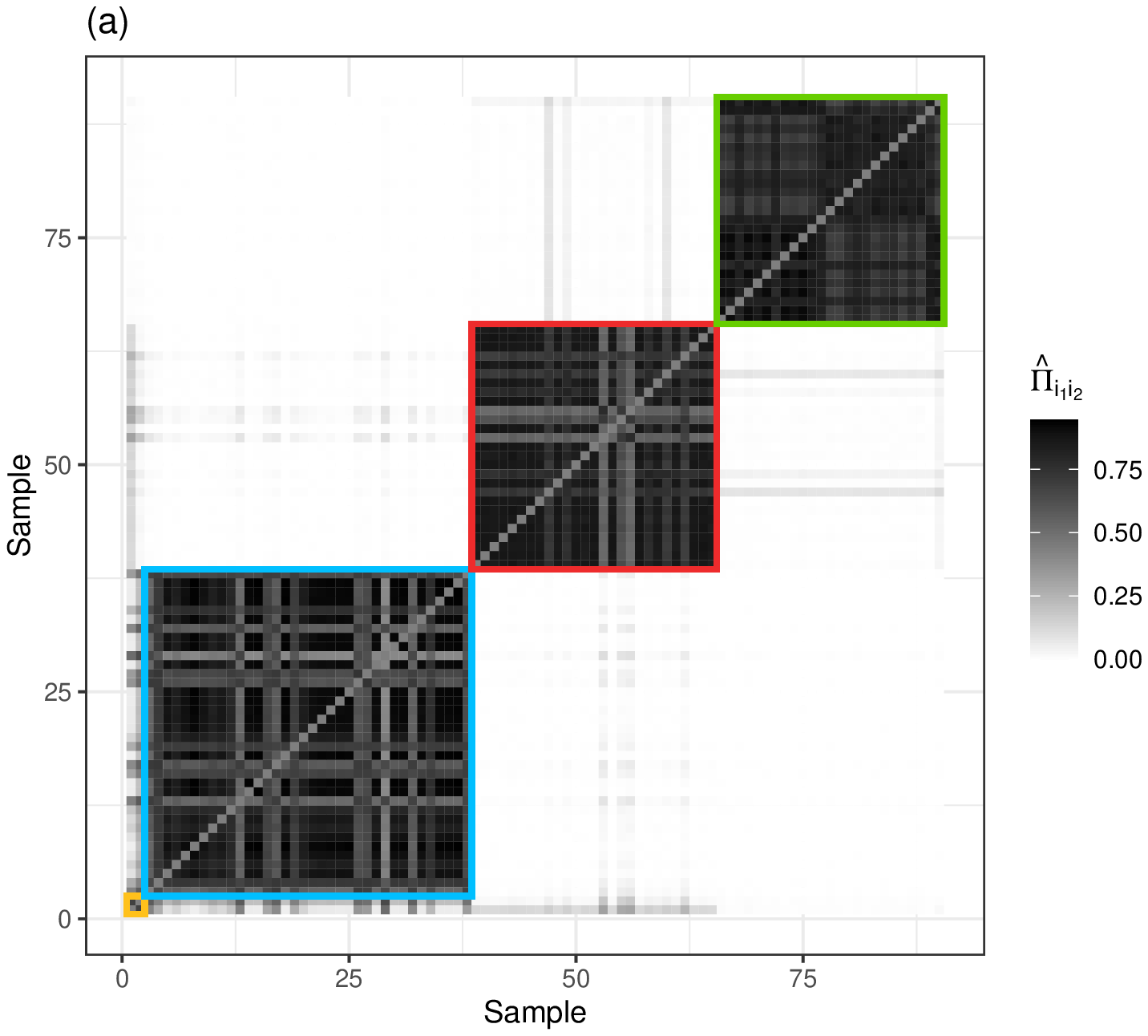} 
 \end{minipage}
 \begin{minipage}[b]{0.235\textwidth}
{\tiny(b).}
    \centering
   \includegraphics[width = 3.36cm, height =  5.6cm]{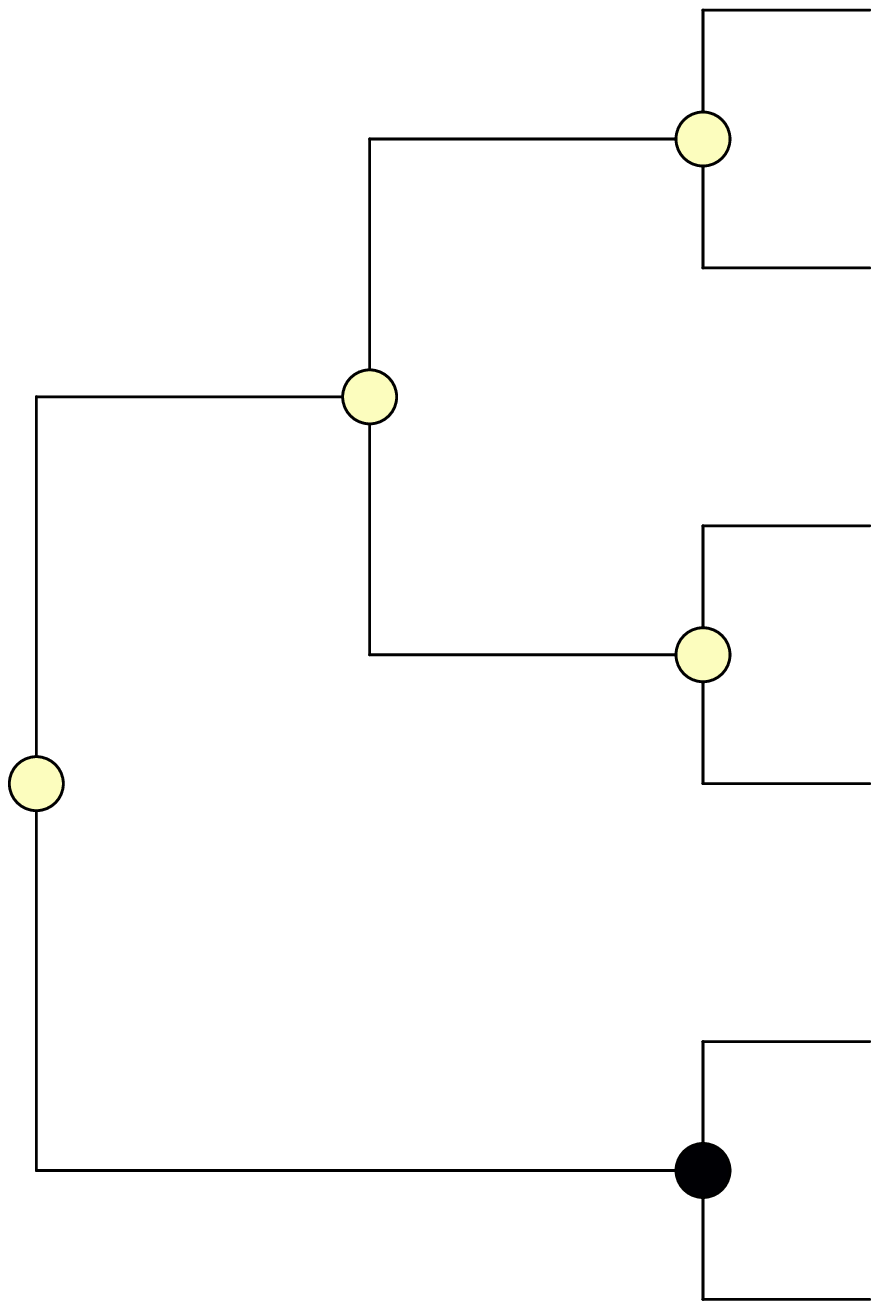} 
   \vspace{0.2cm}
    \end{minipage}\hspace{-0.4cm}
\begin{minipage}[b]{0.25\textwidth}
   \includegraphics[width = 3.5cm, height = 6.125cm]{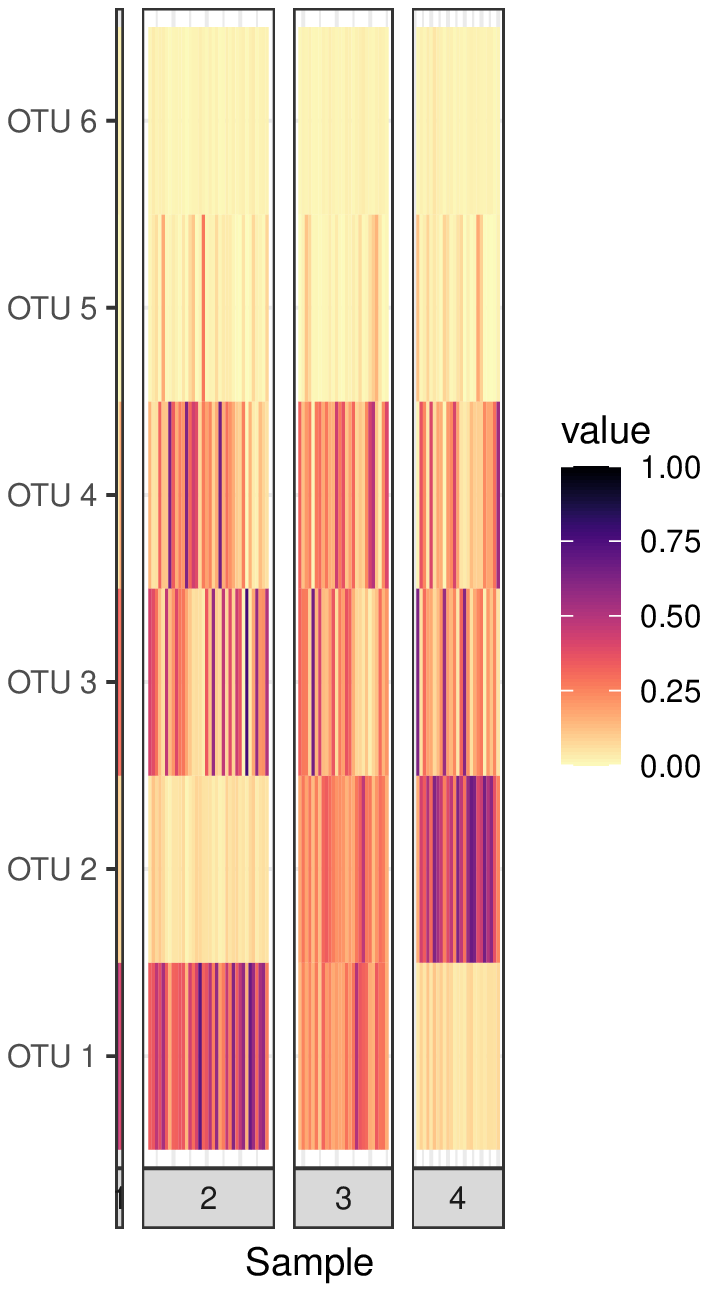}
  \end{minipage}   }
 \caption{Illustrations for an example from simulation scenario $\RN{4}$. (a): Probability of two samples being clustered together by DTMM based on 1000 post-burnin MCMC samples. The samples are ordered by their cluster labels from DTMM. The clusters identified by DTMM are highlighted by squares colored as in \ref{fig:sec_3_nmds_single}. (b): An illustration of the node selection property of DTMM. The nodes are colored by their estimated posterior node activation probabilities. The heatmap plots the relative abundance of the samples grouped by their cluster labels from DTMM.}
\label{fig:sec_3_single}
\end{figure}

We next zoom in to an example to further study the properties of DTMM. In this example, we consider a specific simulation round in scenario $\RN{4}$ with the medium noise level ($n=90$). \ref{fig:sec_3_nmds_single} shows the 2D NMDS plot of the samples colored by the clustering obtained by each method. In this example, the clustering is roughly determined only by the first NMDS axis. With the node selection module, DTMM is capable of picking the relevant dimensions and clustering efficiently. As for a representative clustering, DTMM finds 4 clusters, with one falsely identified cluster containing only two samples. This is consistent with the well-known fact that inference based on Dirichlet process mixture models can identify small clusters that do not reflect the true data-generating process \citep{miller2013simple}. One feature that differs DTMM from its competitors is that it not only outputs a representative clustering, but also a whole MCMC trajectory that allows natural uncertainty quantifications. \ref{fig:sec_3_single} (a) shows the probability of two samples being clustered together by DTMM. Clearly, three stable clusters are identified. Although DTMM falsely puts the first two samples in a separate cluster, the uncertainty is large. 
\ref{fig:sec_3_single} (b) shows the relative abundance of the samples as well as the estimated posterior node activation probabilities. In this example, DTMM is able to uncover the internal nodes that are relevant for clustering. We also consider an example from simulation scenario $\RN{5}$. Illustrations similar to \ref{fig:sec_3_nmds_single} and \ref{fig:sec_3_single} can be found in Figure S5 and S6 in Section 2.3 of the supplementary material. 

 
 \subsection{Validation}

Validating the results of unsupervised learning is often challenging. In microbiome clustering analyses, the best practice is to check the resulting clusters with scientists to gain biological insights on a case by case basis. Instead of trying to provide a general solution of how to justify the clusters found by DTMM, we provide an example to show that DTMM can identify biologically meaningful clusters in real microbiome applications. 

\begin{figure}[!ht]
\begin{center}
\includegraphics[width = 0.9\textwidth]{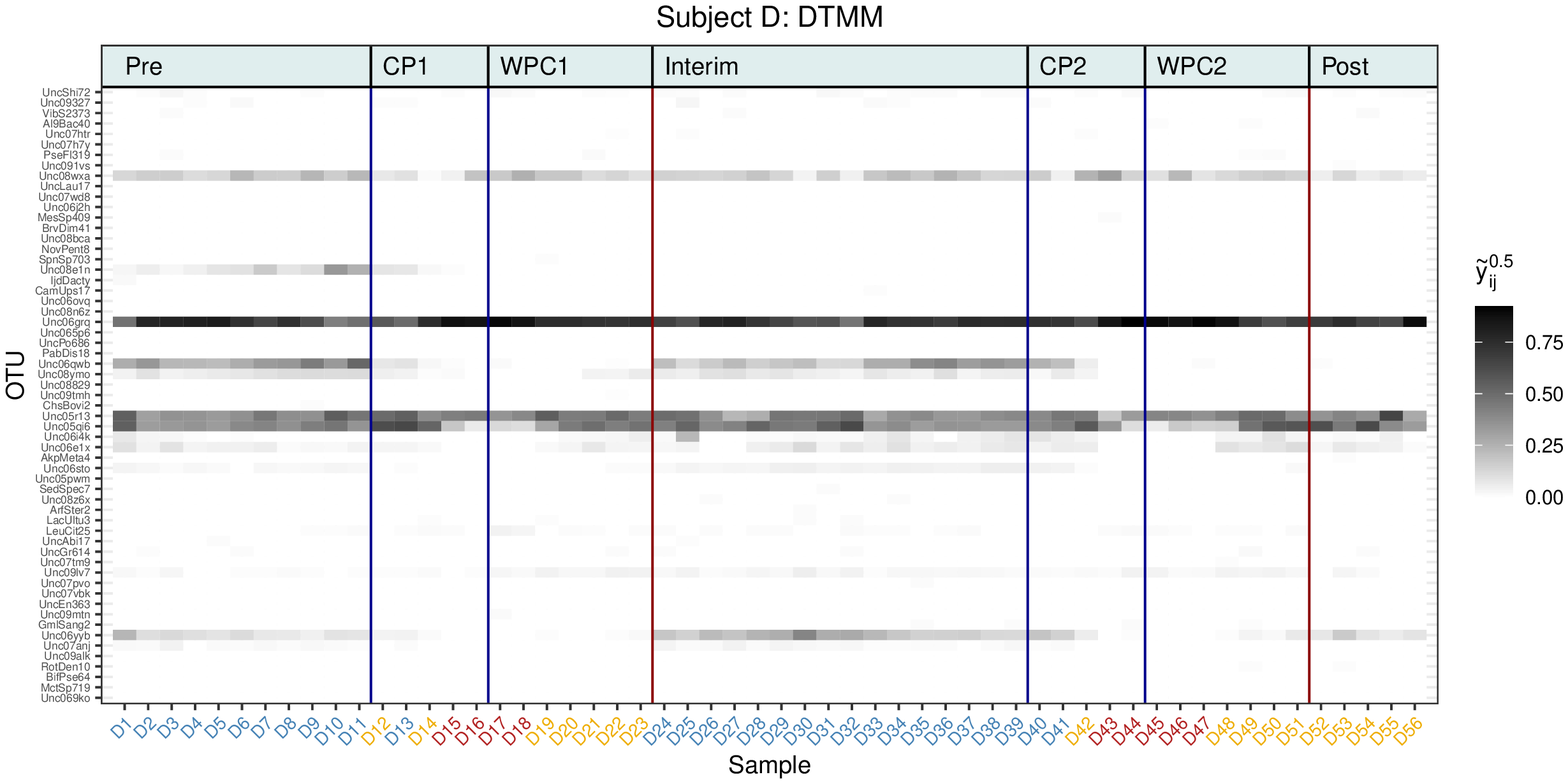} 
\caption{The heatmap of the microbiome samples of patient D (after the square-root transform). Each column represents a specific sample. The columns are ordered by the times the samples were collected. The colors of the $x$-axis labels represent the clustering labels of the samples returned by DTMM. The blue vertical lines mark the two antibiotic treatment courses. "CP" denotes the antibiotic treatment (ciprofloxacin); "WPC" is the week post treatment; "Pre" and "Post" denote the pre-treatment and post-treatment periods.}
\label{fig:sec_val_d}
\end{center}
\end{figure}

\begin{figure}[!ht]
\begin{center}
\includegraphics[width =0.9 \textwidth]{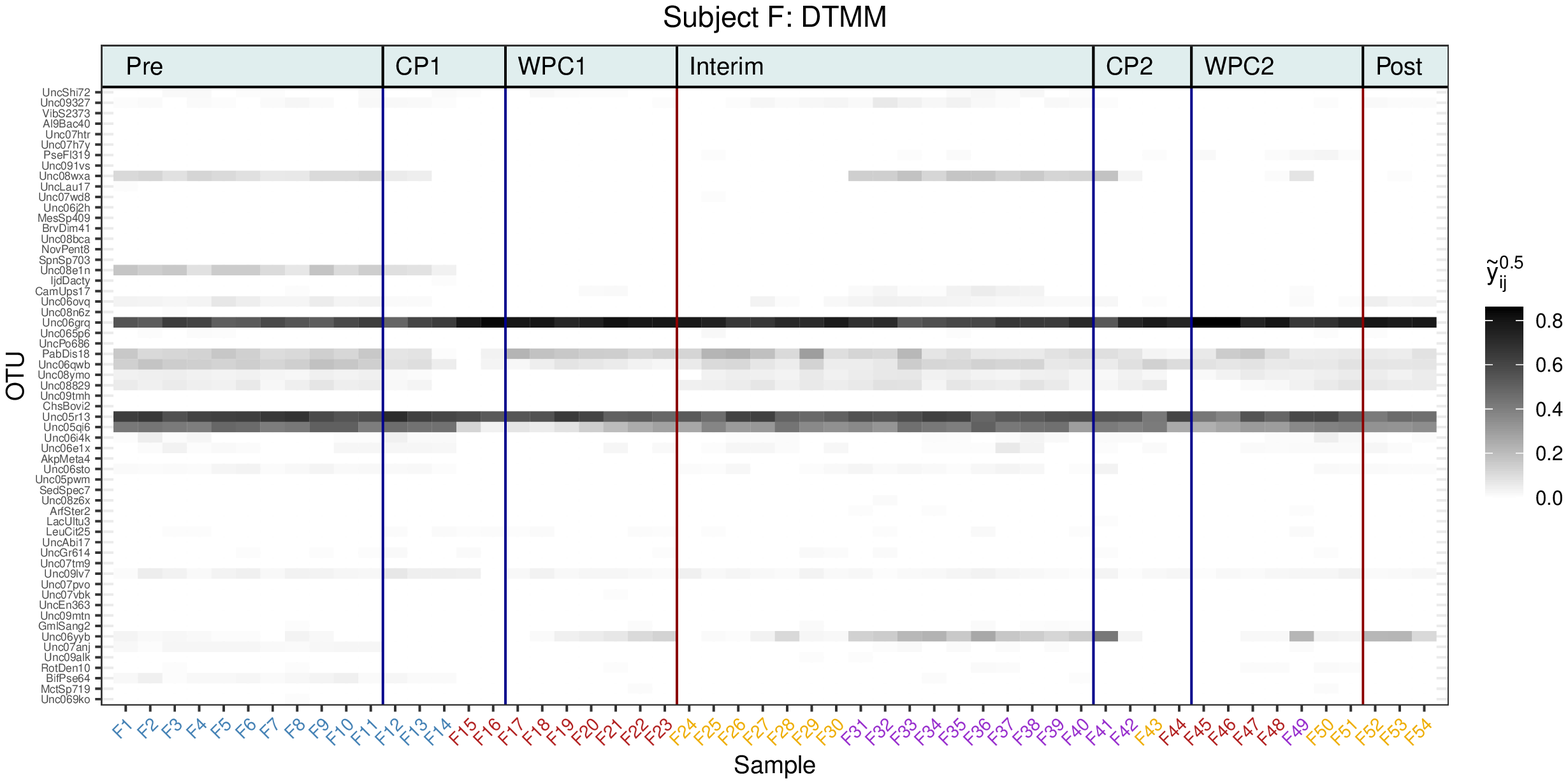} 
\caption{The heatmap of the microbiome samples of patient F (after the square-root transform). Each column represents a specific sample. The columns are ordered by the times the samples were collected. The colors of the $x$-axis labels represent the clustering labels of the samples returned by DTMM. The legends are defined the same way as in \ref{fig:sec_val_d}.}
\label{fig:sec_val_f}
\end{center}
\end{figure}

Specifically, we reanalyze the data in \cite{dethlefsen2011incomplete}, which studies the responses of stable gut microbiota to antibiotic disturbance. In this study, the distal gut microbiome of three patients (patient D, E and F) were monitored over 10 months, including two 5-day antibiotic treatment courses separated by a 5-month interim period. 52 to 56 samples were collected for each patient in the experiment. Samples of patient D and F are shown in \ref{fig:sec_val_d} and  \ref{fig:sec_val_f}, which also illustrate the design of the study. In our analysis, we aggregate the OTU counts to the genus level, which gives 59 OTUs in total. 

As in \cite{dethlefsen2011incomplete}, we analyze the samples from the three patients separately. For each patient, we ignore the time information of when the samples were taken and run DTMM on these samples for 2500 iterations. The first half of the chain was discarded as burn-in. The clustering results for patient D are shown in \ref{fig:sec_val_d} (the $x$-axis labels in these plots are colored by the cluster labels in $\bm{C}_{LS}$ of the samples they represent). For this patient, DTMM identifies three clusters, which can be interpreted as the \textit{stable}, \textit{sterile} and \textit{recover} stages of the microbiota. Based on the clustering results, the gut microbiota of patient D was stable before the treatment. It was able to recover to some stable states from antibiotic treatment within a week after the treatment was finished. However, although the microbiota was able to fully recover to the pre-treatment state after the first antibiotic treatment course, it never made a full recovery to the original state after the second (repeated) antibiotic treatment.

Similarly, the results for patient F are provided in \ref{fig:sec_val_d}. For patient F, DTMM identifies four clusters corresponding to the \textit{sterile}, \textit{recover} and two different \textit{stable} stages, respectively. Like patient D, the gut microbiota of patient F was stable before the treatment and was able to recover from the treatments. Unlike patient D, it did not recover to the pre-treatment state even after the first treatment course. Moreover, it took longer for patient F to recover than patient D. We note that these findings are all consistent to the findings in \cite{dethlefsen2011incomplete}, where the time and design information was used to get these results. 

As a comparison, the clustering results for these two patients under DMM are shown in Figure S7 and Figure S8 in the supplementary material 2.3. For both patients, DMM returns two clusters that roughly represent the \textit{stable} and \textit{unstable} stages of the microbiota. In this example, DTMM is able to discover more interesting latent structures among samples than DMM. It is worth noting that in each analysis, microbiome samples were collected from the same patient. Thus the level of cross-sample variations in this study is much smaller than microbiome studies with multiple subjects. In those cases, we expect DTMM to benefit more from its improved flexibility over DMM and discover even more interpretable structures than the latter.

 
\section{Case studies}
\label{sec:app}

The American Gut project \citep{mcdonald2015context,mcdonald2018american} aims at building an open-source and open-access reference microbiome dataset for general scientific use based on 16S rRNA sequencing and the QIIME pipeline \citep{caporaso2010qiime}. It collects mouth, skin and feces samples over a large variety of US participants on a voluntary basis. The participants send their microbiome samples to UC San Diego for sequencing and complete a questionnaire that covers the dietary habits, lifestyle, and health history. 

We apply DTMM to the July 2016 version of the fecal data from the AGP to construct enterotypes for two groups of samples: firstly, we consider participants who have been diagnosed with inflammatory bowel disease (IBD); secondly, we consider participants who have been diagnosed with diabetes. The diagnoses are made by a medical professional (a doctor or a physician assistant). The specific version of the AGP dataset contains an OTU table of 27774 OTUs. We focus on the top 75 OTUs based on total counts to reduce noises in the dataset and control for the sequencing errors. The top 75 OTUs on average retain 2/3 of the total counts in a sample. We filter the samples by only considering participants with at least 500 counts on the top 75 OTUs. This filtering ends up with 189 samples diagnosed with IBD and 106 samples diagnosed with diabetes.

In the following sections, we fit DTMM to each or the two datasets with the priors and hyperparameters set to the recommended choices in Section~\ref{subsec:node}. In each analysis, we run the Gibbs sampler in Section~\ref{subsec:inference} for $5000$ iterations and discard the first half of the chain as burn-in. The cluster labels are initiated by running the PAM algorithm with $K=5$.

Key findings from our analyses of these datasets are summarized as follows: (i) enterotypes of the two disease-diagnosed groups are determined by a large number of OTUs jointly in a sophisticated manner instead of by a few OTUs; (ii) OTUs from genera \textit{Bacteroides, Prevotella} and \textit{Ruminococcus} are typically important in identifying those enterotypes, which is consistent to the findings in previous works \citep{arumugam2011enterotypes}; (iii) the number of enterotypes and the OTUs that characterize each enterotype can differ across datasets, and (iv) DMM tends to find larger clusters that are unions of clusters found by DTMM.

\subsection{IBD}
\label{sec:app_ibd}

We first consider samples from participants that are diagnosed with IBD. Figure S9 
in Section 3 in the supplementary material shows the traceplots of some one dimensional parameters or summaries of the posterior samples. The Markov chain stabilizes and mixes reasonably well after about 750 iterations. Figure S9 (a) and Figure S9 (b) show the traceplots of the Dirichlet process precision parameter $\beta$ and the prior global activation probability $\lambda$. The posterior means of these parameters are $0.87$ and $0.53$, respectively.  Traceplot of the sampled number of internal nodes with $\gamma(A) = 1$ is shown in Figure S9 (c). On average, $39$ out of $74$ nodes are marked as relevant to the clustering process, indicating that the clustering process is determined by various OTUs jointly in a complicated way instead of being dominated by a few OTUs that are dominant in terms of counts. Figure S9 (d) shows the cumulative proportion of samples in the largest one, two, three, four and five clusters for each iteration. DTMM tends to assign samples into 5 clusters.

\begin{figure}[!ht]
\begin{center}\resizebox {0.9\textwidth} {!}{
 \begin{minipage}[b]{0.6\textwidth}
    \centering
   \includegraphics[width = 7.45cm]{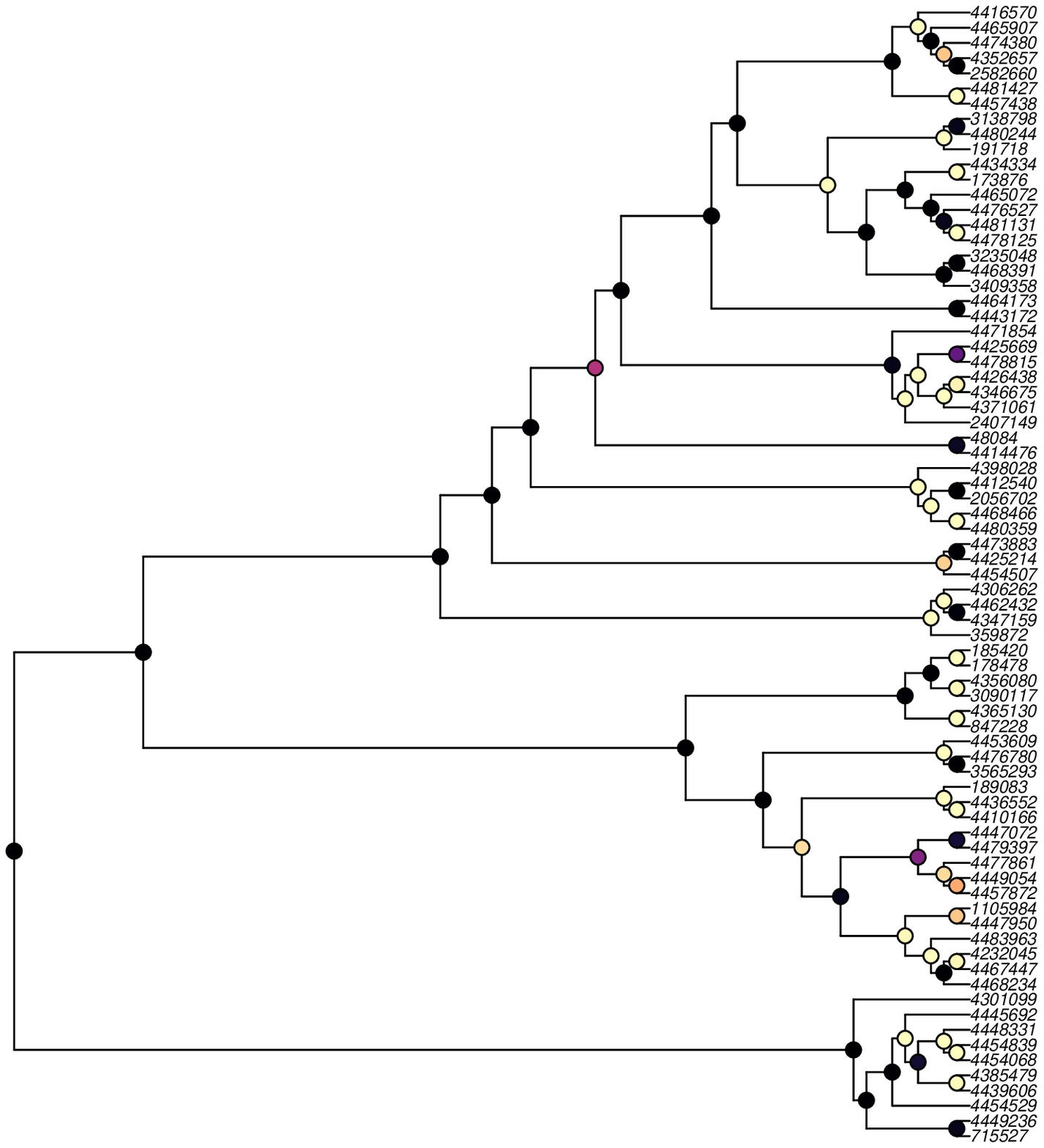} 
   \vspace{0.0cm}\vspace{-0.22cm}
    \end{minipage}\hspace{-1.23cm}
\begin{minipage}[b]{0.4\textwidth}
   \includegraphics[width = 4.7cm]{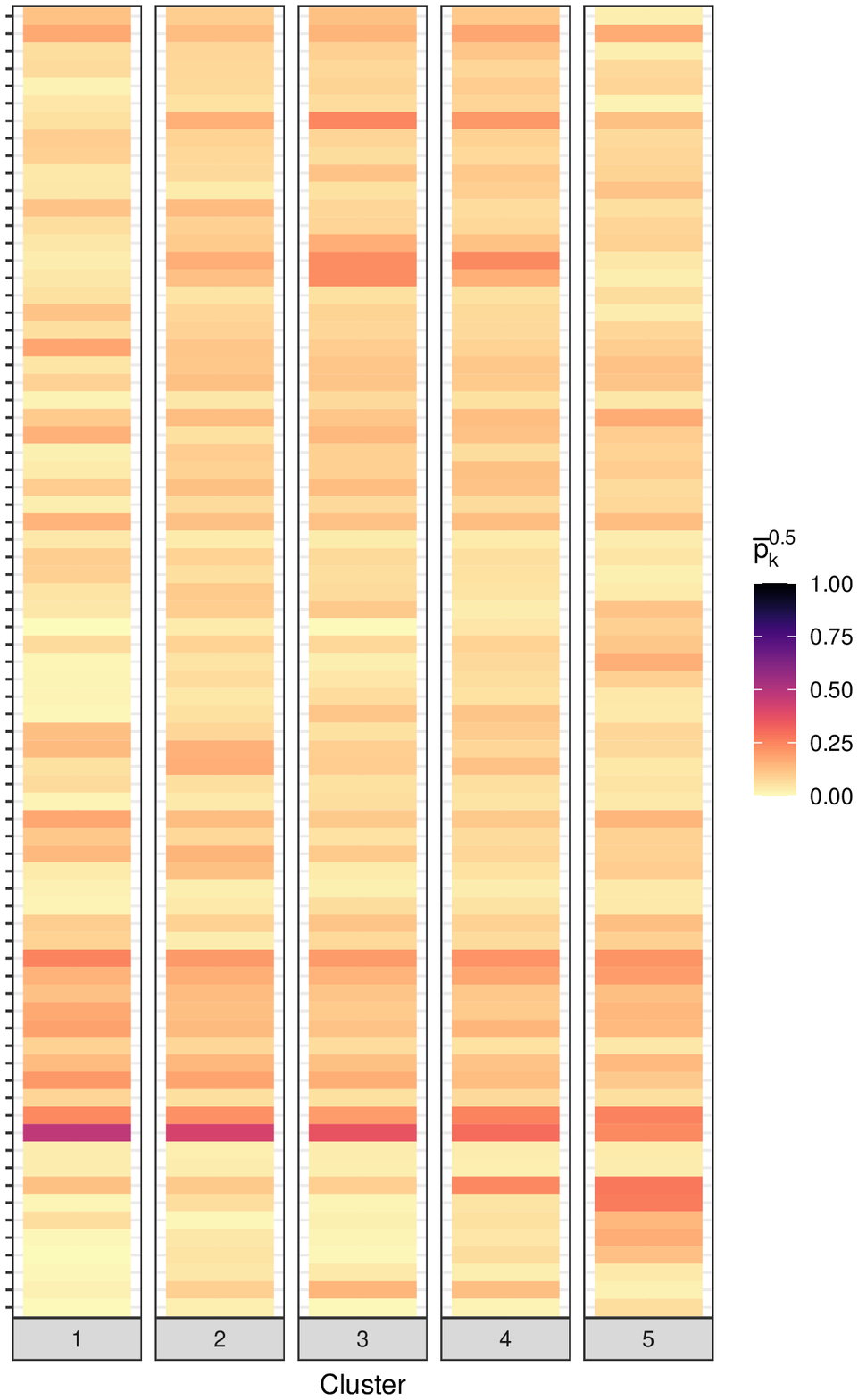}
  \end{minipage}  }
  \caption{Left: Estimated posterior means of the activation indicators at each node of the phylogenetic tree. Right: The estimated centroids of the five clusters in $\bm{C}_{LS}$ (after the square-root transform). }
  \label{fig:sec_4_centroid}
\end{center}
\end{figure}

\begin{figure}[!ht]
\begin{center}
\includegraphics[width = 0.4\textwidth]{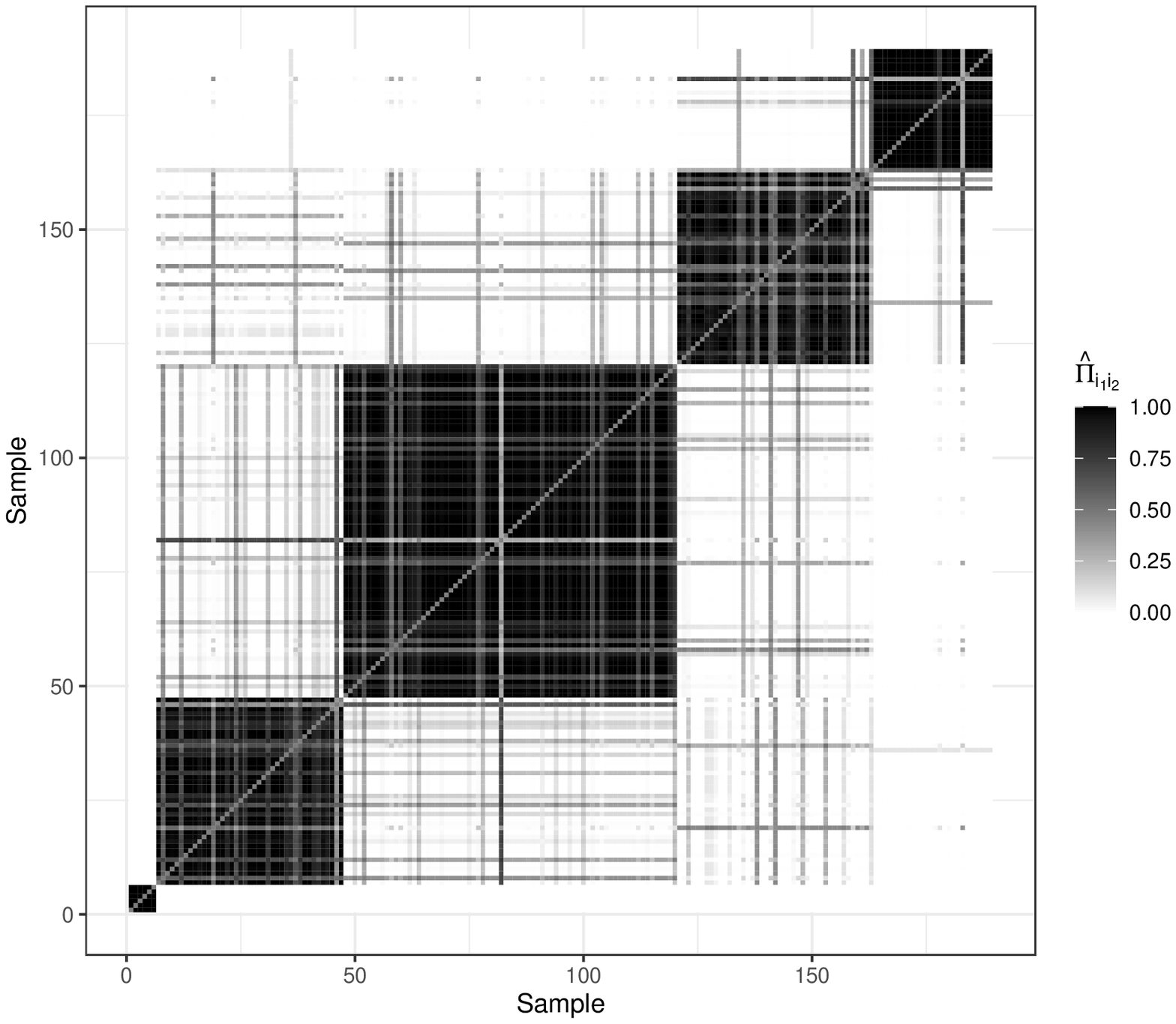}  \includegraphics[width  = 0.4\textwidth]{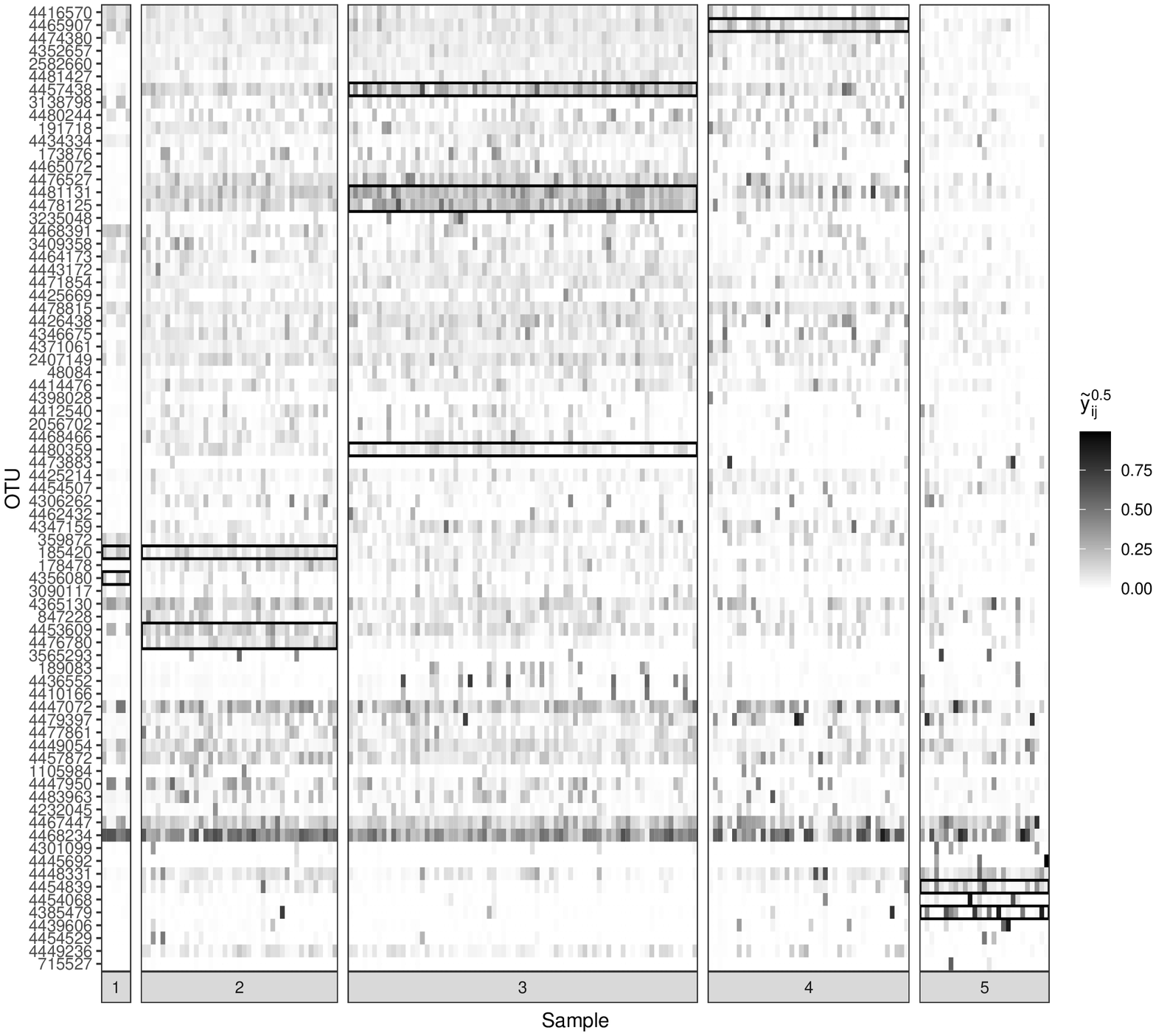}  
\caption{Left: Estimated pairwise co-clustering probabilities. Right: Heatmap of the samples (after the square-root transform) grouped by their labels in $\bm{C}_{LS}$. The black boxes illustrate the characteristic OTUs of each cluster.}
\label{fig:sec_4_co}
\end{center}
\end{figure}

We find $\bm{C}_{LS}$ as defined in (\ref{eq:cls}), which corresponds to $\bc^{(t_0)}$ with $t_0=2717$. $\bm{C}_{LS}$ assigns the samples into five clusters, with size 6, 41, 73, 42 and 27, respectively. The estimated centroids of the five clusters are shown in \ref{fig:sec_4_centroid}, which also shows the estimated posterior means of the activation indicator $s(A)$ at $A\in\mI$. Most internal nodes that are irrelevant to clustering are close to the leaves of the tree. Nodes that are more ``global'' (have more descendant OTUs) generally contribute to the clustering. This indicates that the clustering process is determined by most OTUs jointly in a complicated manner. \ref{fig:sec_4_co} (left) shows the estimated pairwise clustering probability matrix $\hat\Pi$ with the rows and columns ordered by the labels in $\bm{C}_{LS}$. There are noticeable uncertainties in the clustering, especially between cluster 2, 3 and 4. The similarities of these three clusters can also be seen from \ref{fig:sec_4_co} (right), where we plot the heatmap of the samples (after the square-root transform) grouped by their labels in $\bm{C}_{LS}$. 
\ref{fig:sec_4_co} (right) also shows that the within-cluster variations among samples are large.

To see which OTUs are more important in determining $\bm{C}_{LS}$, we consider the following heuristic measure of OTU importance: for $1\leq j\leq M$, let
 \begin{equation}
\begin{aligned}
\vartheta_j = \frac{SSB_j}{SSW_j} = \frac{ \sum\limits_{c\in\bm{C}_{LS}} n_c( \bar y_{cj}-\bar y_j)^2}{\sum\limits_{c\in\bm{C}_{LS}} \sum\limits_{c_i = c}(y_{ij} - \bar y_{cj})^2 },
\end{aligned}
\label{eq:cluster_measure}
\end{equation}
where $\bar y_j$ is the overall mean of $y_{ij}$, $\bar y_{cj}$ the mean of $y_{ij}$ for samples with $c_i = c$. \ref{tab:otu} shows the top 10 OTUs in determining $\bm{C}_{LS}$ in terms of $\vartheta_j$ as well as their compositions in each cluster centroid. Overall, $\bm{C}_{LS}$ is jointly determined by multiple OTUs in a complicated way. Note that OTUs that are important for clustering are not necessarily those with abundant counts. For example, OTU-4468234 and OTU-4447072 (both are \textit{Bacteroides}) are the two OTUs with the most counts in the dataset. However, these two OTUs are prevalent in most samples and thus have limited roles in the clustering.

\begin{table}[]
 \caption{Estimated cluster-specific compositions of the top 10 OTUs in determining $\bm{C}_{LS}$ in terms of $\vartheta_j$. Values of the OTU compositions are shown in the percentage scale.}
\label{tab:otu}
\centering \resizebox {\textwidth} {!}{
\begin{tabular}{@{}lllllllll@{}}
\toprule
OTU & Family & Genus & $\vartheta_j$ & C1 & C2 & C3 & C4 & C5 \\ \midrule
185420 & Bacteroidaceae & Bacteroides  & 0.43 & 2.01 & 2.62 & 0.98 & 0.66 & 0.54 \\ 
  4478125 & Ruminococcaceae & Faecalibacterium &0.43 & 0.22 & 1.71 & 5.72 & 2.71 & 0.10 \\ 
  4356080 & Barnesiellaceae &  -- & 0.33 & 0.53 & 0.41 & 0.37 & 0.42 & 0.28 \\ 
  4476780 &Rikenellaceae & -- & 0.32 & 0.14 & 1.70 & 0.15 & 0.33 & 1.04 \\ 
  4453609 & Rikenellaceae & -- & 0.26 & 2.08 & 2.43 & 1.15 & 0.71 & 0.85 \\ 
  4480359 & Ruminococcaceae & -- &0.22 & 0.22 & 1.02 & 1.22 & 0.11 & 1.48 \\ 
  4465907 &Lachnospiraceae &  Blautia  & 0.21 & 3.32 & 1.82 & 2.40 & 3.47 & 3.04 \\ 
  4481131 & Ruminococcaceae & Faecalibacterium & 0.18 & 0.11 & 2.92 & 5.62 & 6.11 & 0.22 \\ 
  4457438 & Lachnospiraceae & -- & 0.19 & 0.36 & 2.72 & 6.43 & 4.67 & 1.68 \\ 
  4385479 & Enterobacteriaceae & Proteus & 0.17 & 0.02 & 0.21 & 0.04 & 0.24 & 2.91 \\  \bottomrule
\end{tabular} }
\end{table}

\begin{figure}[!ht]
\begin{center} \resizebox {0.85\textwidth} {!}{
 \begin{minipage}[b]{0.19\textwidth}
    \centering
   \includegraphics[width = 3cm]{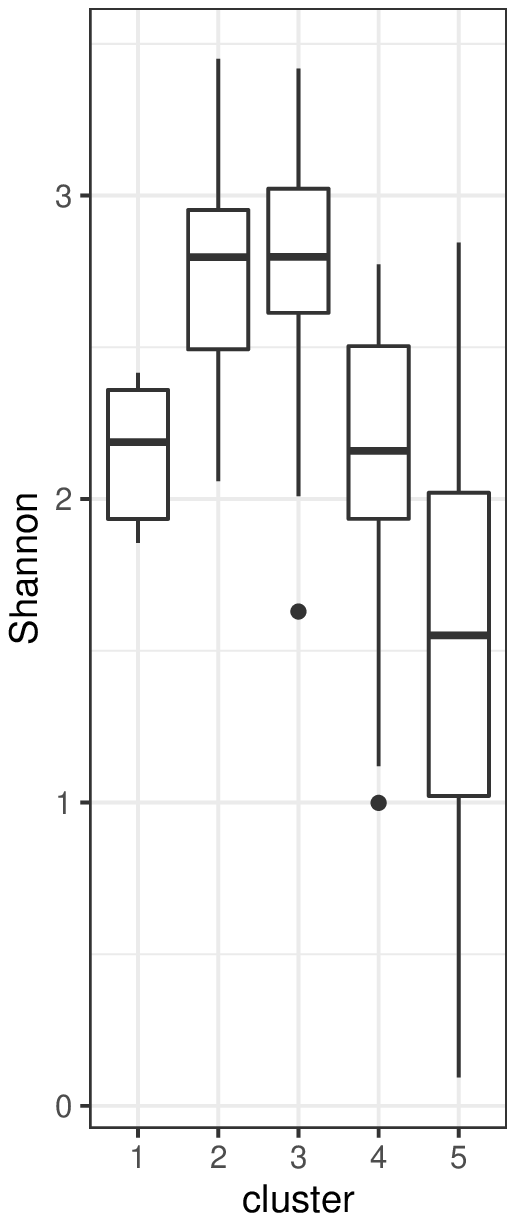} 
    \end{minipage}\hspace{0pt}
    \begin{minipage}[b]{0.8\textwidth}
   \includegraphics[width = 14cm]{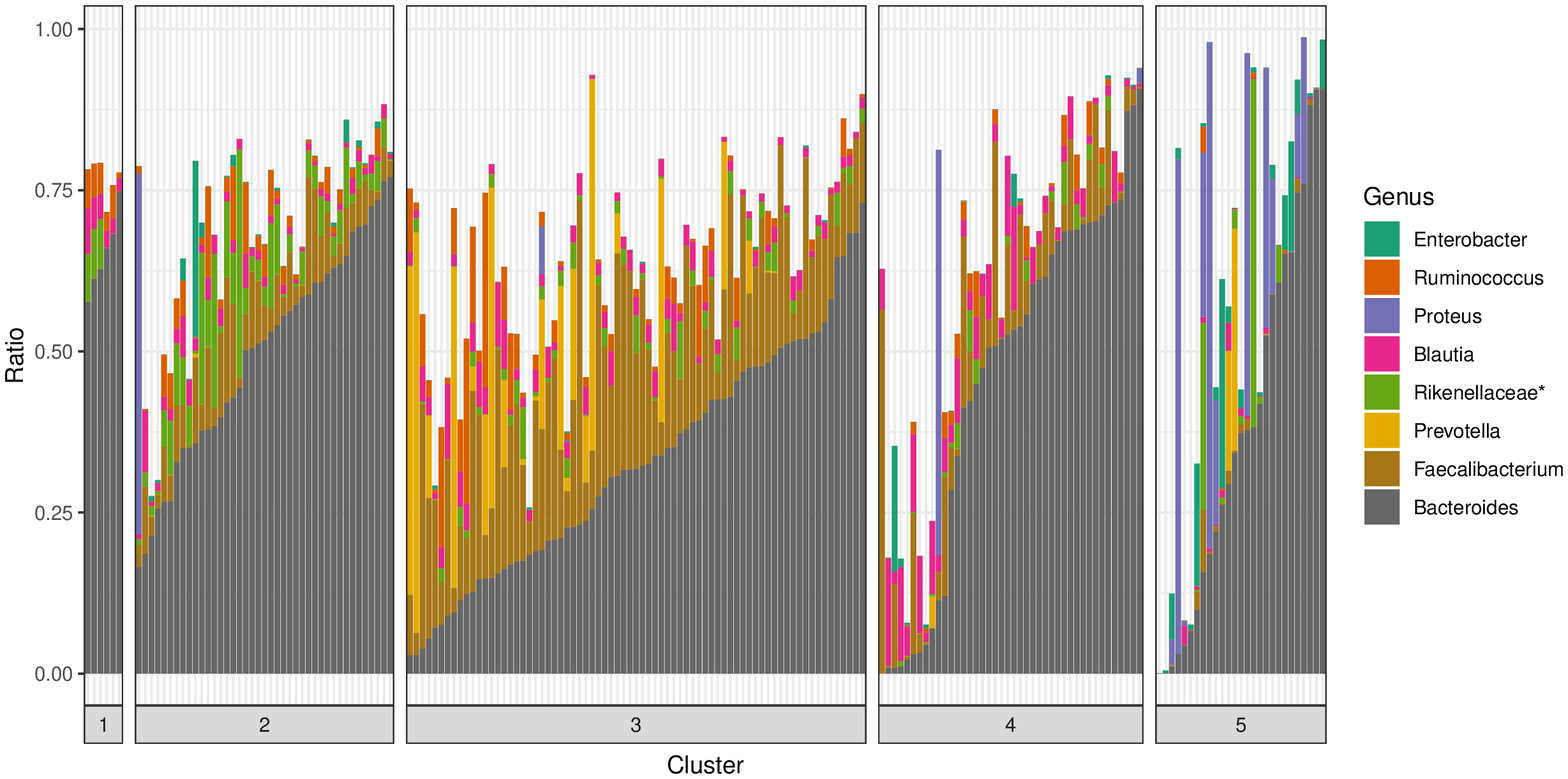}
  \end{minipage}   }
  \caption{Left: Boxplot of the Shannon diversity of samples in each cluster. Right: Relative abundance of 8 genera for each sample. A genus is chosen if its descendant OTUs have large $\vartheta^c_j$ for some $c$. For the 3 OTUs with unavailable genera information, their family is shown instead (indicated by \textit{Rikenellaceae*}). The samples are grouped by their cluster labels in $\bm{C}_{LS}$.}
  \label{fig:sec_4_cluster_sum}
\end{center}
\end{figure}

We next compare the five resulting clusters in more details. \ref{fig:sec_4_cluster_sum} shows the boxplot of the Shannon diversity of samples in the five clusters, respectively. Samples from cluster 2 and 3 tend to have more evenly distributed counts across OTUs compared to those from cluster 1, 4 and 5. Similar to (\ref{eq:cluster_measure}), we can define a heuristic measure of OTU importance in characterizing each of the five clusters. Specifically, for $c=1,\ldots,5$, let
{\small
\begin{equation}
\begin{aligned}
\vartheta^c_j = \frac{SSB^c_j}{SSW^c_j} = \frac{ n_c( \bar y_{cj}-\bar y_j)^2 +  n_{-c}( \bar y_{-cj}-\bar y_j)^2}{ \sum\limits_{c_i = c}(y_{ij} - \bar y_{cj})^2 + \sum\limits_{c_i \not = c}(y_{ij} - \bar y_{-cj})^2  },
\end{aligned}
\label{eq:cluster_measure2}
\end{equation} }
where $n_{-c}$ is the number of samples that are not in cluster $c$, $\bar y_{-cj}$ the mean of $y_{ij}$ for samples with $c_i\not = c$. (\ref{eq:cluster_measure2}) is equivalent to merging the four clusters other than cluster $c$ in (\ref{eq:cluster_measure}). The boxes in \ref{fig:sec_4_co} (right) indicate the top OTUs in terms of $\vartheta^c_j$ for each $c$ (only OTUs with $\vartheta^c_j > 0.1$ are shown). 

Based on these results, we can characterize each cluster by a few OTUs with the top $\vartheta^c_j$. For example, samples from cluster 2 tend to have more counts from the \textit{Rikenellaceae} family (represented by OTU-4453609 and OTU-4476780). Cluster 3 is characterized by having more abundance in the \textit{Faecalibacterium} (represented by OTU-4478125 and OTU-4481131) and the \textit{Lachnospiraceae} family (represented by OTU-4481127 and OTU-4457438). \cite{arumugam2011enterotypes} proposed three enterotypes in human gut microbial communities that are characterized by the variation in the levels of one of the three
genera: \textit{Bacteroides, Prevotella} and \textit{Ruminococcus}. Our analysis suggests that enterotypes of the IBD patients are determined by a sophisticated mechanism involving more genera. Although OTUs from the \textit{Bacteroides, Prevotella} and \textit{Ruminococcus} genera are not always those with the largest $\vartheta^c_j$, they are playing important roles in identifying the five clusters. For example, OTUs from the \textit{Prevotella} genus have large $\vartheta^3_j$ and are thus crucial in determining cluster 3 while OTUs from the \textit{Ruminococcus} genus have large $\vartheta^1_j$ and $\vartheta^2_j$ and are thus important in identifying cluster 1 and 2. This can also be seen from \ref{fig:sec_4_cluster_sum} (right), where relative abundance of 8 genera picked by $\vartheta^c_j$ are shown for each sample.

\subsection{Diabetes}
 \label{sec:app_diabetes}
 
Similar to Section~\ref{sec:app_ibd}, we apply DTMM to samples from diabetes patients. Results for this application are provided in Section 3 in the supplementary material. For example, counterparts of Figure S9 
and \ref{fig:sec_4_co} are shown in Figure S10 
and Figure S11. 
In this example, DTMM finds three clusters with $\bm{C}_{LS}$. Figure S12 
shows the estimated centroids of the three clusters as well as the estimated posterior means of the activation indicators at each node of the phylogenetic tree. Figure S13 
(right) shows for each sample the relative abundance of 6 genera selected based on the importance of their descendant OTUs in identifying the three clusters. Compared with the IBD example, enterotypes in this case can be associated with individual OTUs in a simpler manner. For example, samples in cluster 3 tend to have significantly lower abundance in \textit{Faecalibacterium} and \textit{Bacteroides}, which are the dominating genera in most samples. Compared with cluster 2, cluster 1 is identified with relatively more counts from OTU-173876 and the \textit{Prevotella} family.

On average, 21 out 75 internal nodes of $\T$ are estimated as relevant to clustering. Compared with the IBD example in Section~\ref{sec:app_ibd}, fewer nodes are involved, suggesting that clusters in the diabetes example are determined by fewer OTUs (genera). As shown in Figure S11 
(right), a few OTUs play crucial roles in determining multiple clusters. As a comparison, as shown in \ref{fig:sec_4_co} (right), each cluster in the IBD example is determined by a unique set of OTUs. Since DTMM marks an internal node as relevant if it is relevant in determining \textit{any} cluster, more nodes are selected in the IBD example.

\subsection{DTMM vs. DMM}

We also apply DMM to the two examples in this section and compare it with DTMM. DMM reports two clusters in both examples (see Figure S14 
in Section 3 in the supplementary material). For the IBD dataset, \ref{fig:sec_4_dmm} (left) shows the two-dimensional NMDS plot of the data, colored by the cluster labels reported by DMM. In comparison, \ref{fig:sec_4_dmm} (right) shows the same NMDS plot colored by the cluster labels in $\bm{C}_{LS}$ reported by DTMM. Similar NMDS plots for the diabetes example are shown in Figure S15 in Section 3 in the supplementary material.

\begin{figure}[!ht]
\begin{center}
\includegraphics[width = 0.45\textwidth]{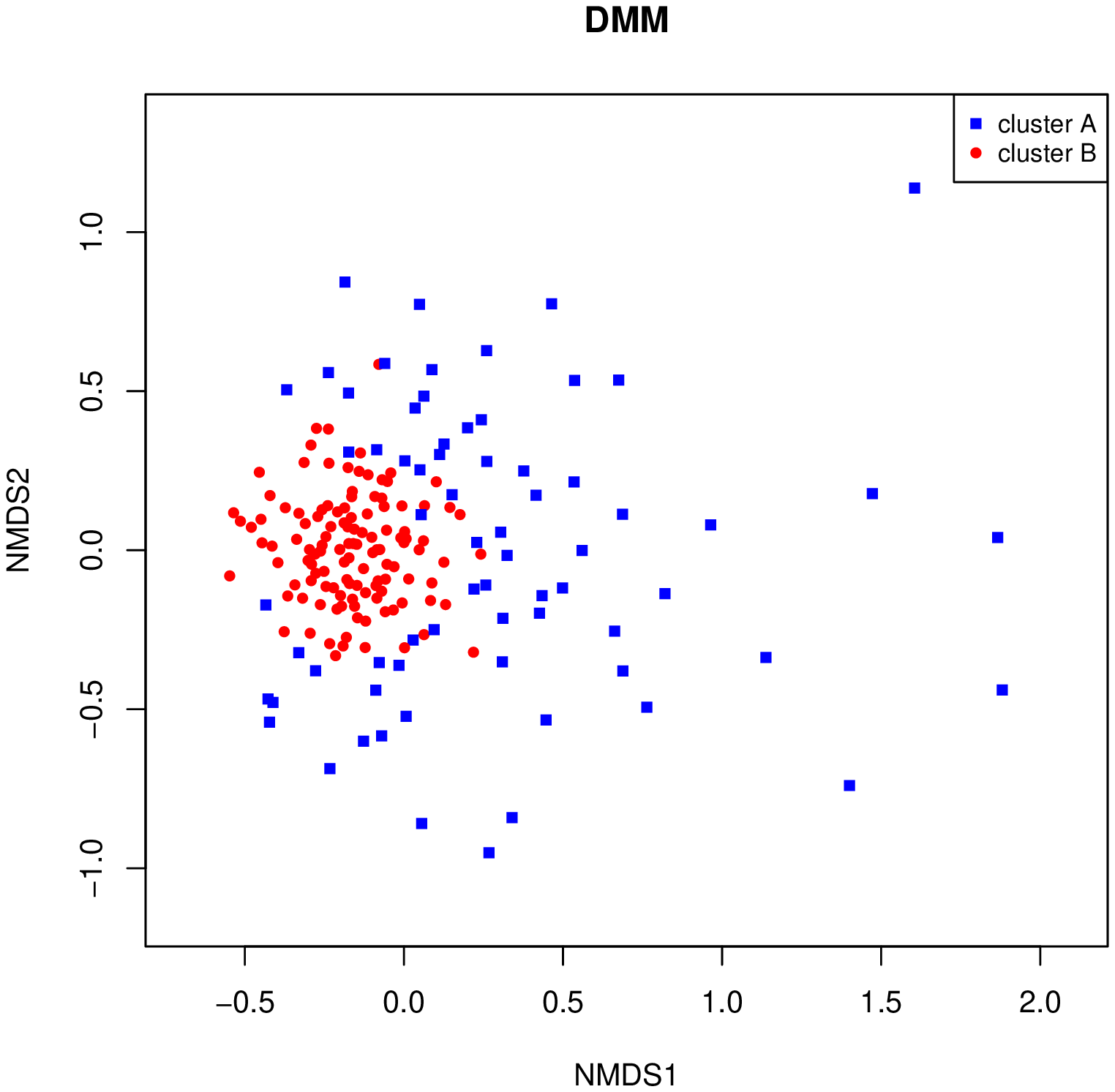}\includegraphics[width = 0.45\textwidth]{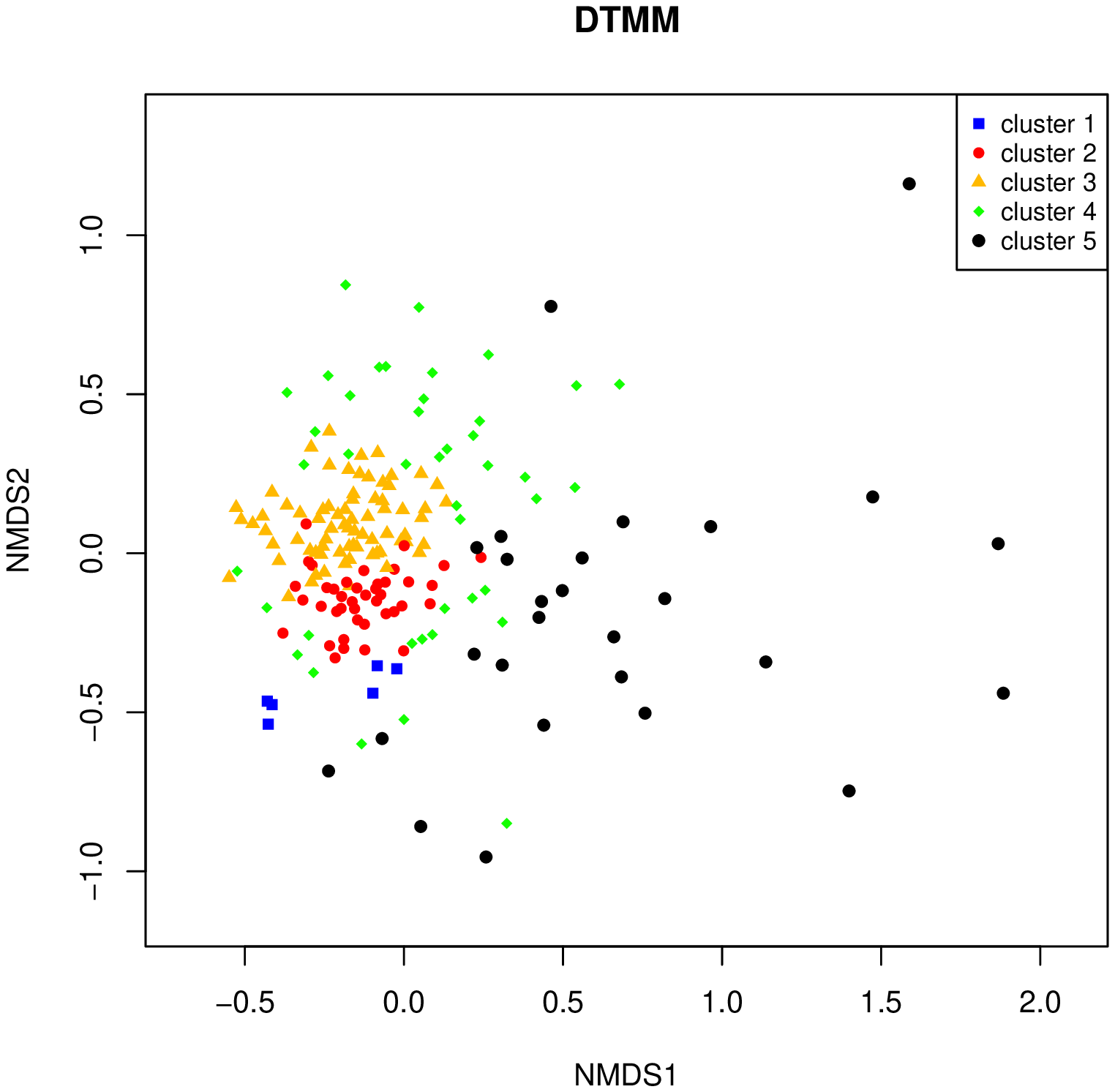} 
\caption{Two-dimensional NMDS plots for the AGP IBD dataset. Points are colored and shaped by the clustering given by DMM (left) or DTMM (right).}
\label{fig:sec_4_dmm}
\end{center}
\end{figure}

For the IBD application, the five clusters reported by DTMM can be seen as refinements of the two clusters reported by DMM. Roughly, cluster B identified by DMM is the union of cluster 2 and 3 from DTMM while cluster A found by DMM is the union of cluster 1, 4 and 5 from DTMM. As shown in \ref{fig:sec_4_co} (right) and \ref{fig:sec_4_cluster_sum} (right), those subclusters from DTMM are not differentiated by OTUs with dominant counts. Thus it is very unlikely for DMM to make further splits. Moreover, based on \ref{fig:sec_4_co} (right), samples within each cluster from DTMM tend to show different levels of heterogeneities across OTUs, making the underlying Dirichlet-multinomial model of DMM unrealistic. For example, counts of OTUs from the \textit{Prevotella} genus tend to show large within cluster variations among samples in cluster 3. To capture this level of variation, DMM has to push the cluster-specific dispersion parameter very large, essentially loose its ability to effectively find those subclusters.


\section{Concluding remarks}
\label{sec:discussion}
We have introduced DTMM as a model-based framework for clustering the amplicon sequencing data in microbiome studies. By directly incorporating the phylogenetic tree, DTMM differs from the popular DMM in three directions: first, it offers a more flexible covariance structure among different OTUs; second, it provides a way for selecting a subset of internal nodes in the phylogenetic tree that is relevant for clustering; moreover, it allows simple and efficient algorithms for posterior inference.

Although the covariance structure offered by DT is richer than that of the Dirichlet distribution, it is still limited compared to the logistic-normal family (LN). In a case with $K$ OTUs, DT models the covariance among OTU counts with $(K-1)$ dispersion parameters in the series of beta distributions while LN uses $K(K-1)/2$ parameters in modeling the covariance matrix. It is interesting to further generalize the covariance structure provided by DTMM without making the inference too complicated. When selecting a subset of internal nodes in the phylogenetic tree that are relevant to clustering, DTMM selects a node if it is relevant in identifying \textit{any} cluster. Intuitively, DTMM first selects a subspace in the node space and performs clustering in that space. An alternative direction worth exploring is to allow the nodes selected to be cluster-dependent such that each cluster can deviate from the ``mean'' cluster at different internal nodes.



\section*{Software}

For DMM, we use the \texttt{R} package \texttt{DirichletMultinomial}. For PAM, we use the \texttt{pam} function in the \texttt{R} package \texttt{cluster}. For spectral clustering, we use the \texttt{specc} function in the \texttt{R} package \texttt{kernlab}. For $k$-means and hierarchical clustering, we use the \texttt{kmeans} and \texttt{hclust} functions in \texttt{R}. We provide an \texttt{R} package (\texttt{DTMM}: {\color{blue}https://github.com/MaStatLab/DTMM}) implementing the proposed method in this paper. We also provide code and a guide to reproduce all simulations and data analyses in Section 3 and Section 4 

\noindent({\color{blue}{https://github.com/MaStatLab/DTMM\_reproducible}}).

\section*{Acknowledgment}
LM's research is partly supported by NIGMS grant 1R01GM135440 and NSF grant DMS-1749789. Part of this work was completed when JM was supported by a Duke Forge Graduate Fellowship in health data science.


\bibliography{DTMM}

\clearpage
\appendix
 
\pagenumbering{arabic}
\renewcommand*{\thepage}{S\arabic{page}}
 
 \beginsupplement
 
\section*{Supplementary Materials}

\renewcommand{\thesection}{\arabic{section}}

\renewcommand{\thefigure}{S\arabic{figure}}
\renewcommand{\theequation}{S\arabic{equation}}

\section{Computational strategies }
\label{subsec:comp}
 
\subsection{Gibbs sampler for DTMM}

A blocked Gibbs sampler for DTMM is summarized in Algorithm~\ref{alg:gibbs}. We first provide details on the full conditionals.

{\bf{The full conditional of} $\bgam$:} For each $A\in\mI$, the Bayes factor comparing $\gamma(A) = 1$ versus $\gamma(A) = 0$ given $\bc$ can be written as
\begin{equation}
\begin{aligned}
M_{10}(A\mid \bgam^{-A},\bc, \beta,\lambda) =M_{10}(A\mid\bc) = \ddfrac{\prod_{c\in \bc^*} \mL^A_1(\bY_c)}{\int \mL^A_0(\bY  \mid \tilde\bpsi(A))d\Pi( \tilde\bpsi(A) )}.
\end{aligned}
\end{equation}
It follows that
\begin{equation}
\begin{aligned}
\Pr(\gamma(A) = 1 \mid \bY, \bgam^{-A}, \bc, \beta,\lambda) = \frac{\lambda M_{10}(A\mid\bc)}{(1 - \lambda)  + \lambda M_{10}(A\mid\bc)}.
\end{aligned}
\label{eq:gamma}
\end{equation}

{\bf{The full conditional of $\bc$:}} For $i=1,\ldots,n$, let $\bc_{-i} = \bc\setminus\{c_i\}$. Following the discussion in \cite{neal2000markov}, we can write the prior conditional distribution of $c_i$ given $\bc_{-i}$ as
\begin{equation}
\begin{aligned}
\Pr(c_i = c\text{ for some }c\in \bc_{-i}\mid \bc_{-i}, \bgam, \beta, \lambda) & = \frac{n_{-i,c}}{n-1+\beta}  \\
\Pr(c_i \not= c_j\text{ for all } j \not= i\mid \bc_{-i},\bgam, \beta, \lambda) & = \frac{\beta}{n - 1 + \beta},
\end{aligned}
\end{equation}
where $n_{-i, c}$ represents the number of samples in the cluster with label $c$ excluding sample $i$. After conditioning on the data, these probabilities become
\begin{equation}
\begin{aligned}
\Pr(c_i = c\text{ for some }c\in c_{-i}\mid \bc_{-i}, \bY,  \bgam,\beta,\lambda ) &\propto n_{-i,c}\times \frac{\mL_1(\by_i, \bY^{-i}_c\mid \bgam )}{\mL_1(\bY_c^{-i}\mid \bgam)}  \\
\Pr(c_i \not= c_j\text{ for all } j \not= i\mid \bc_{-i}, \bY,  \bgam,\beta,\lambda ) &\propto \beta\times \mL_1(\by_i \mid \bgam),
\end{aligned}
\label{eq:c}
\end{equation}
where for any $\by_i,\ldots, \by_l$ in the same cluster,
\begin{equation}
\begin{aligned}
 \mL_1(\by_1,\ldots,\by_l \mid \bgam)  = &  \prod\limits_{\{A\in\mI:\gamma(A) = 1 \}} \mL_1^A(\by_1,\ldots,\by_l).
\end{aligned}
\end{equation}
If $\gamma(A) = 0$ for every $A\in\mI$, we let $ \mL_1(\by_1,\ldots,\by_l) =1$.
Note that given the coupling status $\bgam$, the posterior conditional distribution of $c_i$ only depends on the likelihoods of the data at nodes with $\gamma(A)=1$. Therefore, to update the cluster label for any observation $\by_i$, we only need to focus on the nodes with $\gamma(A)=1$ and at each of these nodes compute: (i) the marginal likelihood of $\by_i$ and (ii) for each $c\in\bc_{-i}$, the conditional likelihood of $\by_i$ given $\bY_c^{-i}$. All nodes with $\gamma(A) = 0$ and thus the parameters at these nodes are essentially ``nuisance'' for the cluster labels.    

{\bf{The full conditional of} $\beta$:} Instead of fixing $\beta$, one can put a prior on it and incorporate it into the Gibbs sampler. For example, when gamma priors are used, \cite{escobar1995bayesian} update $\beta$ using a data augmentation trick. When arbitrary priors are used, $\beta$ can be updated by reparameterizing $b = \frac{\beta}{\beta + 1}$ \citep{hoff2006model}. Specifically, let $\pi(b)$ be the induced prior on $b$, we have
\begin{equation}
\begin{aligned}
\pi(b \mid \bY,\bc,\bgam, \lambda) \propto \pi(b) \times \left( \frac{b}{1-b}\right)^{|\bc^*|} \frac{\Gamma(b/(1-b))}{\Gamma(b/(1-b)+n)}.
\end{aligned}
\label{eq:b}
\end{equation}

{\bf{The full conditional of} $\lambda$:} By the beta-binomial conjugacy, 
\begin{equation}
\begin{aligned}
\lambda \mid \bY, \bc, \bgam, \beta \sim \text{Beta}\left(a_0 + \sum\limits_{A\in\mI} \gamma(A), b_0 + \sum\limits_{A\in\mI}(1- \gamma(A)) \right ).
\end{aligned}
\label{eq:lambda}
\end{equation}

 \vspace{2em}
 We next provide details on the derivation of these full conditionals. Recall that for $\bY^I_c=\{\by_i: c_i = c,i\in I \}$, the marginal likelihood of $\bY^I_c$ at node $A$ given $\gamma(A) = 1$ or $\gamma(A) = 0$ after marginalizing out the sample-specific parameter and the cluster-specific parameter can be written as
{\small\begin{equation}
\begin{aligned}
 \mL_1^A(\bY^I_c) & :=   \iint   \mL^A(\bY^I_c\mid \bpsi^*_c(A), \gamma(A) = 1, \tilde\bpsi(A) ) d\Pi(\bpsi^*_c(A) \mid \gamma(A)=1, \tilde\bpsi(A)) \\
&  =   \iint    \prod\limits_{\{i\in I:c_i=c\}}  {y_i(A) \choose y_i(A_l)} \frac{B(\theta(A)\tau(A) + y_i(A_l), (1 - \theta(A))\tau(A) + y_i(A_r))}{B(\theta(A)\tau(A), (1 - \theta(A))\tau(A))}  \\
& \hspace{1.3cm} \times  \frac{\theta(A)^{\theta_0(A)\nu_0(A) - 1} (1 - \theta(A))^{(1 - \theta_0(A))\nu_0(A) - 1}}{B(\theta_0(A)\nu_0(A), (1 - \theta_0(A))\nu_0(A))}d\theta(A) dF^A(\tau)  ,
\end{aligned}
\label{eq:m1}
\end{equation}
\begin{equation}
\begin{aligned}
 \mL_0^A(\bY^I_c \mid \tilde\bpsi(A)) & :=    \iint   \mL^A(\bY^I_c\mid \bpsi^*_c(A), \gamma(A) = 0, \tilde\bpsi(A) ) d\Pi(\bpsi^*_c(A) \mid \gamma(A)=0, \tilde\bpsi(A)) \\
 & =\prod\limits_{\{i\in I:c_i=c\}}  {y_i(A) \choose y_i(A_l)} \frac{B(\tilde\theta(A)\tilde\tau(A) + y_i(A_l), (1 - \tilde\theta(A))\tilde\tau(A) + y_i(A_r))}{B(\tilde\theta(A)\tilde\tau(A), (1 - \tilde\theta(A))\tilde\tau(A))} .
\end{aligned}
\end{equation}}

\begin{algorithm}[h]
\caption{Gibbs sampler for DTMM}\label{gibbs}
\begin{algorithmic}
   \vspace{0.5em}
\Procedure{GIBBS}{$B, T, \{\by_1,\ldots,\by_n\}$} \Comment{$B:$ burn-in; $T$: total number of iterations.}
   \vspace{0.5em}
  \State Initialize $\bc^{(0)}, \bgam^{(0)}, \beta^{(0)}, \gamma^{(0)}$.
  \vspace{0.5em}
   \For{$t = 1, 2, \ldots, T$}
   \vspace{0.5em}
   \State {\textbf{[2] Update the coupling indicators:}}
    \Indent
       \For{$A\in\mI$}    
           \State Compute $M^{(t-1)}_{10}(A\mid\bc^{(t-1)})$ as defined by (\ref{eq:gamma}).
           \State Draw a new value for $\gamma^{(t)}(A)\sim \text{Binom}\left(1, \frac{ \lambda^{(t-1)}M^{(t-1)}_{10}(A\mid\bc^{(t-1)})}{(1 - \lambda^{(t-1)})  +  \lambda^{(t-1)}M^{(t-1)}_{10}(A\mid\bc^{(t-1)})}\right)$. 
       \EndFor    
   \EndIndent 

 \vspace{0.5em}
   \State {\textbf{[1] Update the cluster labels:}}
   \Indent
       \For{$i=1,2,\ldots, n$}   
           \State Draw a new value for $c^{(t)}_i$ from 
           $$c_i\mid c_{1}^{(t)}, \ldots, c_{i-1}^{(t)}, c_{i+1}^{(t-1)}, \ldots, c_{n}^{(t-1)}, \bY, \bgam^{(t)},\beta^{(t-1)}$$
           \hspace{2.4cm} as defined by (\ref{eq:c}).
       \EndFor    
   \EndIndent

   \vspace{0.5em}
   \State {\textbf{[3] Update the Dirichlet process precision parameter:}}
   \Indent
       \State Draw value $b^{\text{new}}$ from $b\sim\pi(b\mid \bY,\bc^{(t)},\bgam^{(t)},\lambda^{(t-1)})$ as defined in (\ref{eq:b}).
       \State Let $\beta^{(t)} = \frac{b^{\text{new}}}{1-b^{\text{new}}}$.
    \EndIndent
 
    \vspace{0.5em}
   \State {\textbf{[4] Update the prior coupling probabilities:}}
   \Indent
   	\State Draw $\lambda^{(t)} $ from $\lambda\mid \bY, \bc^{(t)}, \bgam^{(t)}, \beta^{(t)}$ as defined in (\ref{eq:lambda}).
   \EndIndent

\EndFor
 \vspace{0.5em}
 
 \State\Return $\big[\{\bc^{(B+1)},\bgam^{(B+1)},\beta^{(B+1)},\lambda^{(B+1)}\}, \ldots, \{\bc^{(T)},\bgam^{(T)},\beta^{(T)},\lambda^{(T)}\} \big]$.
 
  \vspace{0.5em}
  
  \EndProcedure
\end{algorithmic}
\label{alg:gibbs}
\end{algorithm}

\begin{enumerate}
\item[(i).]
{\bf{The full conditional of $\bgam$.}} For each $A\in\mI$, let $\bY(A) = \{\by_i(A): i\in [n]  \}$ and $\bY_c(A) = \{\by_i(A): c_i=c, i\in [n]  \}$. We have
\begin{equation}
\begin{aligned}
& \hspace{15pt} M_{10}(A\mid \bgam^{-A},\bc, \beta,\lambda) \\
& = \frac{\mL(\bY \mid \gamma(A) = 1,\bgam^{-A}, \bc, \beta,\lambda)}{\mL(\bY \mid \gamma(A) = 0, \bgam^{-A}, \bc, \beta,\lambda)}\\
& =   \frac{\mL(\bY(A_l) \mid \bY(A), \gamma(A) = 1, \bc, \beta,\lambda)}{\mL(\bY(A_l) \mid \bY(A), \gamma(A) = 0,  \bc, \beta,\lambda)}\\
& = \ddfrac{\prod_{c\in \bc^*}\int \mL(\bY_c(A_l) \mid \bY_c(A), \bpsi^*_c(A), \gamma(A) = 1) d\Pi(\bpsi^*_c(A)\mid\gamma(A) = 1)}{\int \mL(\bY(A_l) \mid \bY(A), \tilde\bpsi(A), \gamma(A) = 0) d\Pi(\tilde\bpsi(A))} \\
& = \ddfrac{\prod_{c\in \bc^*} \mL^A_1(\bY_c)}{\int \mL^A_0(\bY  \mid \tilde\bpsi(A))d\Pi( \tilde\bpsi(A) )}.
\end{aligned}
\end{equation}
 
\item[(ii).]
{\bf{The full conditional of $\bc$.}} For $i\in [n]$, let $\bY^{-i}$ denote the set of samples with sample $i$ excluded and let $\bc^*_{-i} = \{ \bc_{-i}\}$ be the set of distinct values of $\bc_{-i}$. For $c\in\bc^*$, we have 
{\small\begin{equation}
\begin{aligned}
&\hspace{14pt} \Pr(c_i = c\mid \bc_{-i}, \bY,  \bgam,\beta,\lambda ) \\
& \propto \Pr(c_i = c\mid \bc_{-i}, \bgam,\beta,\lambda ) \mL(\by_i \mid \bY^{-i}, c_i = c, \bc_{-i}, \bgam, \beta,\lambda)\\
& \propto n_{-i,c} \times \int \mL(\by_i \mid \bY^{-i}, c_i = c, \bc_{-i}, \bgam, \beta,\lambda, \tilde \bpsi)d\Pi(\tilde\bpsi)\\
& \propto n_{-i,c} \times \int \left[  \int  \mL(\by_i \mid \bY^{-i}, \bpsi^*_c, \bc_{-i}, \bgam, \beta,\lambda, \tilde \bpsi)  d\Pi(\bpsi^*_c\mid \tilde\bpsi, \bgam) \right ]  d\Pi(\tilde\bpsi) \\
& \propto n_{-i,c} \times \prod\limits_{\{ A\in\mI: \gamma(A) = 1 \}}	\frac{\mL^A_1(\by_i,\bY^{-i}_c) }{\mL^A_1(\bY^{-i}_c)} \times \left[ \int \prod\limits_{\{ A\in\mI: \gamma(A) = 0 \}} \frac{\mL^A_0(\bY \mid \tilde\bpsi(A))}{\mL^A_0(\bY^{-i} \mid \tilde\bpsi(A))} d\Pi(\tilde\bpsi)	\right] \\
& \propto n_{-i,c} \times \frac{\mL_1(\by_i,\bY^{-i}_c \mid \bgam) }{\mL_1(\bY^{-i}_c \mid \bgam )} \times \left[ \int \prod\limits_{\{ A\in\mI: \gamma(A) = 0 \}} \frac{\mL^A_0(\bY \mid \tilde\bpsi(A))}{\mL^A_0(\bY^{-i} \mid \tilde\bpsi(A))} d\Pi(\tilde\bpsi)	\right] ,
\end{aligned}
\label{eq:full_c1}
\end{equation}}
where we use $\mL(\by_i \mid -)$ to denote the conditional likelihood of $\by_i$ given certain parameters or other samples. 
Similarly, we have 
\begin{equation}
\begin{aligned}
&\hspace{14pt} \Pr(c_i \not =c_j \text{ for all } j\not = i \mid \bc_{-i}, \bY,  \bgam,\beta,\lambda ) \\
& \propto  \Pr(c_i \not \in \bc^*_{-i}\mid \bc_{-i}, \bgam,\beta,\lambda )  \mL(\by_i \mid \bY^{-i}, c_i \not\in \bc^*_{-i}, \bc_{-i}, \bgam, \beta,\lambda)\\
& \propto \beta \times \int \mL(\by_i \mid \bY^{-i}, c_i \not\in \bc^*_{-i} , \bc_{-i}, \bgam, \beta,\lambda, \tilde \bpsi)d\Pi(\tilde\bpsi)\\
& \propto \beta\times  \int \left[  \int  \mL(\by_i \mid \bY^{-i}, \bpsi^*_{c_i}, \bc_{-i}, \bgam, \beta,\lambda, \tilde \bpsi)  d\Pi(\bpsi^*_{c_i}\mid \tilde\bpsi, \bgam) \right ]  d\Pi(\tilde\bpsi)  \\
& \propto \beta \times \prod\limits_{\{ A\in\mI: \gamma(A) = 1 \}}\mL^A_1(\by_i) \times \left[ \int \prod\limits_{\{ A\in\mI: \gamma(A) = 0 \}} \frac{\mL^A_0(\bY \mid \tilde\bpsi(A))}{\mL^A_0(\bY^{-i} \mid \tilde\bpsi(A))} d\Pi(\tilde\bpsi)	\right] \\
& \propto \beta \times \mL_1(\by_i\mid\bgam ) \times \left[ \int \prod\limits_{\{ A\in\mI: \gamma(A) = 0 \}} \frac{\mL^A_0(\bY \mid \tilde\bpsi(A))}{\mL^A_0(\bY^{-i} \mid \tilde\bpsi(A))} d\Pi(\tilde\bpsi)	\right] .
\end{aligned}
\label{eq:full_c2}
\end{equation}
Putting together (\ref{eq:full_c1}) and (\ref{eq:full_c2}), we have 
\begin{equation}
\begin{aligned}
\Pr(c_i = c\text{ for some }c\in c_{-i}\mid \bc_{-i}, \bY,  \bgam,\beta,\lambda ) &\propto n_{-i,c}\times \frac{\mL_1(\by_i, \bY^{-i}_c\mid \bgam )}{\mL_1(\bY_c^{-i}\mid \bgam)}  \\
\Pr(c_i \not= c_j\text{ for all } j \not= i\mid \bc_{-i}, \bY,  \bgam,\beta,\lambda ) &\propto \beta\times \mL_1(\by_i \mid \bgam).
\end{aligned}
\label{eq:full_c}
\end{equation}
Note that the integral term in (\ref{eq:full_c1}) and (\ref{eq:full_c2}) represents the marginal likelihood of $\by_i$ at the nodes with $\gamma(A)=0$ conditioning on all other samples. This likelihood does not depend on $\bc_{-i}$ since the cluster-specific parameters $\bpsi^*_c$ share the same value at these nodes regardless of $c$.

\end{enumerate}


\subsection{Numerical approximations to the marginal likelihoods}
To fully specify the Gibbs sampler, we need to numerically approximate the following marginal likelihoods at $A\in\mI$:
\begin{enumerate}
\item
The marginal likelihoods of a set of samples from the same cluster given $\gamma(A) = 1$: $\mL^A_1(\bY^I_c)$;
\item
The marginal likelihoods of the samples given $\gamma(A) = 0$:
\begin{equation}
\begin{aligned}
\int \mL^A_0(\bY  \mid \tilde\bpsi(A))d\Pi( \tilde\bpsi(A) ).
\end{aligned}
\label{eq:ml_0}
\end{equation}
\end{enumerate}
Note that (\ref{eq:ml_0}) is a one-dimensional integral that can be easily approximated by quadrature. For $\mL^A_1(\bY^I_c)$, let 
\begin{equation}
\begin{aligned}
g(\theta, \tau) & =  \prod\limits_{\{i\in I:c_i=c\}}  {y_i(A) \choose y_i(A_l)} \frac{B(\theta\tau + y_i(A_l), (1 - \theta)\tau + y_i(A_r))}{B(\theta\tau, (1 - \theta)\tau)} \\
h(\theta) & = \frac{\theta^{\theta_0(A)\nu_0(A) - 1} (1 - \theta)^{(1 - \theta_0(A))\nu_0(A) - 1}}{B(\theta_0(A)\nu_0(A), (1 - \theta_0(A))\nu_0(A))}.
\end{aligned}
\end{equation}
Then we have
\begin{equation}
\begin{aligned}
\mL^A_1(\bY_c^I) & = \int \left[ \int  g(\theta, \tau) dF^A(\tau)\right]h(\theta) d\theta.
\end{aligned}
\label{eq:int_num}
\end{equation}
Both integrals in (\ref{eq:int_num}) are one-dimensional. In our software, we approximate both integrals with quadratures. 

 \subsection{Compute the clustering centroids}

We can also portray the cluster centroids given any representative clustering and the corresponding coupling indicators. Suppose that $\bc_{rep} = \bm{g}^{(t_{0})}$ for $B< t_0 \leq T$ is one of the representative clusterings and let $\bgam_{rep} = \bm{s}^{(t_0)}$. For the $k$-th resulting cluster defined by $\bc_{rep}$, $1\leq k\leq |\bc_{rep}^*|$, the posterior mean of the branching probability at $A\in\mI$ given the coupling status $\gamma_{rep}(A)$ can be written as
\begin{equation}
\begin{aligned}
\mathbb{E}[\theta^*_k(A) \mid \bY, \gamma_{rep}(A) = 1] &= \iint \theta^*_k(A)\times\pi(\theta^*_k(A) ,\tau^*_k(A) \mid \bY_k) d\theta^*_k(A) d\tau^*_k(A)\\
\mathbb{E}[\tilde\theta(A) \mid \bY, \gamma_{rep}(A) = 0] &= \iint \tilde\theta(A)\times\pi(\tilde\theta(A) ,\tilde\tau(A) \mid \bY) d\tilde\theta(A) d\tilde\tau(A)\\
\end{aligned}
\label{eq:centroid_expt}
\end{equation}
where 
\begin{equation}
\begin{aligned}
\pi(\theta^*_k(A) ,\tau^*_k(A)\mid \bY_k) &\propto  \prod\limits_{\by_i\in\bY_K}  {y_i(A)\choose y_i(A_l)}\frac{B(\theta^*_k(A)\tau^*_k(A) + y_i(A_l), (1 - \theta^*_k(A))\tau^*_k(A) + y_i(A_r))}{B(\theta^*_k(A)\tau^*_k(A), (1 - \theta^*_k(A))\tau^*_k(A))}\\
&\hspace{15mm} \times  \pi(\theta^*_k(A) ,\tau^*_k(A))\\
\pi(\tilde\theta(A) ,\tilde\tau(A) \mid \bY) & \propto \mL^A_0(\bY\mid \tilde\bpsi(A)) \times \pi(\tilde\theta(A) ,\tilde\tau(A)).
\end{aligned}
\label{eq:sec_2_centroid}
\end{equation}
Note that (\ref{eq:centroid_expt}) involves two-dimensional integrals that can be numerically approximated accurately. Let $\bar\bp_k = (\bar p_{k1},\ldots, \bar p_{kM})$ be the posterior mean of $\bp^*_k$, which is the centroid of the $k$-th cluster. For $\omega_j\in\Omega$, let $A^j_0\rightarrow A^j_1\rightarrow\cdots\rightarrow A^j_{l_j}\rightarrow\omega_j$ be the unique path in $\T$ connecting $\Omega$ and $\omega_j$. Then
\begin{equation}
\begin{aligned}
\bar p_{kj} = \prod\limits_{l=0}^{l_j} \hspace{5pt} & \mathbb{E}[\theta^*_k(A^j_l) \mid \bY, \gamma_{rep}(A^j_l) = 1]^{\mathbbm{1}(\gamma_{rep}(A^j_l) = 1)}  \\
\times & \mathbb{E}[\tilde\theta(A^j_l) \mid \bY, \gamma_{rep}(A^j_l) = 0]^{\mathbbm{1}(\gamma_{rep}(A^j_l) = 0)}.
\end{aligned}
\end{equation}
Note that $\bar\bp_k$ only characterizes the centroid of the $k$-th cluster. To characterize the within-cluster dispersion, we need to look at the posterior distribution of $\tau^*_k(A)$ or $\tilde\tau(A)$ for $A\in\mI$, which is available through marginalizing out $\theta^*_k(A)$ or $\tilde\theta(A)$ in (\ref{eq:sec_2_centroid}).

\subsection{Sample classification for microbiome compositions with DTMM}

The DTMM framework can also be used in the supervised setting to achieve sample classification based on a training microbiome dataset. Without loss of generality, suppose that the training dataset contains microbiome samples from $K$ classes: $\{ (\by_1, c_1), (\by_2, c_2),\ldots, (\by_n, c_n) \}$, $1\leq c_i \leq K$. We consider the following generative model of $\bY$:
\begin{equation}
\begin{aligned}
\by_i \mid c_i = k & \overset{\rm{ind}}{\sim}  \text{DT}_{\T}(\bthe^*_k, \btau^*_k) \\
(\bthe^*_k, \btau^*_k)  & \overset{\rm{iid}}{\sim} G_0(\bthe, \btau\mid\bgam,\tilde\bthe,\tilde\btau)\\
\Pr(c_i = k) & = \pi_k,
\end{aligned}
\label{eq:class}
\end{equation}
where $G_0$ and the hyperparameters of $G_0$ are specified as in (\ref{eq:model_base}) and (\ref{eq:model_hyper}). 

Let $\tilde\by$ be a new microbiome sample from (\ref{eq:class}). It follows that 
{\small\begin{equation}
\begin{aligned}
 \Pr(\tilde c = k \mid \tilde\by, \bY) & \propto \pi_k \mL(\tilde\by \mid \tilde c = k, \bY) \\
&\propto \sum\limits_{\bgam} \pi(\bgam\mid\bY) \iint \mL(\tilde\by \mid \bpsi^*_k, \bgam,\tilde\bpsi, \bY_k) d\Pi(\bpsi^*_k\mid \bY_k, \bgam, \tilde\bpsi) d\Pi(\tilde\bpsi\mid\bY).
\end{aligned}
\label{eq:class_inference}
\end{equation}}
Note that (\ref{eq:class_inference}) can be numerically evaluated in a way similar to the marginal likelihood evaluation in DTMM. 
 
Similar to the computations in (\ref{eq:full_c1}), we have
{\small\begin{equation}
\begin{aligned}
& \Pr(\tilde c = k \mid \tilde\by, \bY)\\
& \propto \pi_k \mL(\tilde\by \mid \tilde c = k, \bY) \\
&\propto\pi_k \sum\limits_{\bgam} \pi(\bgam\mid\bY) \iint \mL(\tilde\by \mid \bpsi^*_k, \bgam,\tilde\bpsi, \bY_k) d\Pi(\bpsi^*_k\mid \bY_k, \bgam, \tilde\bpsi) d\Pi(\tilde\bpsi\mid\bY) \\
& \propto\pi_k \sum\limits_{\bgam} \pi(\bgam\mid\bY) \int \left[ \int \mL(\tilde\by \mid \bpsi^*_k, \bgam,\tilde\bpsi, \bY_k) d\Pi(\bpsi^*_k\mid\bgam, \tilde\bpsi) \right] d\Pi(\tilde\bpsi) \\
& \propto \pi_k \sum\limits_{\bgam} \pi(\bgam\mid\bY) \prod\limits_{\{ A\in\mI: \gamma(A) = 1 \}}	\frac{\mL^A_1(\tilde\by,\bY_k) }{\mL^A_1(\bY_k)} \times \left[ \int \prod\limits_{\{ A\in\mI: \gamma(A) = 0 \}} \frac{\mL^A_0(\tilde\by, \bY \mid \tilde\bpsi(A))}{\mL^A_0(\bY \mid \tilde\bpsi(A))} d\Pi(\tilde\bpsi)	\right] \\
&  \propto \pi_k \sum\limits_{\bgam} \pi(\bgam\mid\bY) \prod\limits_{\{ A\in\mI: \gamma(A) = 1 \}}	\frac{\mL^A_1(\tilde\by,\bY_k) }{\mL^A_1(\bY_k)} \times  \prod\limits_{\{ A\in\mI: \gamma(A) = 0 \}} \int \frac{\mL^A_0(\tilde\by, \bY \mid \tilde\bpsi(A))}{\mL^A_0(\bY \mid \tilde\bpsi(A))} d\Pi(\tilde\bpsi(A)),
\end{aligned}
\label{eq:class_inference_supp}
\end{equation}}
where
\begin{equation}
\begin{aligned}
\pi(\bgam\mid\bY) & \propto \pi(\bgam) \mL(\bY\mid \bgam) \\
& \propto \pi(\bgam)  \prod\limits_{\{ A\in\mI: \gamma(A) = 1 \}} \prod\limits_{1\leq k\leq K}\mL^A_1(\bY_k) \times \prod\limits_{\{ A\in\mI: \gamma(A) = 0 \}}\int \mL^A_0(\bY  \mid \tilde\bpsi(A))d\Pi( \tilde\bpsi(A) ).
\end{aligned}
\label{eq:class_inference_supp2}
\end{equation}
Note that (\ref{eq:class_inference_supp}) and (\ref{eq:class_inference_supp2}) can be evaluated numerically.


\clearpage

\section{Additional materials for the numerical examples}


\subsection{Details for the simulation setups}

In this section, we provide details of the data generating process for the simulation studies in Section 3.1.

\begin{enumerate}
\item [$\RN{1}$.] {\textbf{Dirichlet-tree kernel.}} We first fix $\T=\T_6$ as in \ref{fig:sec_3_phylotree} and let $H_k(\bp_i \mid \bbet_k) = {\rm{DT}}_{\T}(\bthe_k, \btau_k)$ such that DTMM is the ``true'' model. The parameters $(\bthe_k, \btau_k)$ are chosen such that the branching probabilities at the 5 internal nodes have the Beta distributions as shown in \ref{fig:sec_3_dt}, where $\bnu_1 = (10\alpha, 2\alpha)$, $\bnu_2 = (6\alpha, 6\alpha)$, $\bnu_3 = (2\alpha, 10\alpha)$ and $\gamma = 0.1$. We write this specific family of Dirichlet-tree distributions as $\text{DT}_{\T_6}(\bnu_k;\alpha;\tau)$, $k=1,2,3$. Note that when $\alpha = \gamma$, the Dirichlet-tree distribution becomes the Dirichlet distribution. In this case, only the branching probabilities at node C contribute to the clustering. The signals are thus local to a single node. Three signal levels are considered by letting $\alpha = 1,3,6$.

\vspace{0.5em}
 \begin{minipage}{\textwidth}   
  \begin{minipage}[b]{0.45\textwidth}
    \centering
   \includegraphics[width = 7.1cm]{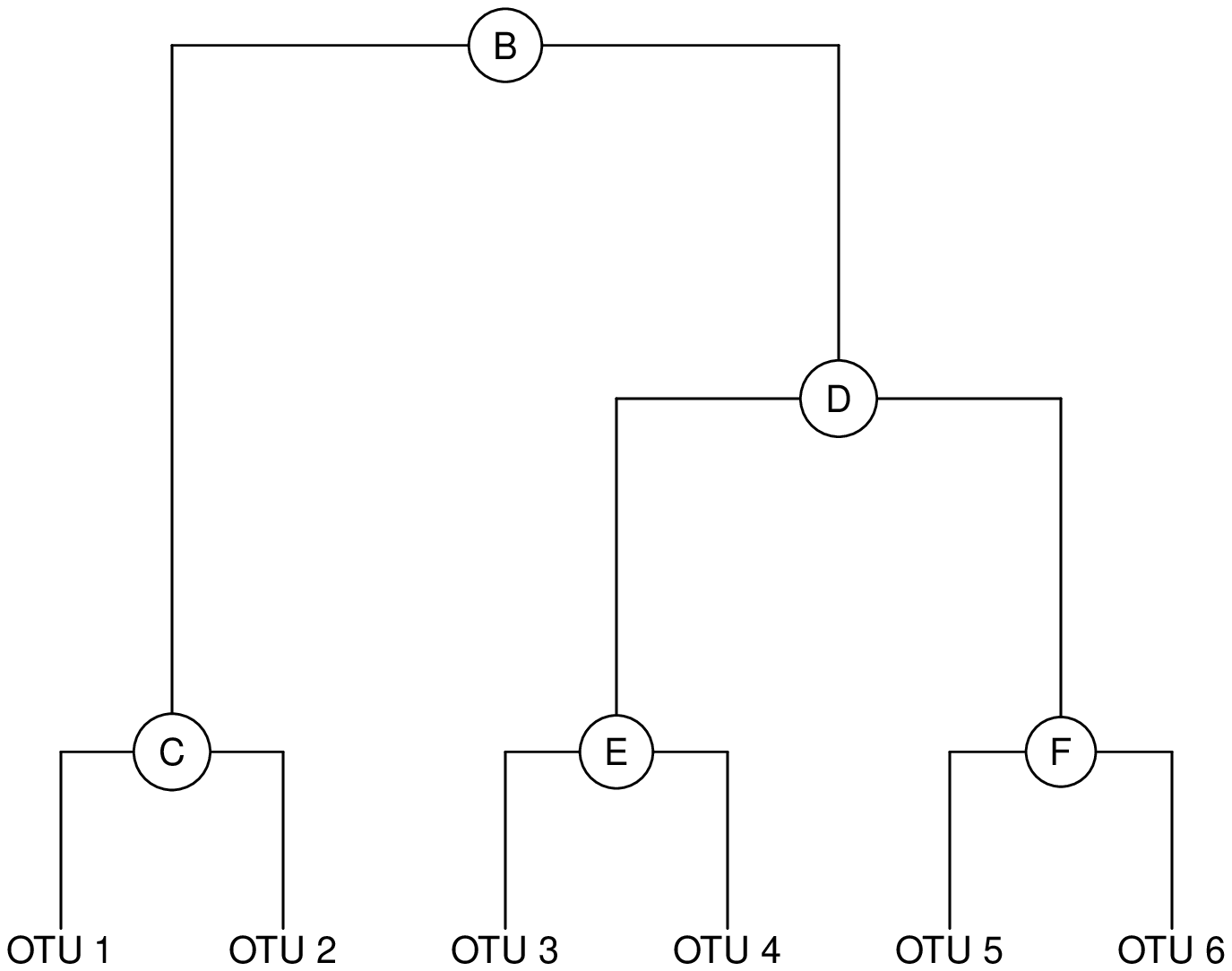}
     \captionof{figure}{The phylogenetic tree for the simulated examples.}
\label{fig:sec_3_phylotree}
  \end{minipage}
  \hspace{0.5cm}
   \begin{minipage}[b]{0.45\textwidth}
    \centering
 \resizebox{1\textwidth} {!}{ \begin{tikzpicture}[sibling distance=8em,
  every node/.style = {shape=rectangle, rounded corners,
     align=left, 
    top color=white, bottom color=white}]]
  \node {${\rm{B: Beta}}(12\alpha, 12\alpha)$}
    child { node[draw] { ${\rm{C: Beta}(\bnu_k)}$} }
    child { node {${\rm{D: Beta}}(8\gamma, 4\gamma)$}
      child { node {${\rm{E: Beta}}(4\gamma, 4\gamma)$}}
      child { node {${\rm{F: Beta}}(2\gamma, 2\gamma)$} } };
\end{tikzpicture} }
\vspace{0.3cm}
   
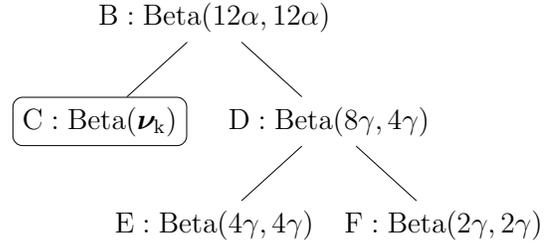
\captionof{figure}{Distributions for the branching probabilities in Scenario $\RN{1}$.}
        \label{fig:sec_3_dt}
  \end{minipage}   
       \hfill
  \end{minipage} 

\item [$\RN{2}$.] {\textbf{Dirichlet kernel.}} In this scenario, we let $H_k(\bp_i \mid \bbet_k) = {\rm{Dir}}(\balp_k)$ such that DMM is the ``true'' model that generates the data. In this case, DTMM is still correct but is over-specified. We let $\balp_1 = (2,2,5,2,3,1)\cdot \alpha_0$, $\balp_2 = (2,4,3,2,1,3)\cdot \alpha_0$ and $\balp_3 = (2,6,1,2,2,2)\cdot \alpha_0$ for $\alpha_0 > 0$ such that all six OTUs are active in differentiating the clusters and the signals are global. We consider three signal levels with $\alpha_0 = 1,3,5$.
\end{enumerate}

In the following three examples, we evaluate the performance of DTMM when the model is misspecified. Let $\bp = (p_{1}, \ldots, p_6) \in \mathbb{S}^5$. We say that $\bp$ has the logistic normal distribution \citep{atchison1980logistic} and denote as $\bp\sim{\text{Logit-Norm}}(\bmu, \bSigma)$ if 
\begin{equation*}
\begin{aligned}
\bx & =  \left(\log \left(\frac{p_{1}}{p_{6}}\right), \ldots, \log\left(\frac{p_{5}}{p_{6}}\right)\right)^\top\\
\bx & \overset{\rm{ind}}{\sim} \text{N}(\bmu, \bSigma). 
\end{aligned}
\end{equation*}
In Scenarios $\RN{3}$, $\RN{4}$ and $\RN{5}$, we let $H_k(\bp_i \mid \bbet_k)={\text{Logit-Norm}}(\bmu_k, \bSigma_k)$. In these examples, we assume that the phylogenetic tree $\T_6$ provides some insights on the covariate structures of the OTUs.  

\begin{enumerate} 
\item [$\RN{3}$.] {\textbf{Logistic-normal approximations to the Dirichlet-tree kernel.}} Consider the Dirichlet-tree kernels $q_k=\text{DT}_{\T_6}(\bnu_k;\alpha;\tau)$ as in Scenario $\RN{1}$ with $\gamma = 0.5$ for $k=1,2,3$. In this example, we let
\begin{equation*}
\begin{aligned}
H_k(\bp_i \mid \bbet_k) = \argmin\limits_{h \in \mathbb{L}^5} \infdiv{q_k}{h}
\end{aligned}
\end{equation*}
where $\infdiv{q_k}{h}$ is the Kullback-Leibler divergence from $h$ to $q_k$, $\mathbb{L}^5$ the set of logistic-normal distributions on $\mathbb{S}^5$. It is shown in Section B.2 in the supplementary material that $H_k(\bp_i \mid \bbet_k) = \text{Logit-Norm}(\bmu_k, \bSigma_k)$, where
\begin{equation*}
\begin{aligned}
\bmu_k = \mathbb{E}_{q_k}\left[\log\left(\frac{\bx_{-6}}{x_6} \right)\right], \quad \bSigma_k = \mathbb{V}_{q_k}\left[\log\left(\frac{\bx_{-6}}{x_6} \right)\right],
\end{aligned}
\end{equation*}
which are available in closed form by the properties of the exponential family. In this scenario, DTMM is not the correct model, but we expect that it is not severely misspecified. Three signal levels are considered with $\alpha = 3, 6, 9$.

\item [$\RN{4}$.] {\textbf{Logistic-normal kernel (single node).}} In this scenario, we let $H_k(\bp_i \mid \bbet_k) = \text{Logit-Norm}(\bmu_k, \bSigma_k)$, where $\bSigma_k = \text{diag}(0.05, 0.05, 1,1,1)$, $\bmu_1 = (3,1,a,b,0)$, $\bmu_2 = (2.43, 2.43,a,b,0)$ and $\bmu_3 = (1,3,a,b,0)$. The $\bmu_k$'s are chosen such that only node C in $\T_6$ is relevant for clustering. The diagonal covariance matrix suggests that DTMM is misspecified (in comparison, in $\RN{3}$ the covariance matrices are dense). We consider three signal levels with $(a,b) = (5,3), (2,2)$ and $(1,1)$. Note that when the relative abundance of $\text{OTU}_3$ and $\text{OTU}_4$ are high, the compositional nature of the data implies that fewer counts are generated for $\text{OTU}_1$ and $\text{OTU}_2$ that determine the clustering, resulting in a low signal-to-noise ratio.

\item [$\RN{5}$.] {\textbf{Logistic-normal kernel (multiple nodes).}} Similar to scenario $\RN{4}$, we let $H_k(\bp_i \mid \bbet_k) = \text{Logit-Norm}(\bmu_k, \bSigma_k)$, where $\bSigma_k = \text{diag}(1,1, 0.05,0.05,0.05)$, $\bmu_1 = (c,d,3.5,3,2.5)$, $\bmu_2 = (c,d,2.5,3.5,3)$ and $\bmu_3 = (c,d,3,2.5,3.5)$. In this case, the $\bmu_k$'s are chosen such that the clusters are determined by the relative abundance of $\text{OTU}_3$, $\text{OTU}_4$ and $\text{OTU}_5$, which are reflected in nodes D, E and F in $\T_6$ in \ref{fig:sec_3_phylotree}. Three signal levels with $(c,d) = (6,6), (3,3)$ and $(1,1)$ are considered.
\end{enumerate}


\subsection{Logistic normal approximations to the Dirichlet-tree distribution}

\begin{lem}
For $d\geq 2$ and $d\in\mathbb{N}^+$, let $Q$ be a distribution on $\mathbb{S}^{d-1}$. Let $\mathbb{L}^{d-1}$ be the set of logistic-normal distributions on $\mathbb{S}^{d-1}$. We have
\[
P = \argmin\limits_{h\in\mathbb{L}^d} \infdiv{Q}{h},
\]
where
\begin{equation*}
\begin{aligned}
P\overset{d}{=}\mathrm{Logit}\text{-}\mathrm{Norm}(\tilde\bmu, \tilde\bSigma),\quad \tilde\bmu = \mathbb{E}_{\bx\sim Q}\left[\log\left(\frac{\bx_{-d}}{x_d} \right)\right], \quad \tilde\bSigma = \mathbb{V}_{\bx\sim Q}\left[\log\left(\frac{\bx_{-d}}{x_d} \right)\right].
\end{aligned}
\end{equation*}
\end{lem}

\proof For $h\sim\mathrm{Logit}\text{-}\mathrm{Norm}(\tilde\bmu, \tilde\bSigma)\in\mathbb{L}^d$, the pdf of $h$ can be written as 
\begin{equation*}
\begin{aligned}
p_h(\bx) & = |2\pi\bSigma|^{-\frac{1}{2}}\left( \prod\limits^{d}_{j = 1}x_j\right)^{-1}e^{-\frac{1}{2}\left\{ \log\left(\frac{\bx_{-d}}{x_d}\right) -\bmu \right\}^\top \bSigma^{-1}\left\{ \log\left(\frac{\bx_{-d}}{x_d}\right) -\bmu \right\}} \\
& = f(\bx) \exp\left\{ \bm{\eta}^\top \bm{T}(\bx) - A(\bm{\eta})\right\}, \\
\end{aligned}
\end{equation*}
where
\begin{equation*}
\begin{aligned}
\bm{\eta} &= \begin{pmatrix}
 \bSigma^{-1}\bmu\\
{\rm{vec}}(-\frac{1}{2}\bSigma^{-1})
\end{pmatrix} \\
\bm{T}(\bx) &= \begin{pmatrix}
\log\left(\frac{\bx_{-d}}{x_d}\right)\\
{\rm{vec}}\left[\log\left(\frac{\bx_{-d}}{x_d}\right)\log\left(\frac{\bx_{-d}}{x_d}\right)^\top \right]
\end{pmatrix} \\
A(\bm{\eta}) &= \frac{1}{2}\bmu^\top\bSigma^{-1}\bmu  - \frac{1}{2}\log\bSigma\\
\end{aligned}
\end{equation*}
Thus $h$ is an exponential family distribution with natural parameter $\bm{\eta}$. Note that
\begin{equation*}
\begin{aligned}
\infdiv{Q}{h} & = \mathbb{E}_Q\left[\log\left(\frac{Q}{h}\right) \right] \\
& = \mathbb{E}_Q\left[\log Q \right] + \mathbb{E}_Q\left[ A(\bm{\eta}) \right] -  \mathbb{E}_Q\left[ \bm{\eta}^\top \bm{T}(\bx) \right] - \mathbb{E}_Q\left[\log f(\bx) \right]
\end{aligned}
\end{equation*}
By the property of the exponential family, we have
\begin{equation*}
\begin{aligned}
\nabla_{\bm{\eta}}\infdiv{Q}{h} = \mathbb{E}_{h}[\bm{T}(\bx)] - \mathbb{E}_Q\left[ \bm{T}(\bx) \right]. 
\end{aligned}
\end{equation*}
Let $h^*$ satisfy $\nabla_{\bm{\eta}}\infdiv{Q}{h} = 0$, if follows that $\mathbb{E}_{h^*}[\bm{T}(\bx)] = \mathbb{E}_Q\left[ \bm{T}(\bx) \right]$.
Consider the second derivative of $\infdiv{Q}{h}$ at $h^*$
\begin{equation*}
\begin{aligned}
\nabla\nabla_{\eta}\infdiv{Q}{h} = \nabla\nabla_{\eta}\mathbb{E}_Q\left[ A(\bm{\eta}) \right]  = \mathbb{V}_{h^*}[\bm{T}(\bx)],
\end{aligned}
\end{equation*}
which is positive semi-definite. Therefore, $h^*$ minimizes $\infdiv{Q}{h}$.  \qedhere

\vspace{0.5cm}

Therefore, for $Q \overset{d}{=} \rm{DT}_{\T}(\bthe,\btau)$, the best approximation to $Q$ in the logistic-normal family is $P\overset{d}{=}\mathrm{Logit}\text{-}\mathrm{Norm}(\tilde\bmu, \tilde\bSigma)$ such that
\begin{equation*}
\begin{aligned}
\tilde\bmu = \mathbb{E}_{\bx\sim Q}\left[\log\left(\frac{\bx_{-d}}{x_d} \right)\right], \quad \tilde\bSigma = \mathbb{V}_{\bx\sim Q}\left[\log\left(\frac{\bx_{-d}}{x_d} \right)\right].
\end{aligned}
\end{equation*}
For the $j$-the OTU $\omega_j$, let $A_{jp}$ be its parent in the phylogenetic tree. Let $\alpha_j = \theta(A_{jp})\tau(A_{jp})$ if $\omega_j$ is the left child of $A_{jp}$ and $\alpha_j = (1-\theta(A_{jp}))\tau(A_{jp})$ otherwise. Similarly, for $A\in \T$, let $\beta_A = \theta(A_p)\tau(A_p) - \theta(A)$ if $A$ is the left child of $A_p$ and $\beta_A = (1-\theta(A_p))\tau(A_p) - \theta(A)$ otherwise. The density function of $Q$ can be written as
\begin{equation*}
\begin{aligned}
p_Q(\bx) = \left[\prod\limits_{A\in\T} B(\theta(A)\tau(A), (1-\theta(A))\tau(A)) \right]^{-1} \prod\limits_{j = 1}^d x_j^{\alpha_j - 1} \prod\limits_{A\in\T} x_A^{\beta_A}, 
\end{aligned}
\end{equation*}
where $x_A = \sum_{j\in A} x_j$. Therefore, the Dirichlet-tree distribution is a member of the exponential family with natural parameter $\bm{\eta}= \{ \{\alpha_j: 1\leq j\leq d\}, \{\beta_A: A\in\T \} \}$. Let 
\begin{equation*}
\begin{aligned}
A(\bm{\eta}) = \sum\limits_{A\in\T}\log\left[ B(\theta(A)\tau(A), (1-\theta(A))\tau(A))  \right].
\end{aligned}
\end{equation*}
Again by the property of the exponential family, $\tilde\bmu$ and $\tilde\bSigma$ can be computed with the digamma and the trigamma functions.


\clearpage 
\subsection{Additional tables and figures for the numerical examples}

In this section, we provide additional results and figures for the numerical examples in Section 3.

\comment{
 \null
 \vfill
\begin{figure}[!h]
\centering
\scalebox{0.95}{
\includegraphics[width = 9.5cm]{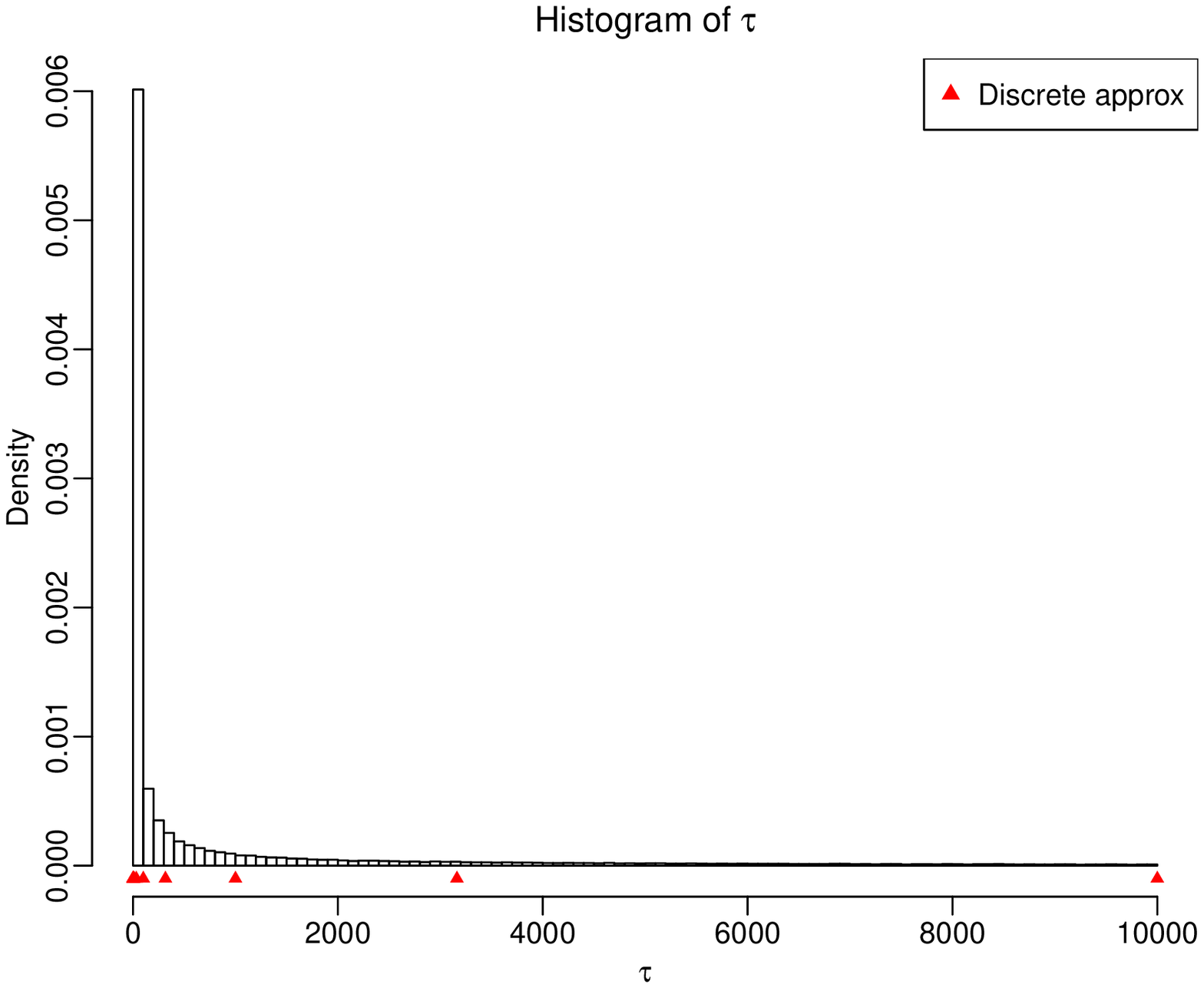}
}
\caption{Histogram of $\tau$. The discrete prior puts equal mass at the red points.}
\label{fig:sec_2_hist_tau}
\end{figure}

\vfill
\clearpage 
}

\null
\vfill
\begin{table}[!ht]
\centering
\caption{RMSE of the Jaccard index (large sample size). Cells with the lowest RMSE in each row are highlighted.}
\label{tab:rmse_180}
\begin{tabular}{@{}llccllp{30pt}p{30pt}p{30pt}p{30pt}p{30pt}c@{}}
\toprule
 \multicolumn{11}{c}{$n = 180$}  \\\toprule
\multicolumn{2}{c}{}                        & \multicolumn{2}{c}{Signal} &  & \multicolumn{6}{c}{Method}       \\ \cmidrule(lr){3-5}\cmidrule(lr){6-11} 
\multicolumn{2}{c}{\multirow{-2}{*}{Expt}}  & Level                & $R^2$ &  & {\color[HTML]{111111} DTMM } & DMM & K-ms & PAM & Hclust & Spec \\ \midrule
&                       &   --   & --    &    & {\bf 0.38}   &  0.75     &   \multicolumn{1}{c}{--}     &     \multicolumn{1}{c}{--}      &    \multicolumn{1}{c}{--}   &    \multicolumn{1}{c}{--}                  \\
&                       & W                    &  0.30   &  &   \bf{0.45}     &   0.65   &    0.67     &   0.72   &  0.65     &       0.71        \\
&                       & M                     &  0.35   &  & {\bf 0.36}     &  0.69   &    0.69     &  0.72   &   0.66     &     0.71          \\
\multirow{-4}{*}{\RN{1}}&\multirow{-4}{*}{DT}  & S                    &  0.36  &  & \bf{0.19}     &    0.72  & 0.69   & 0.71    & 0.65       &    0.70          \\ \midrule
&                       & --    & --    &     &   0.50   & {\bf 0.00}  &   \multicolumn{1}{c}{--}        &   \multicolumn{1}{c}{--}    &      \multicolumn{1}{c}{--}    &    \multicolumn{1}{c}{--}     	   \\
&                       & W                    &  0.34   &  & 0.51     & {\bf 0.46}    &  0.58      &  0.59   &   0.59     &    0.56          \\
&                       & M                    &  0.51   &  &  0.16    &  {\bf 0.12}   &  0.33       &  0.29   &    0.37    &   0.28            \\
\multirow{-4}{*}{\RN{2}}&\multirow{-4}{*}{Dir} & S                    &  0.59  &   & 0.08    & \bf{0.04}    &  0.31       &   0.17   &   0.39     &     0.22          \\ \midrule
&                       & --   & --   &       &  0.67                &  {\bf 0.21}   &   \multicolumn{1}{c}{--}     &   \multicolumn{1}{c}{--}    &     \multicolumn{1}{c}{--}     &       \multicolumn{1}{c}{--}          \\
&                       & W                    &   0.37  &  & 0.56     & 0.55    &   0.52      & {\bf 0.48 }   &    0.56   &   0.51           \\
&                       & M                     &   0.38   &  & {\bf 0.26 }     &   0.55  &     0.49    &  0.40   &    0.56    & 0.44              \\
\multirow{-4}{*}{\RN{3}}&\multirow{-4}{*}{LN-A} & S                 &  0.39   &  & {\bf 0.13 }   &   0.56  & 0.46        &  0.38   &  0.54      &   0.41           \\ \midrule
&                       & --  &  --   &       &  0.74                 & {\bf 0.66}    &    \multicolumn{1}{c}{--}       &    \multicolumn{1}{c}{--}   &    \multicolumn{1}{c}{--}      &           \multicolumn{1}{c}{--}      \\
&                       & W                    &  0.09  &  & {\bf 0.62}     & 0.75   &    0.77     &  0.78   &   0.72    &   0.73        \\
&                      & M                     &  0.40   &  & {\bf 0.46}     & 0.52     &  0.58       &  0.54   &   0.62     &      0.47         \\
\multirow{-4}{*}{\RN{4}}&\multirow{-4}{*}{LN-S} & S                  &   0.59   &  &  {\bf 0.19}      &    0.33  &  0.34       &  0.22   &   0.42  &     0.20          \\ \midrule
&                       & --  & --   &        & {\bf 0.48}                          &   0.69 &    \multicolumn{1}{c}{--}     &   \multicolumn{1}{c}{--}  &  \multicolumn{1}{c}{--}      &       \multicolumn{1}{c}{--}        \\
&                       & W                     &   0.03  &  & {\bf 0.29}     &  0.81   &   0.79      &  0.79   &     0.78   &  0.78              \\
&                       & M                     &   0.22 &  & {\bf 0.24 }    &   0.73  &   0.76      &   0.71    &    0.74    &   0.57            \\
\multirow{-4}{*}{\RN{5}} &\multirow{-4}{*}{LN-M} & S                 &  0.52   &  & \bf{0.17}    &  0.25   &  0.17       & \bf{0.17 }  &   0.24     &    0.17           \\ \bottomrule
\end{tabular}
\end{table}

 \vfill

 \null
\vfill
\begin{figure}[!ht]
\begin{center}
 \begin{minipage}[b]{1\textwidth}
 \centering
\includegraphics[width = 0.33\textwidth]{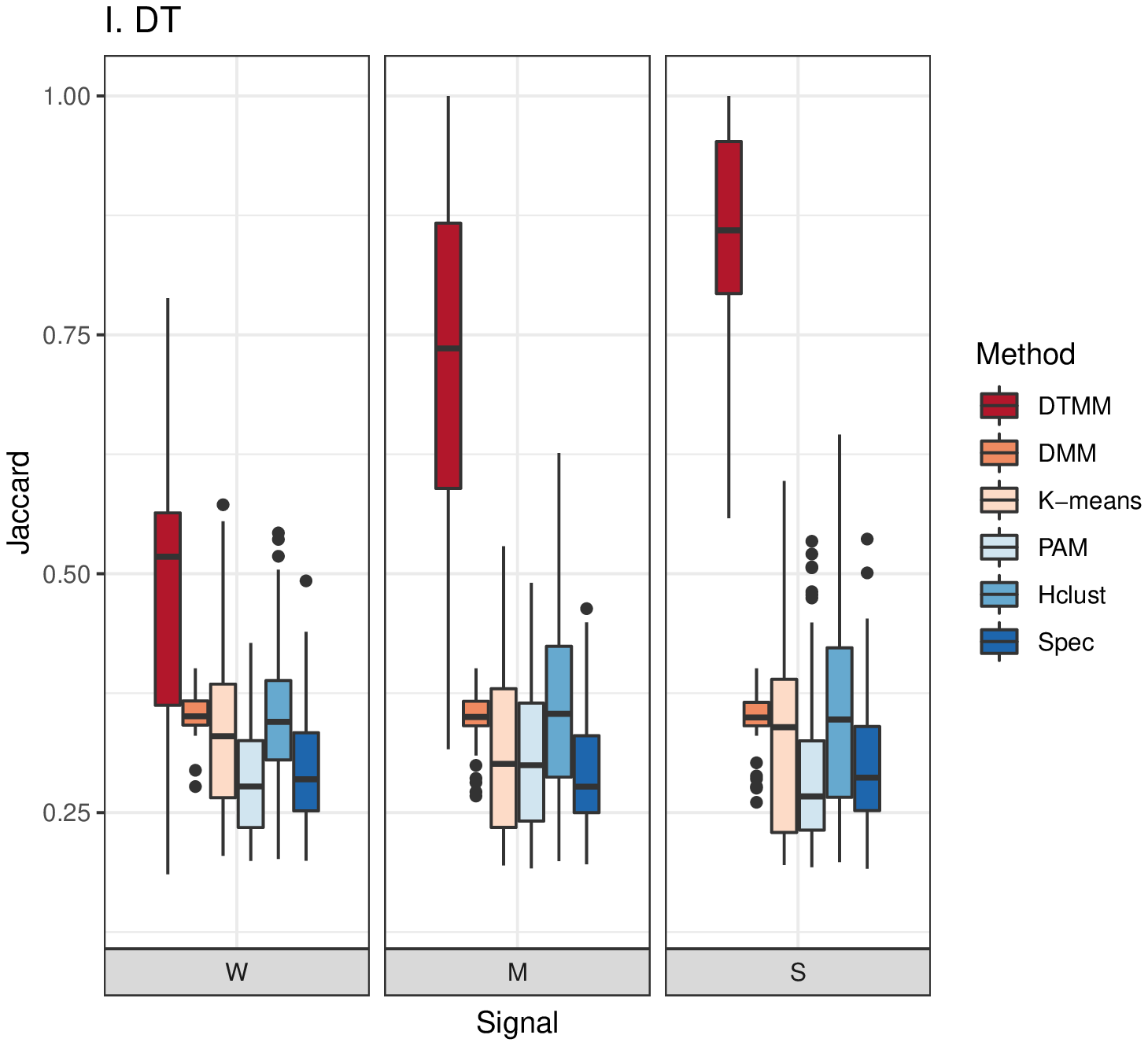}\includegraphics[width = 0.33\textwidth]{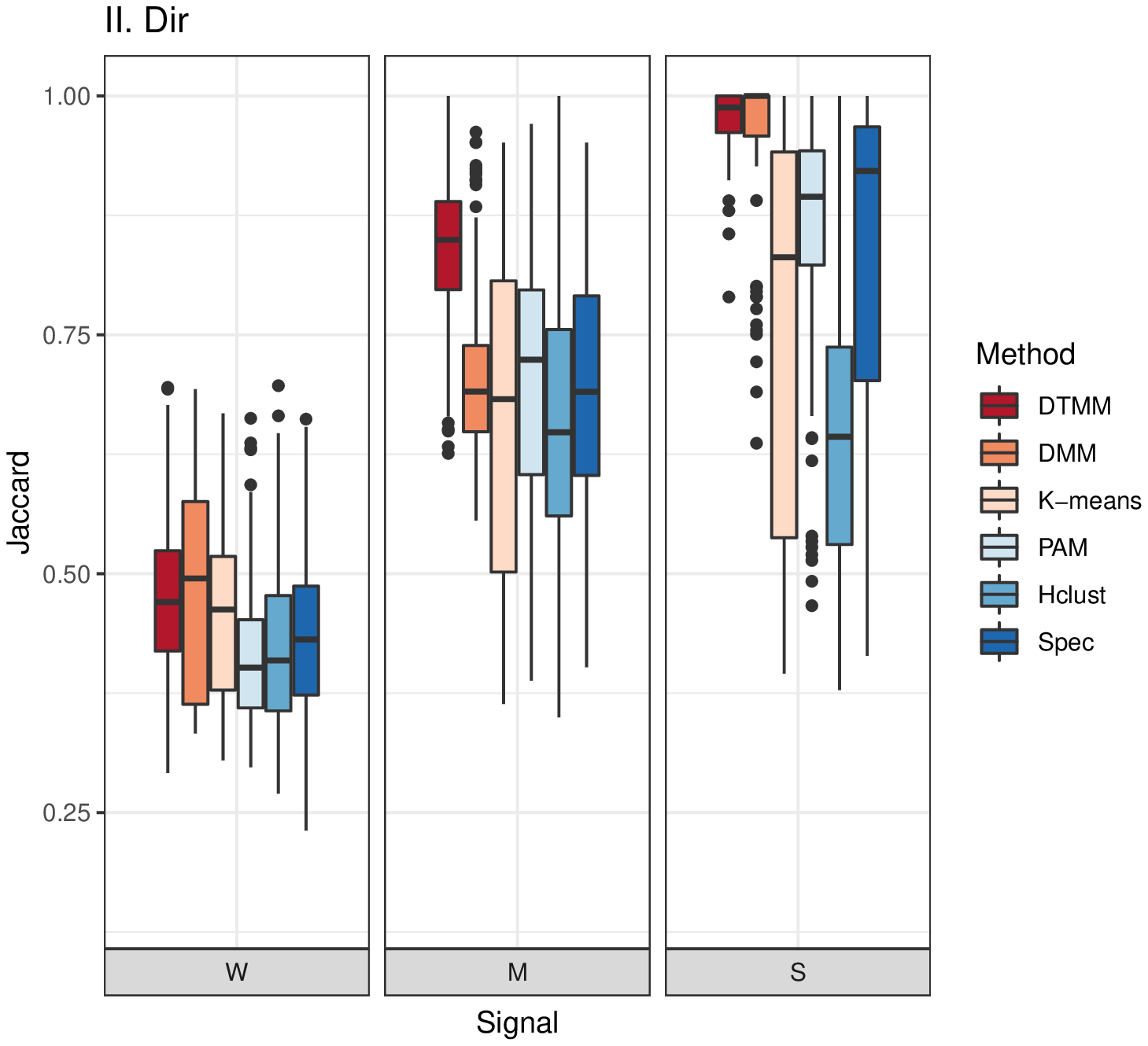}\includegraphics[width = 0.33\textwidth]{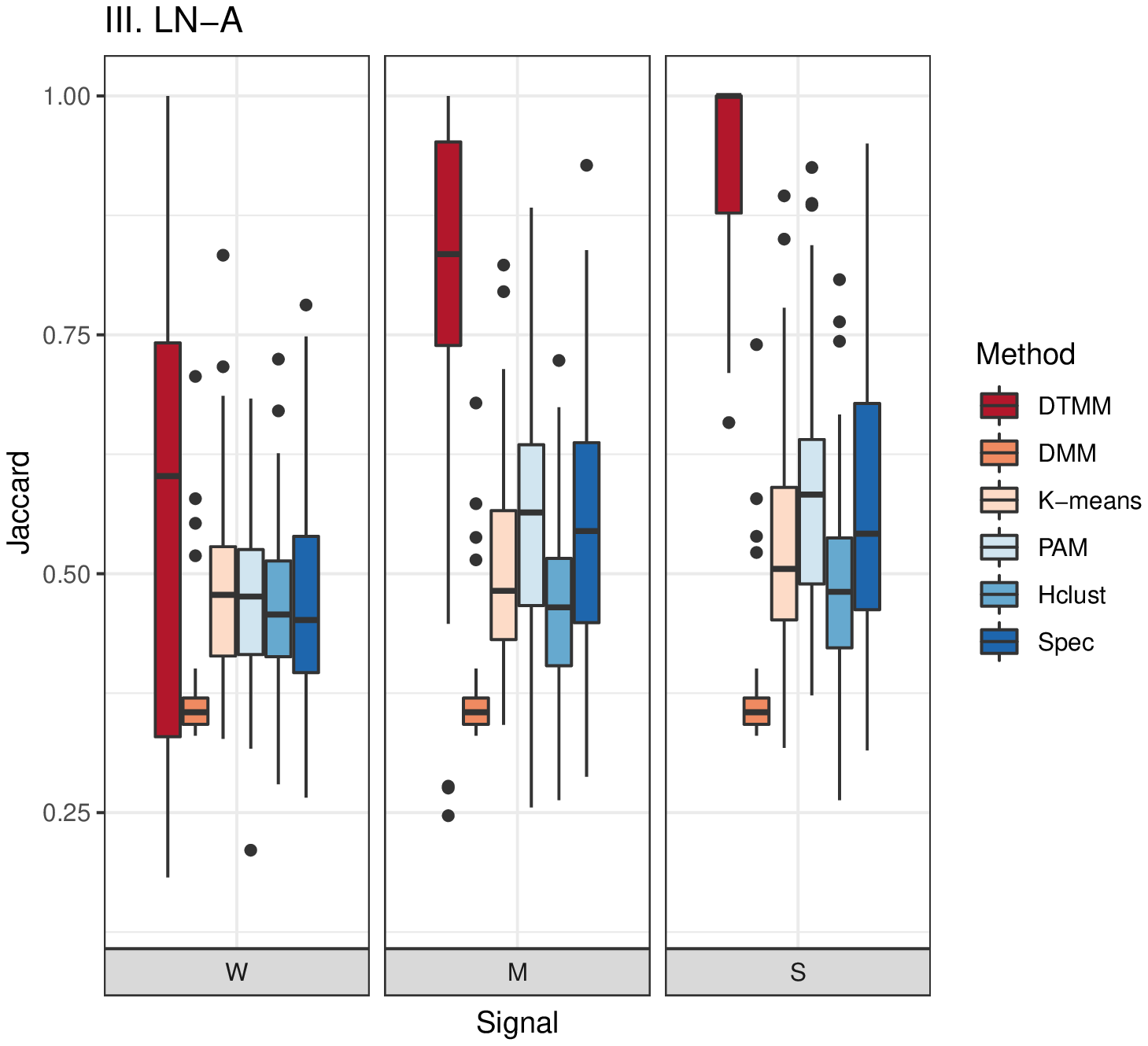}
\includegraphics[width = 0.33\textwidth]{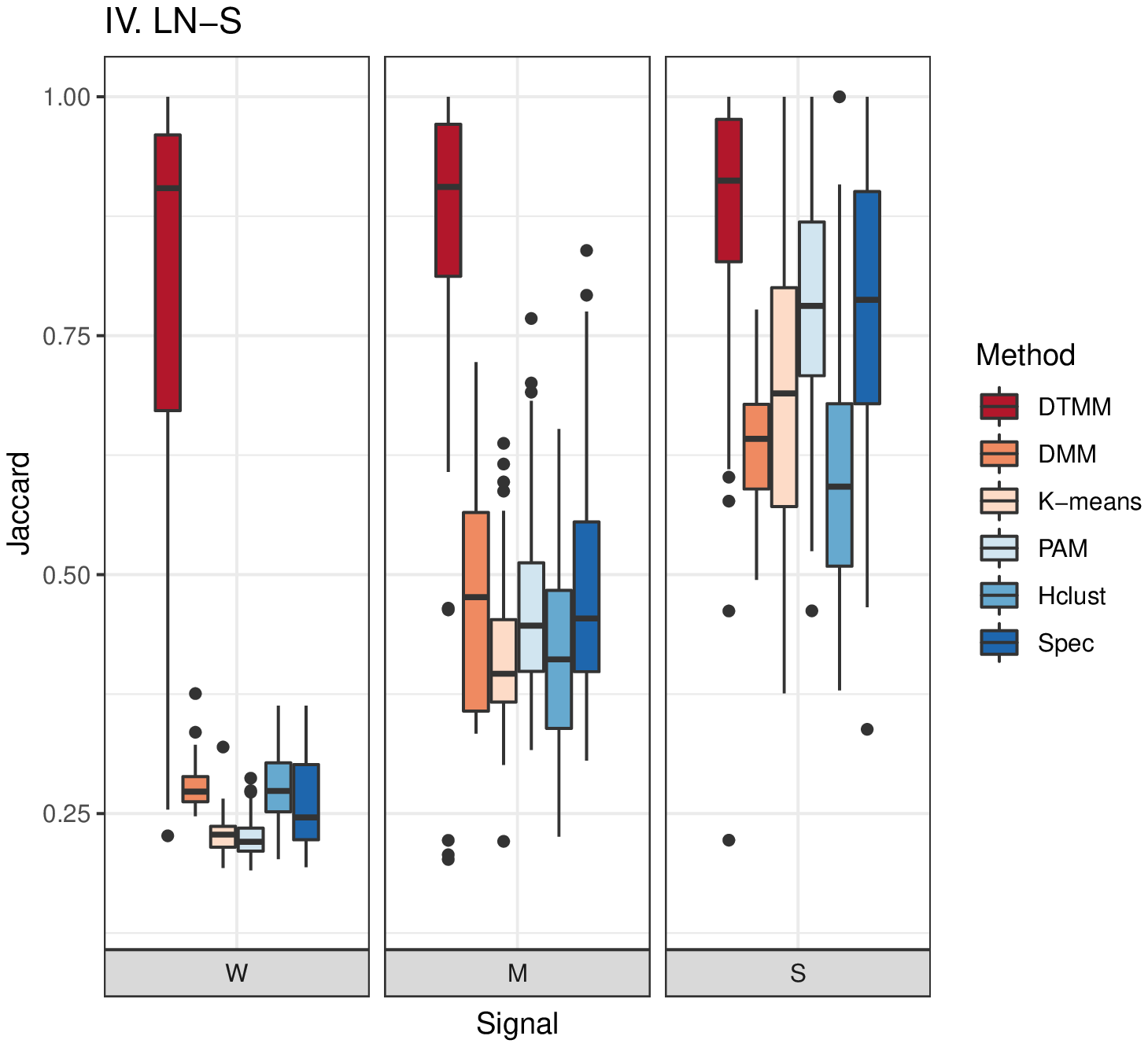}\includegraphics[width = 0.33\textwidth]{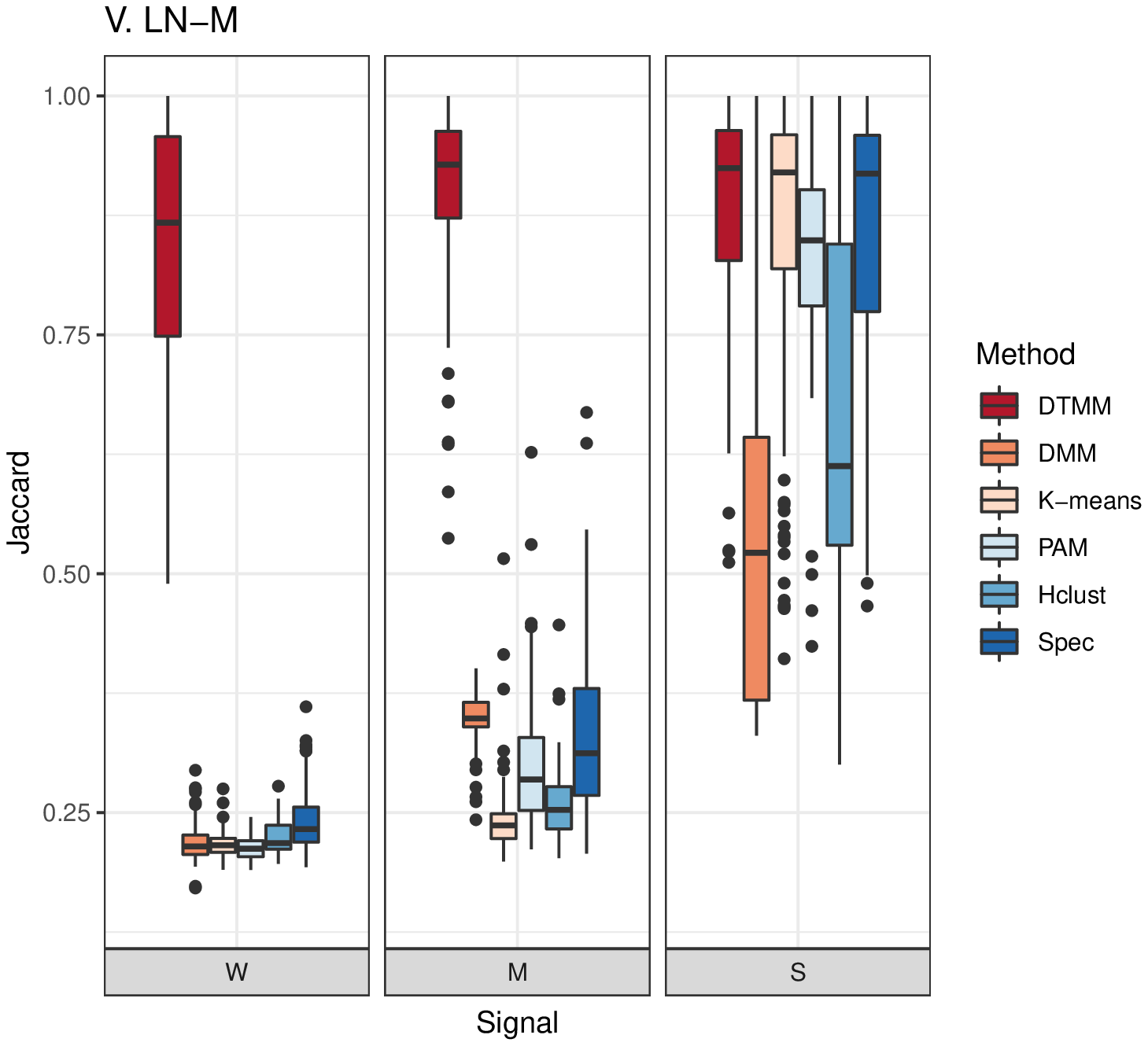}
\caption{Boxplots of the Jaccard index ($n=90$).}
\label{fig:app_sim_box_small}
\end{minipage}
\end{center}
\end{figure}
 \vfill
 
 \clearpage
 \null
 \vfill

\begin{figure}[!ht]
\begin{center}
 \begin{minipage}[b]{1\textwidth}
  \centering
\includegraphics[width = 0.33\textwidth]{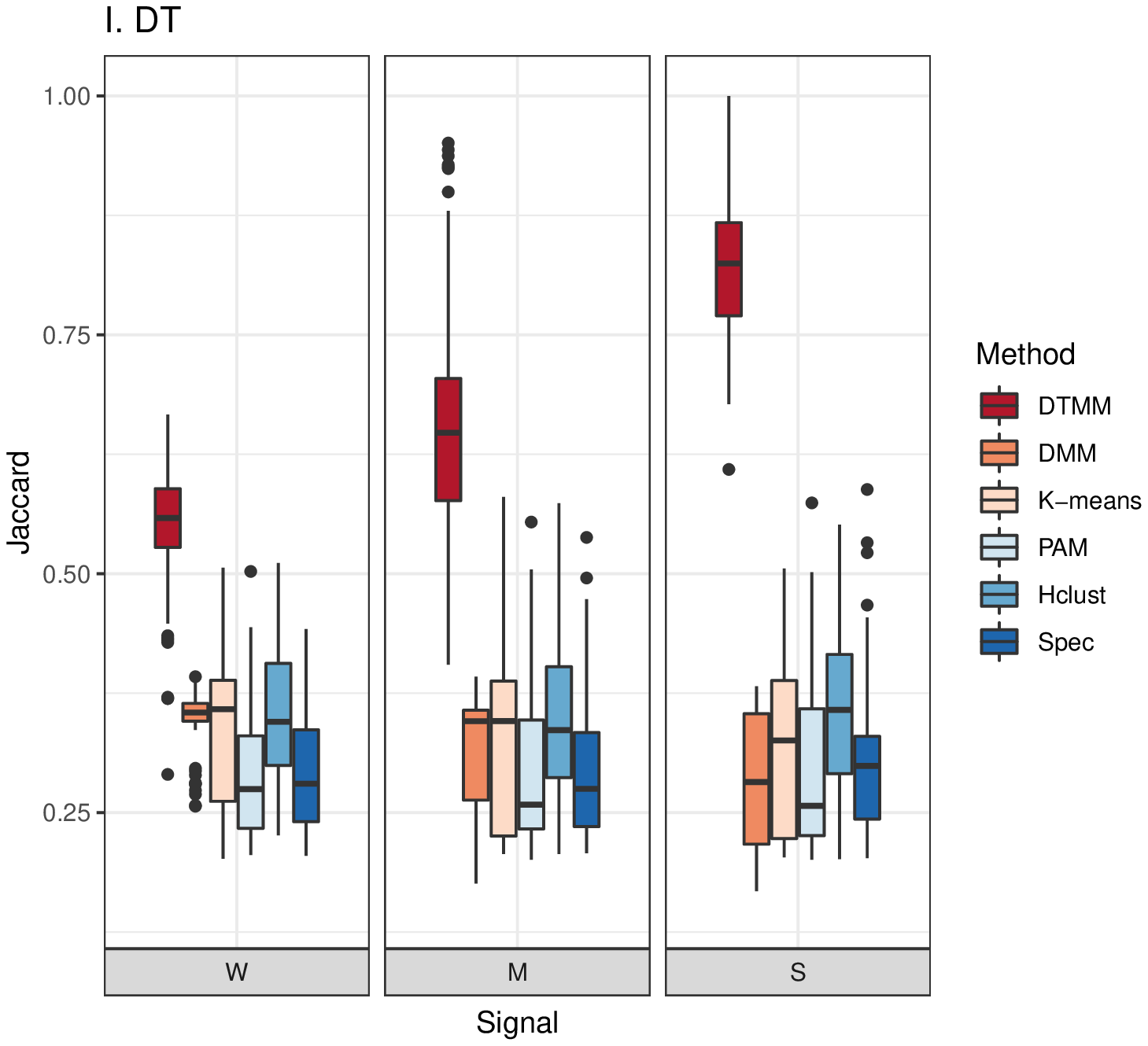}\includegraphics[width = 0.33\textwidth]{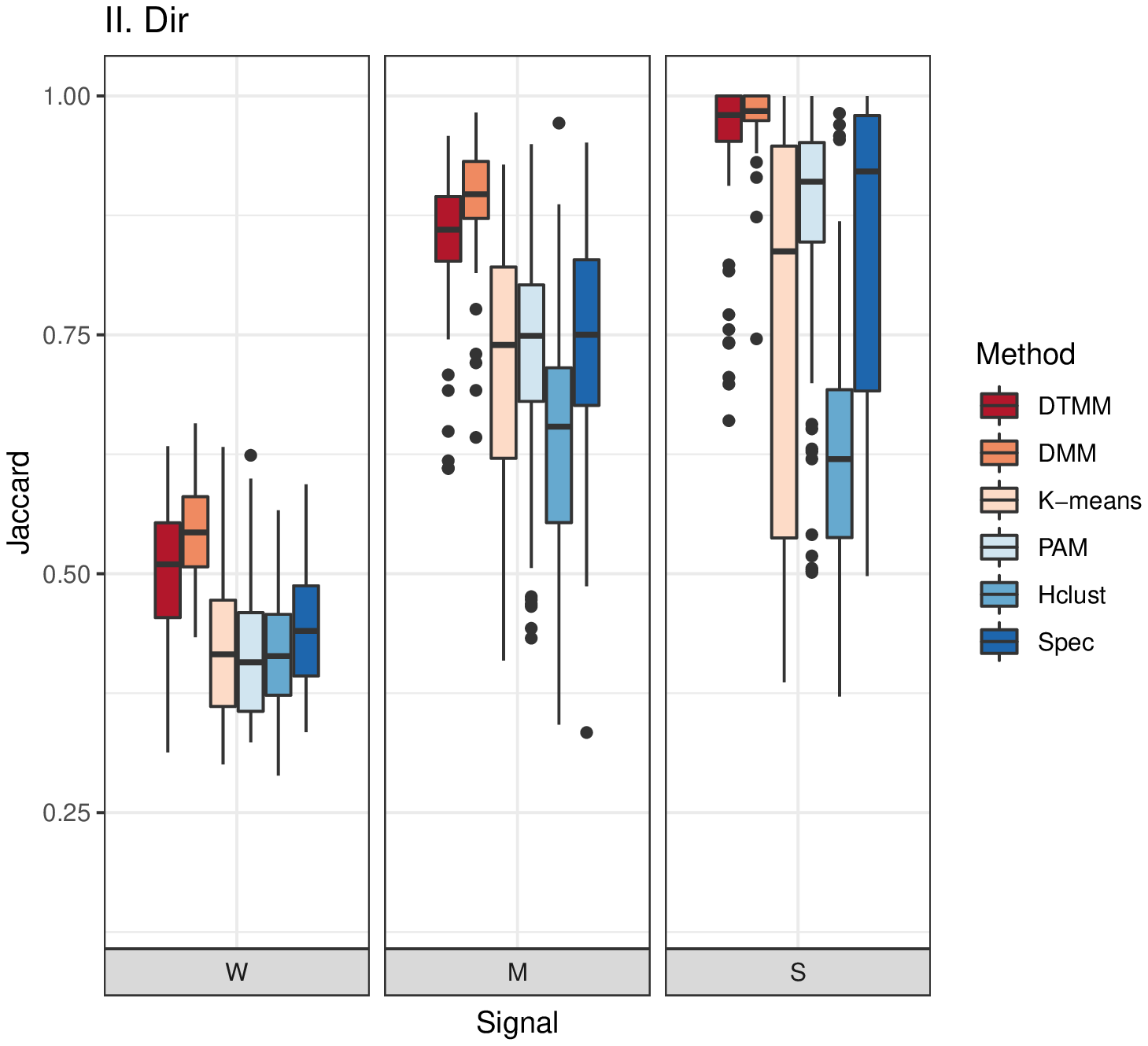}\includegraphics[width = 0.33\textwidth]{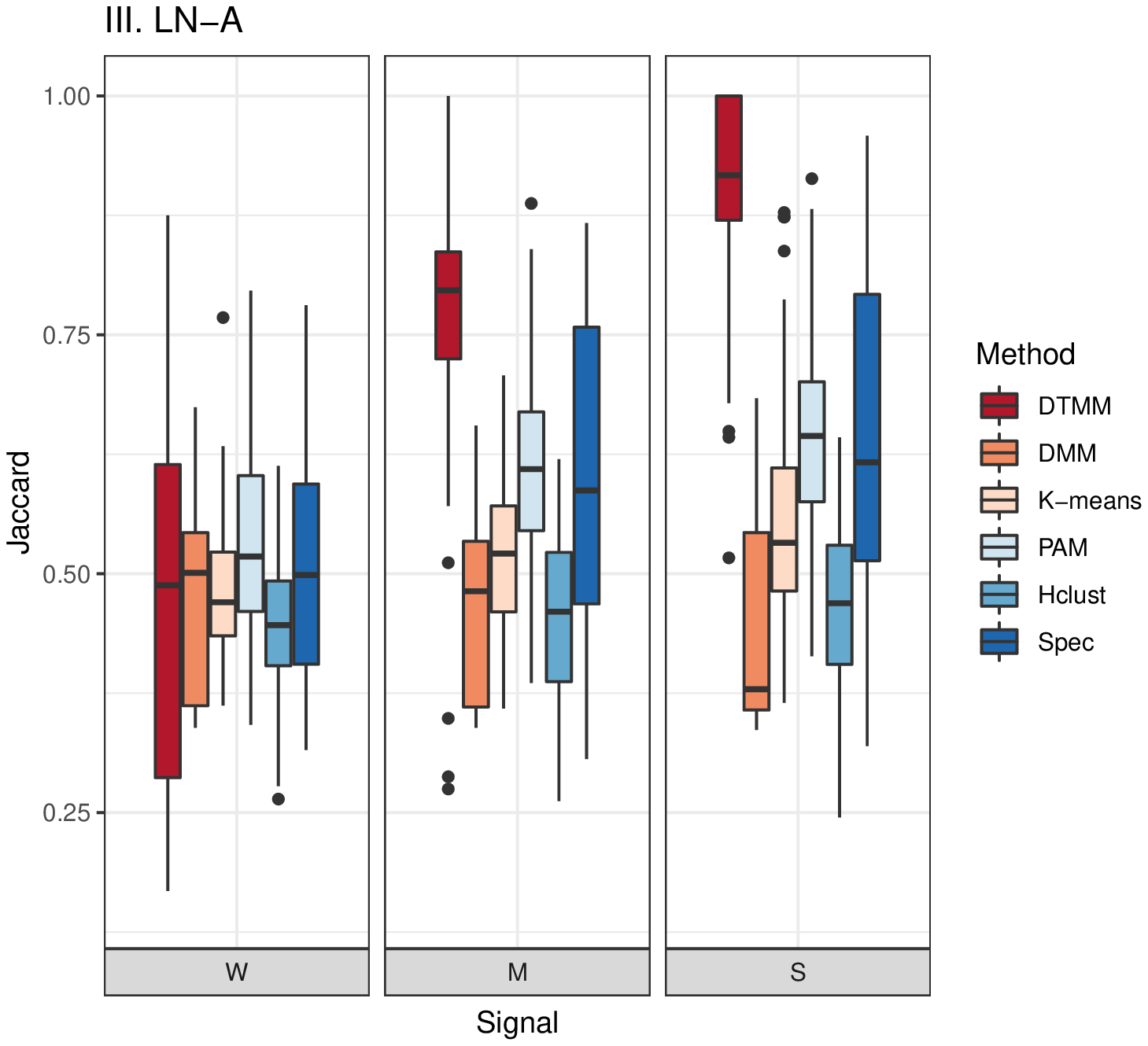}
\includegraphics[width = 0.33\textwidth]{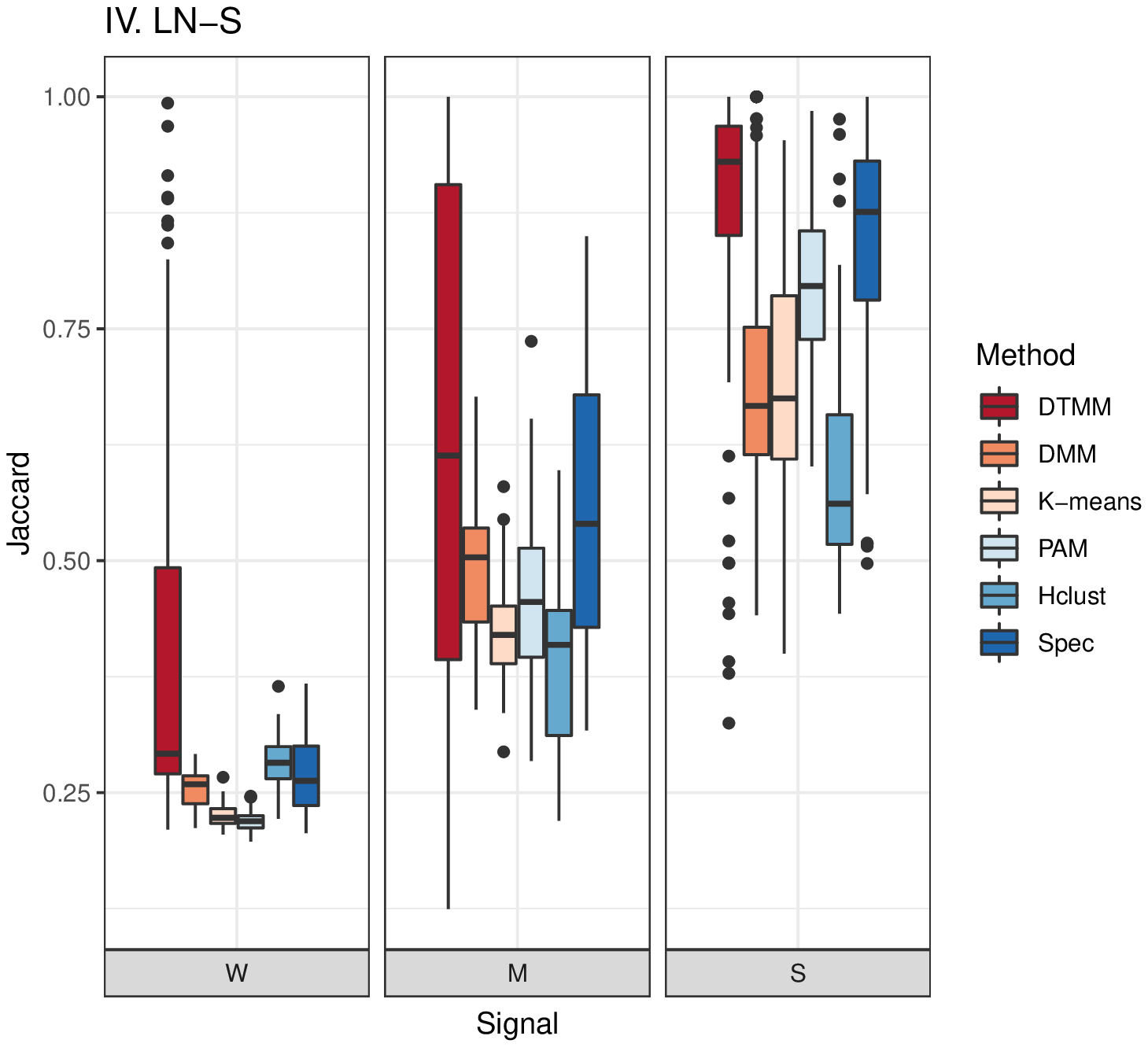}\includegraphics[width = 0.33\textwidth]{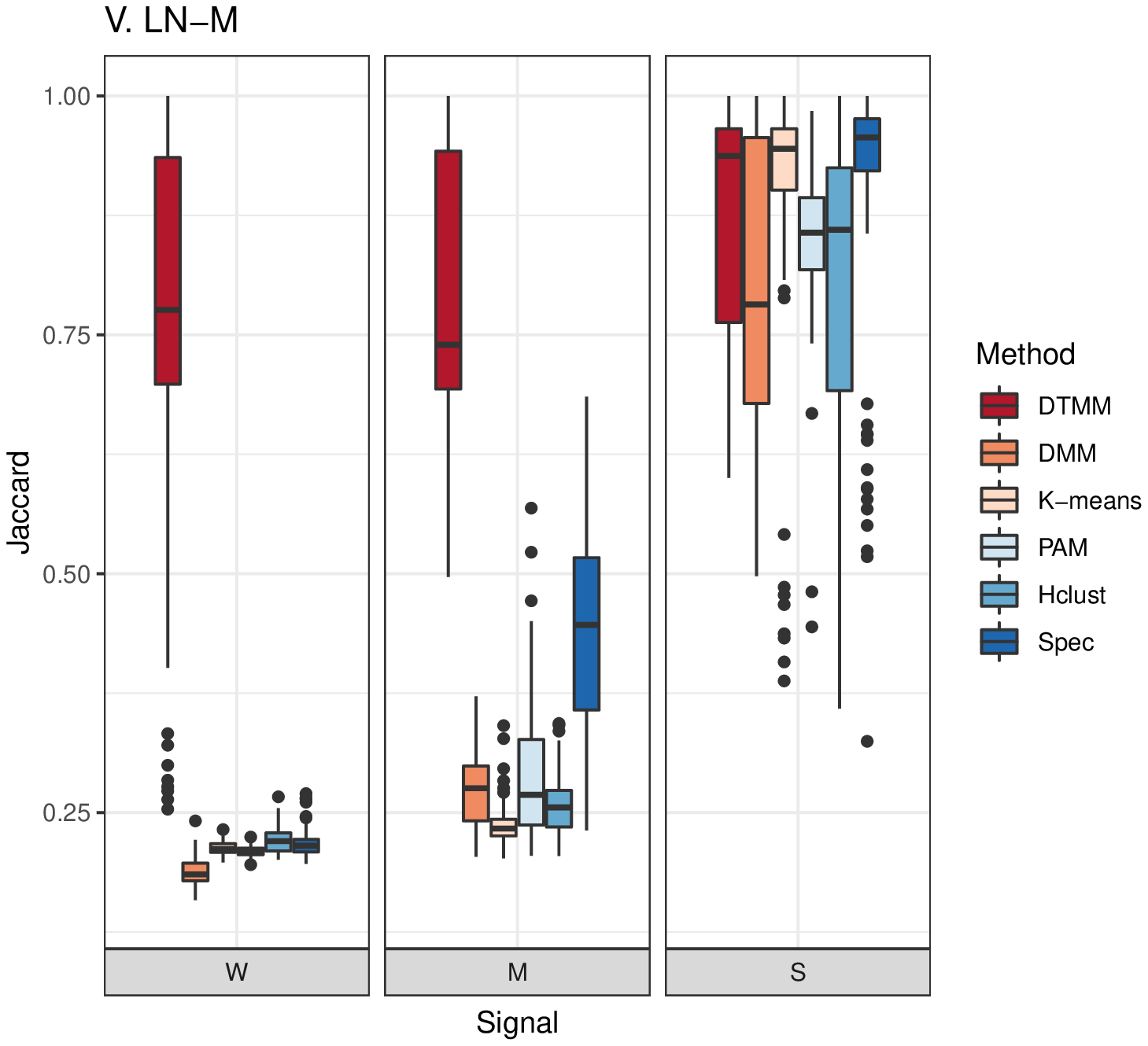}
\caption{Boxplots of the Jaccard index ($n = 180$).}
\label{fig:app_sim_box_large}
\end{minipage}
\end{center}
\end{figure}
\vfill

\clearpage

\null
\vfill
\begin{figure}[!ht]
\begin{center}
\includegraphics[width = \textwidth]{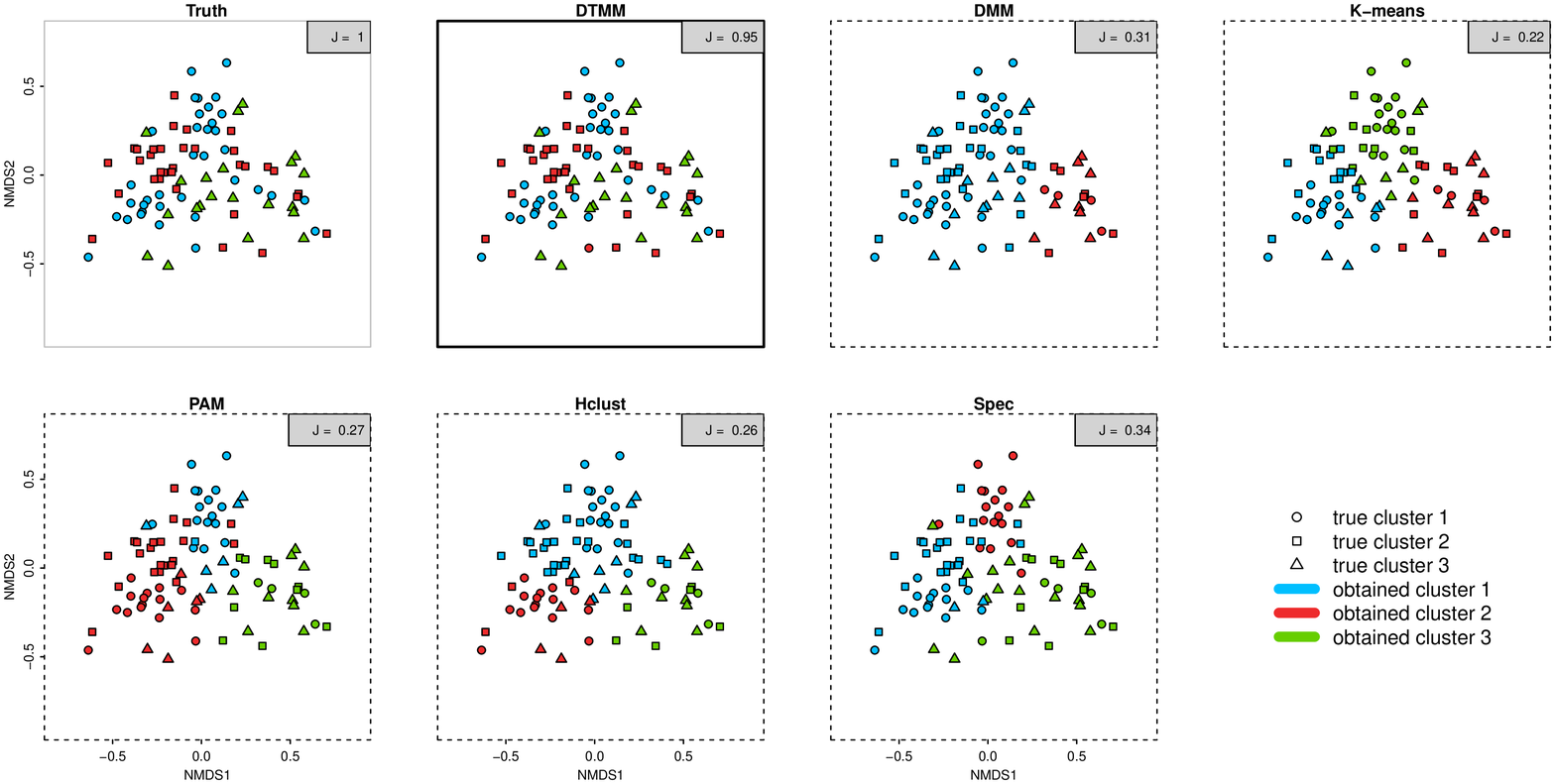}
\caption{2D NMDS plot of samples in a simulation round in scenario $\RN{5}$ ($n = 90$, medium noise level). In each sub-plot, the true clustering is indicated by the shape of the points while the clustering obtained is indicated by the color.}
\label{fig:sec_3_nmds_multi}
\end{center}
\end{figure}
\vfill

\clearpage
\null
\vfill
 \begin{figure}[!ht]
 \centering
\resizebox {1\textwidth} {!}{
  \begin{minipage}[b]{0.52\textwidth}
    \centering
   \includegraphics[width = 1\textwidth]{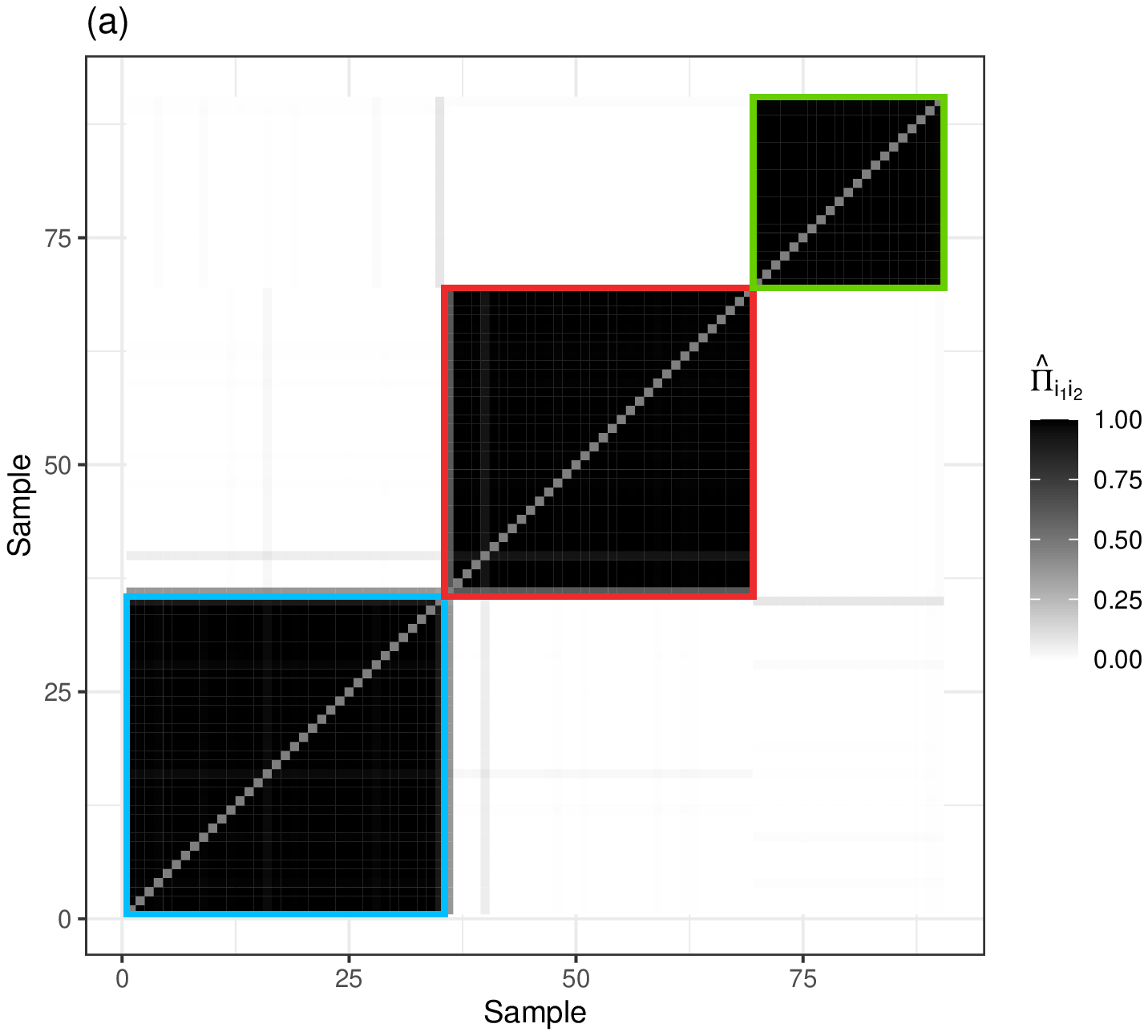} 
 \end{minipage}
 \begin{minipage}[b]{0.235\textwidth}
{\tiny(b).}
    \centering
   \includegraphics[width = 3.36cm, height =  5.6cm]{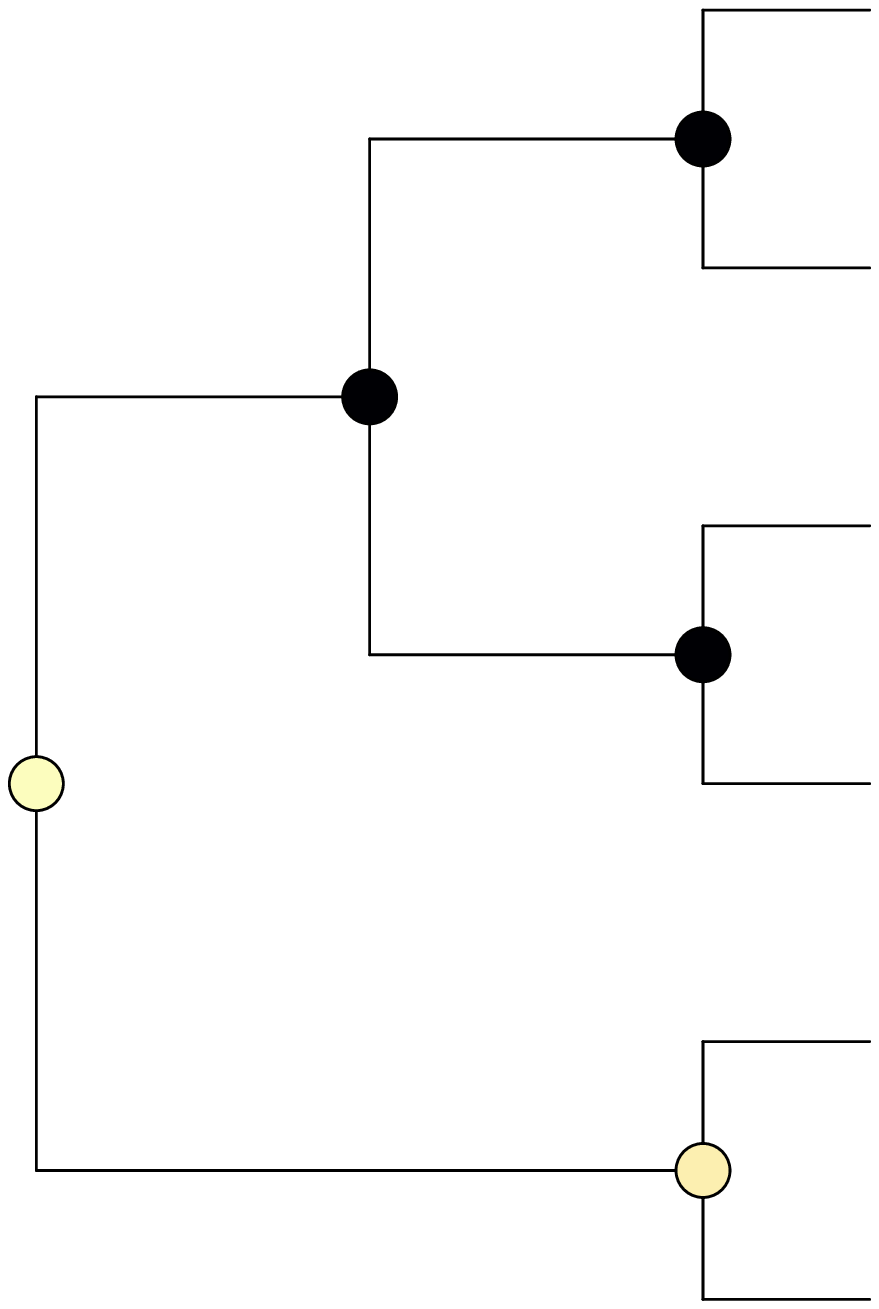} 
   \vspace{0.2cm}
    \end{minipage}\hspace{-0.4cm}
\begin{minipage}[b]{0.25\textwidth}
   \includegraphics[width = 3.5cm, height = 6.125cm]{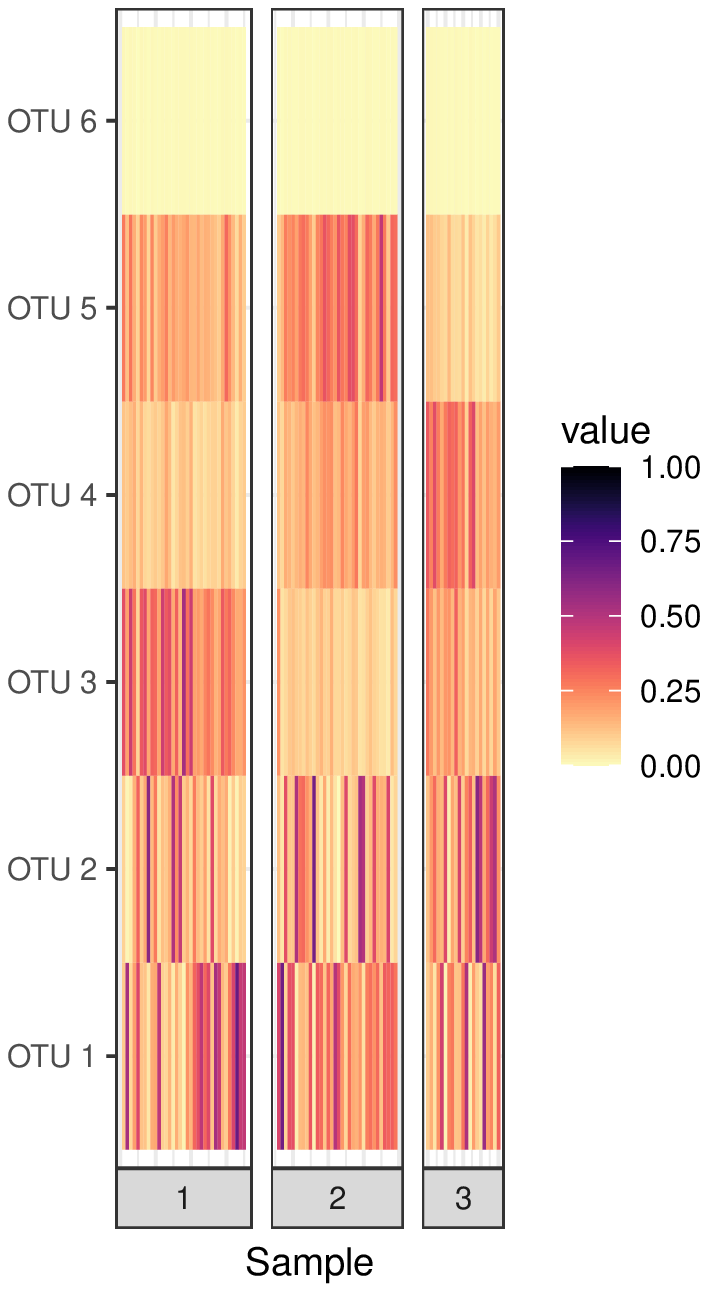}
  \end{minipage}   }
 \caption{Illustrations for an example from simulation scenario $\RN{5}$. (a): Probability of two samples being clustered together by DTMM based on 1000 post-burnin MCMC samples. The samples are ordered by their cluster labels from DTMM. The clusters identified by DTMM are highlighted by squares colored as in \ref{fig:sec_3_nmds_multi}. (b): An illustration of the node selection property of DTMM. The nodes are colored by their estimated posterior node selection probabilities. The heatmap plots the relative abundance of the samples grouped by their cluster labels from DTMM.}
\label{fig:sec_3_multi}
\end{figure}
\vfill
 
 \clearpage

  \null
 \vfill
 \begin{figure}[!ht]
\begin{center}
\includegraphics[width = \textwidth]{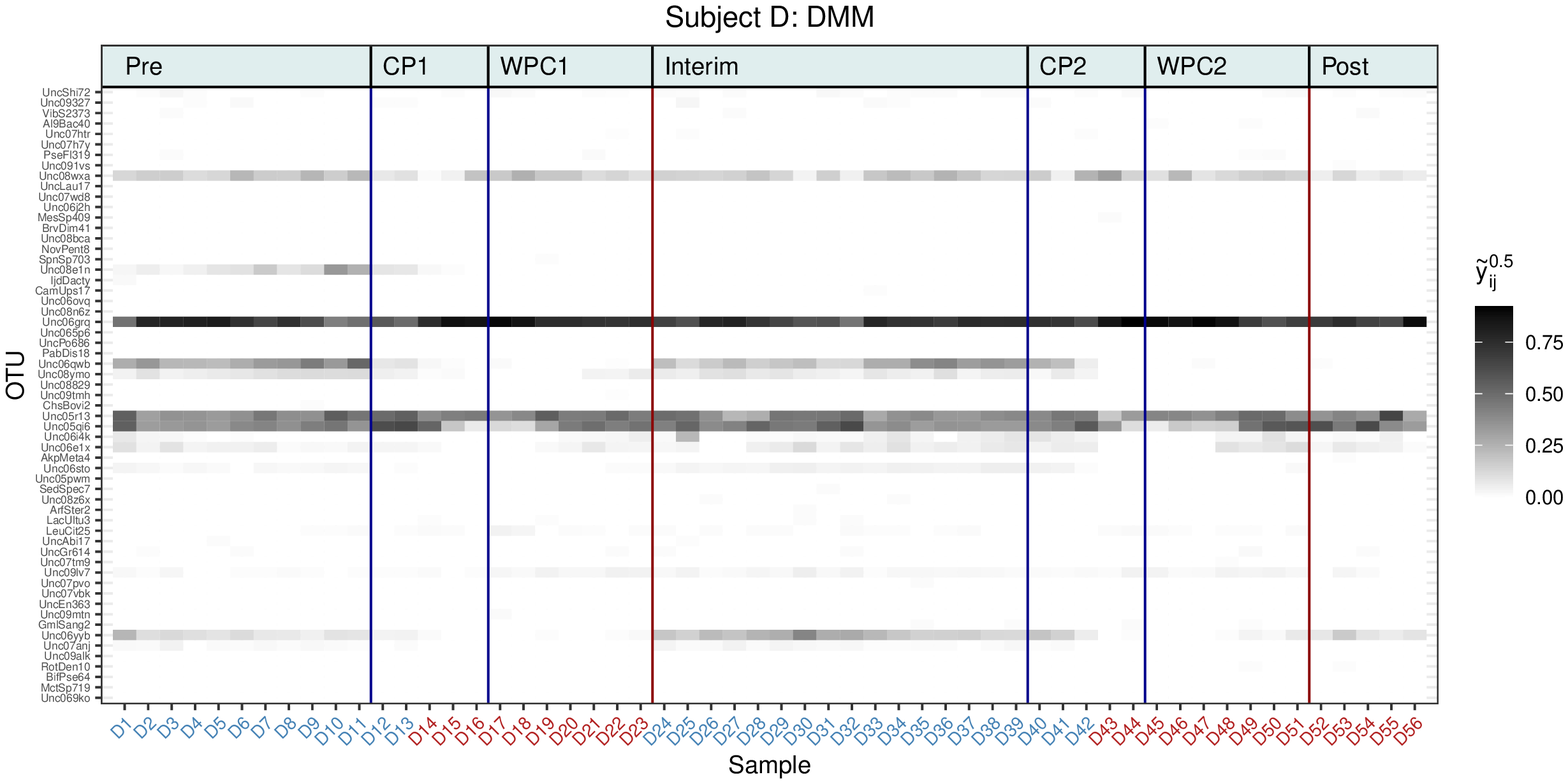} 
\caption{The heatmap of the microbiome samples of patient D (after the square-root transform). Each column represents a specific sample. The columns are ordered by the times the samples were collected. The colors of the $x$-axis labels represent the clustering labels of the samples returned by DMM. The legends are defined the same way as in Figure 8.}
\label{fig:sec_val_d_supp}
\end{center}
\end{figure}
\vfill

\clearpage
 \null
 \vfill
\begin{figure}[!ht]
\begin{center}
\includegraphics[width = \textwidth]{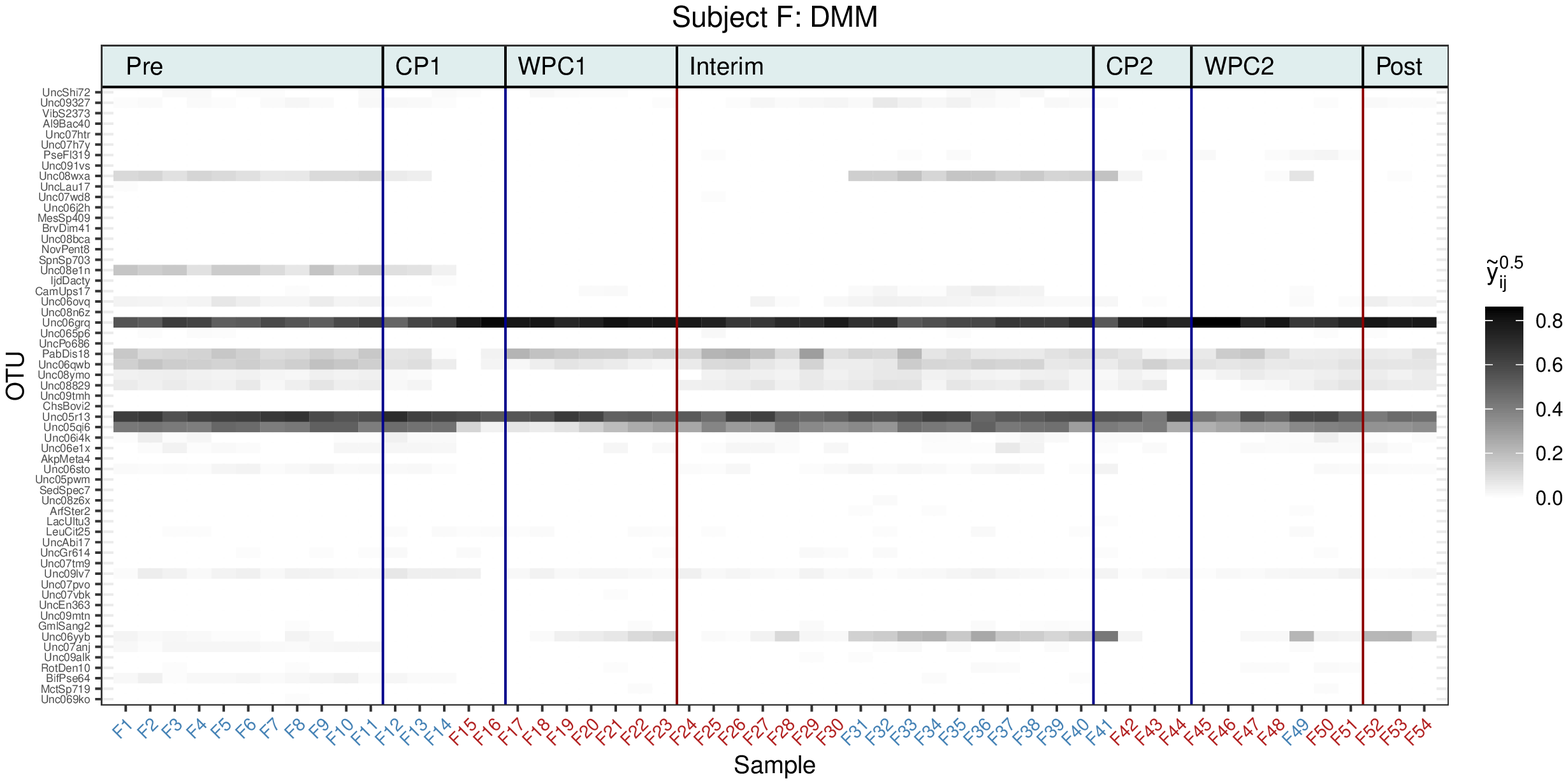} 
\caption{The heatmap of the microbiome samples of patient F (after the square-root transform). Each column represents a specific sample. The columns are ordered by the times the samples were collected. The colors of the $x$-axis labels represent the clustering labels of the samples returned by DMM. The legends are defined the same way as in Figure 9.}
\label{fig:sec_val_f_supp}
\end{center}
\end{figure}
\vfill


\clearpage
\section{Additional materials for the case studies} This section provides additional figures and results for the case studies (Section 4). Most results of the diabetes example (Section 4.2) are presented here.

\null
\vfill
\begin{figure}[!ht]
\begin{center}
\includegraphics[width = 14cm]{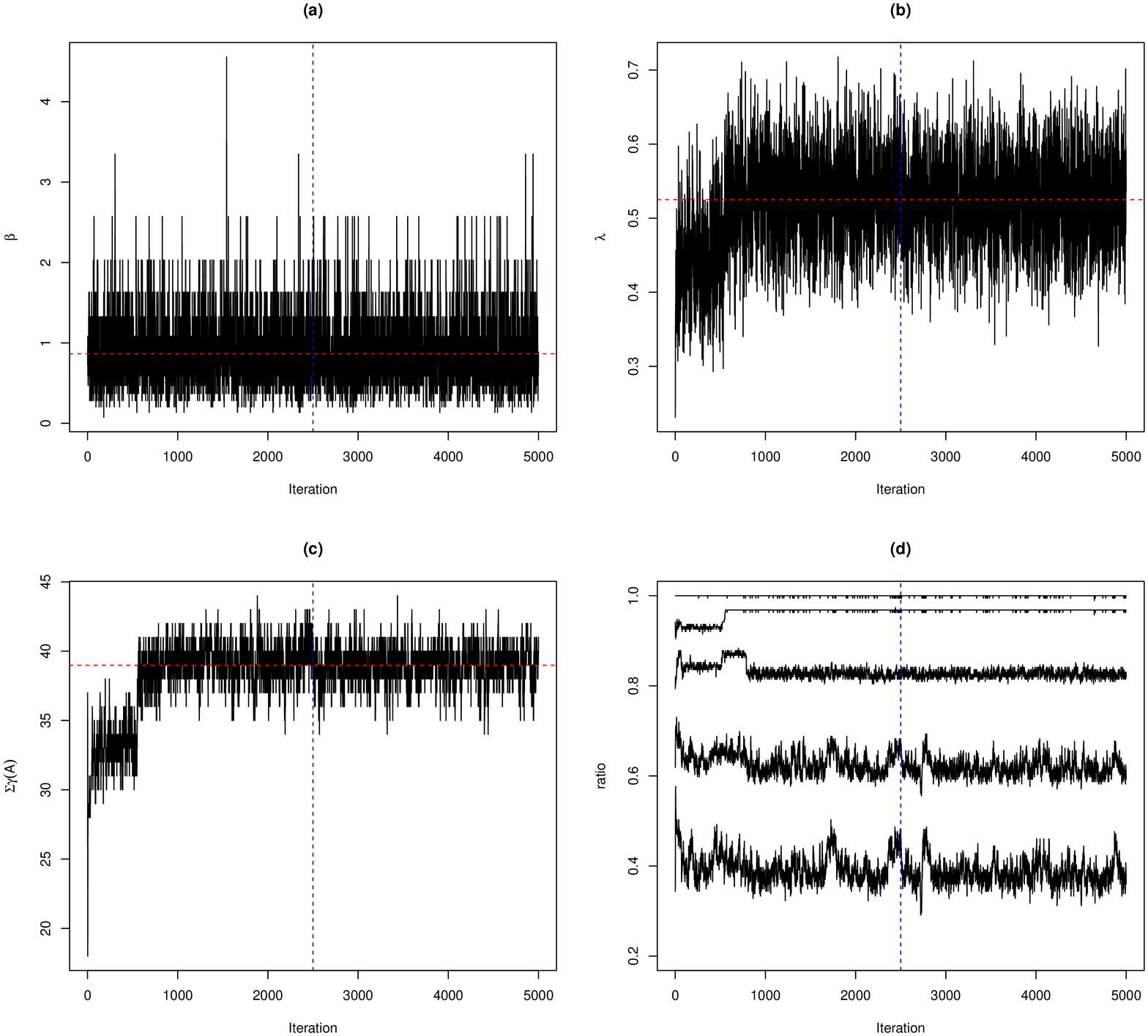} 
\caption{Traceplots of some summary statistics for Section 4.1.}
\label{fig:sec_4_trace}
\end{center}
\end{figure}
\vfill

\null
\vfill
\begin{figure}[!ht]
\begin{center}
\includegraphics[width = 14cm]{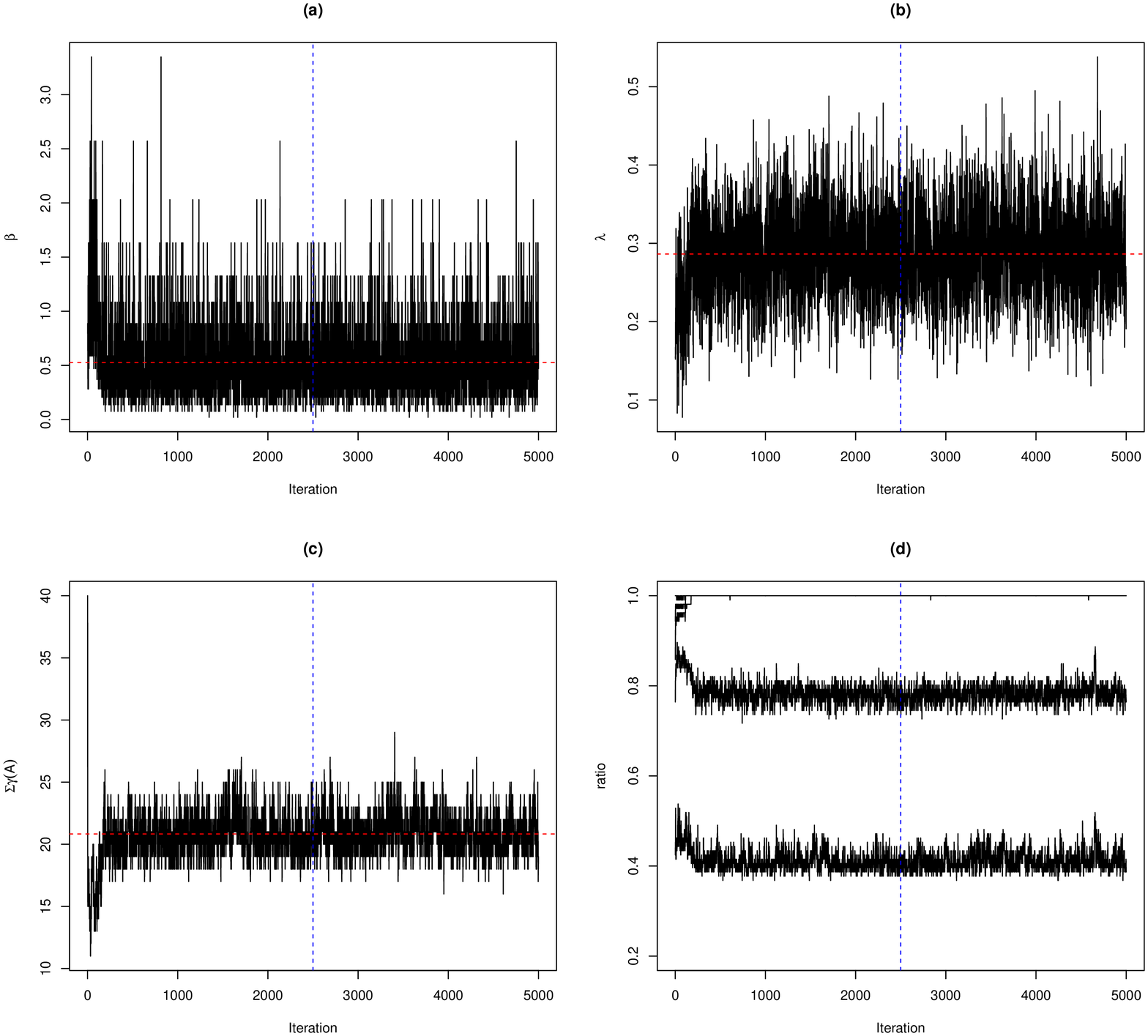} 
\caption{(Diabetes application.) Traceplots of some summary statistics for Section 4.2.}
\label{fig:sec_4_trace_dia}
\end{center}
\end{figure}
\vfill

\begin{figure}[!ht]
\begin{center}
\includegraphics[width = 7cm]{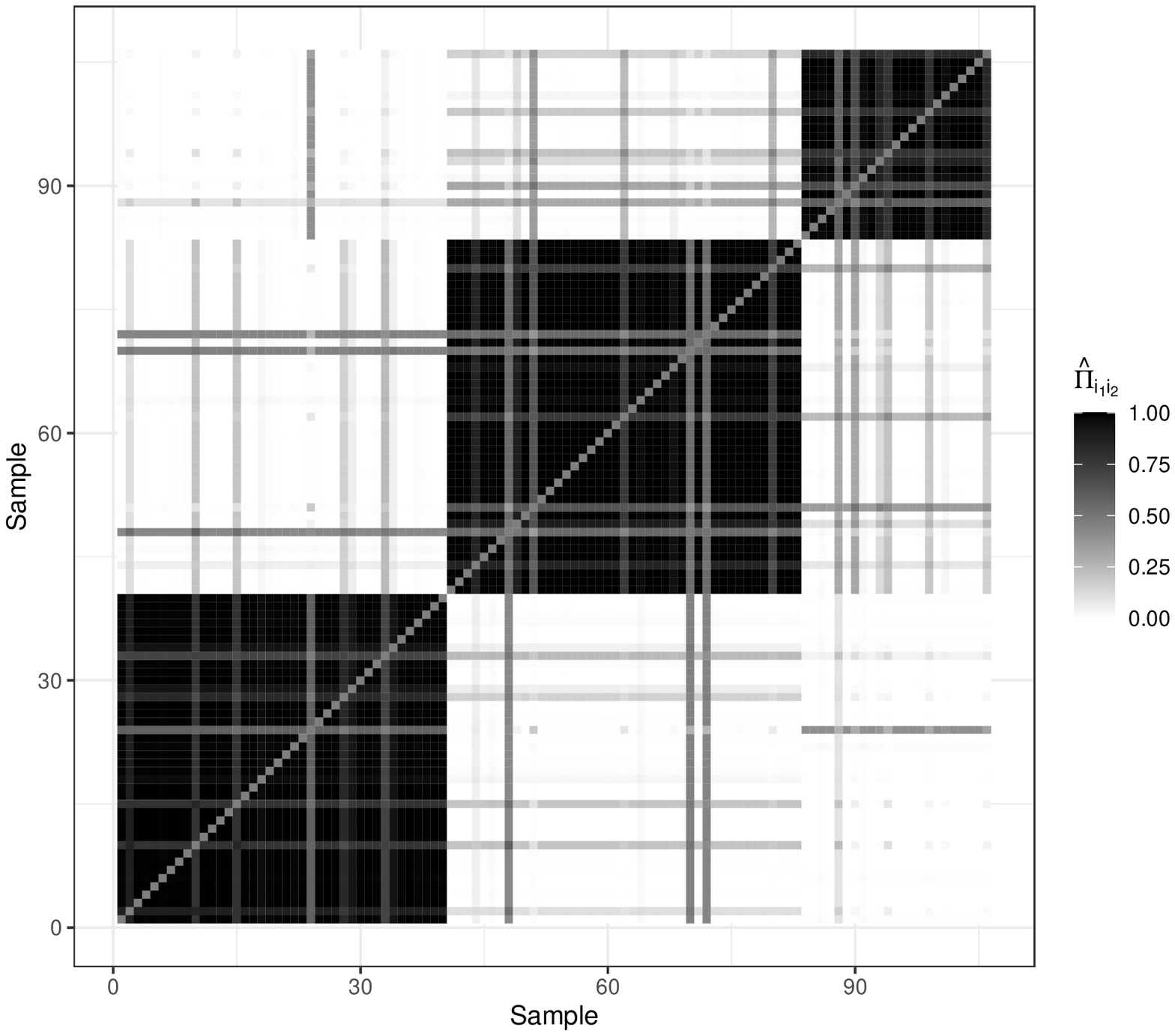}  \includegraphics[width = 7cm]{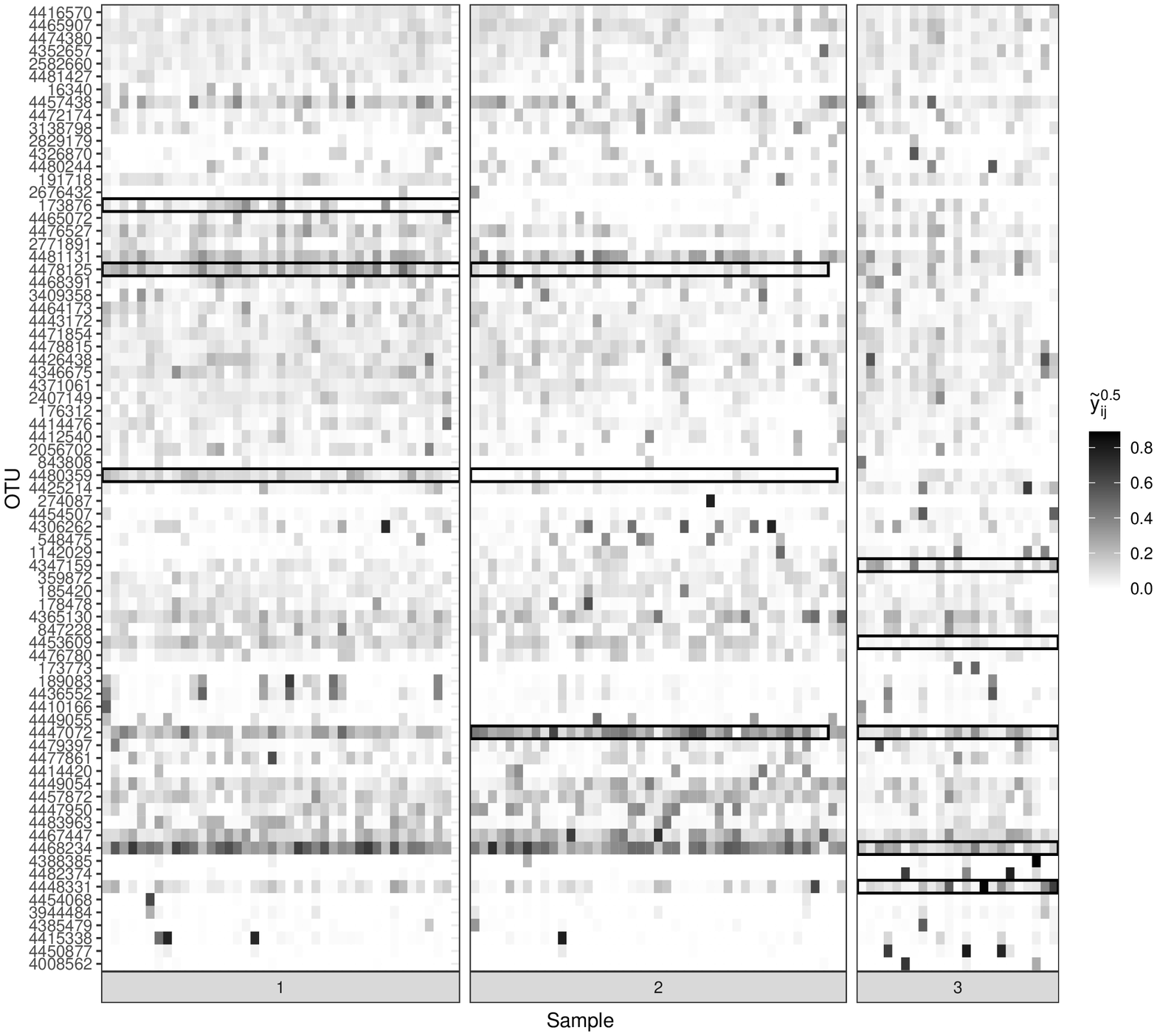}  
\caption{(Diabetes application.) Left: Estimated pairwise co-clustering probabilities. Right: Heatmap of the samples (after the square-root transform) grouped by their labels in $\bm{C}_{LS}$. The black boxes illustrate the characteristic OTUs of each cluster.}
\label{fig:sec_4_co_dia}
\end{center}
\end{figure}

\clearpage
\null
\vfill
\begin{figure}[!ht]
\begin{center}\resizebox {0.9\textwidth} {!}{
 \begin{minipage}[b]{0.6\textwidth}
    \centering
   \includegraphics[width = 7.45cm]{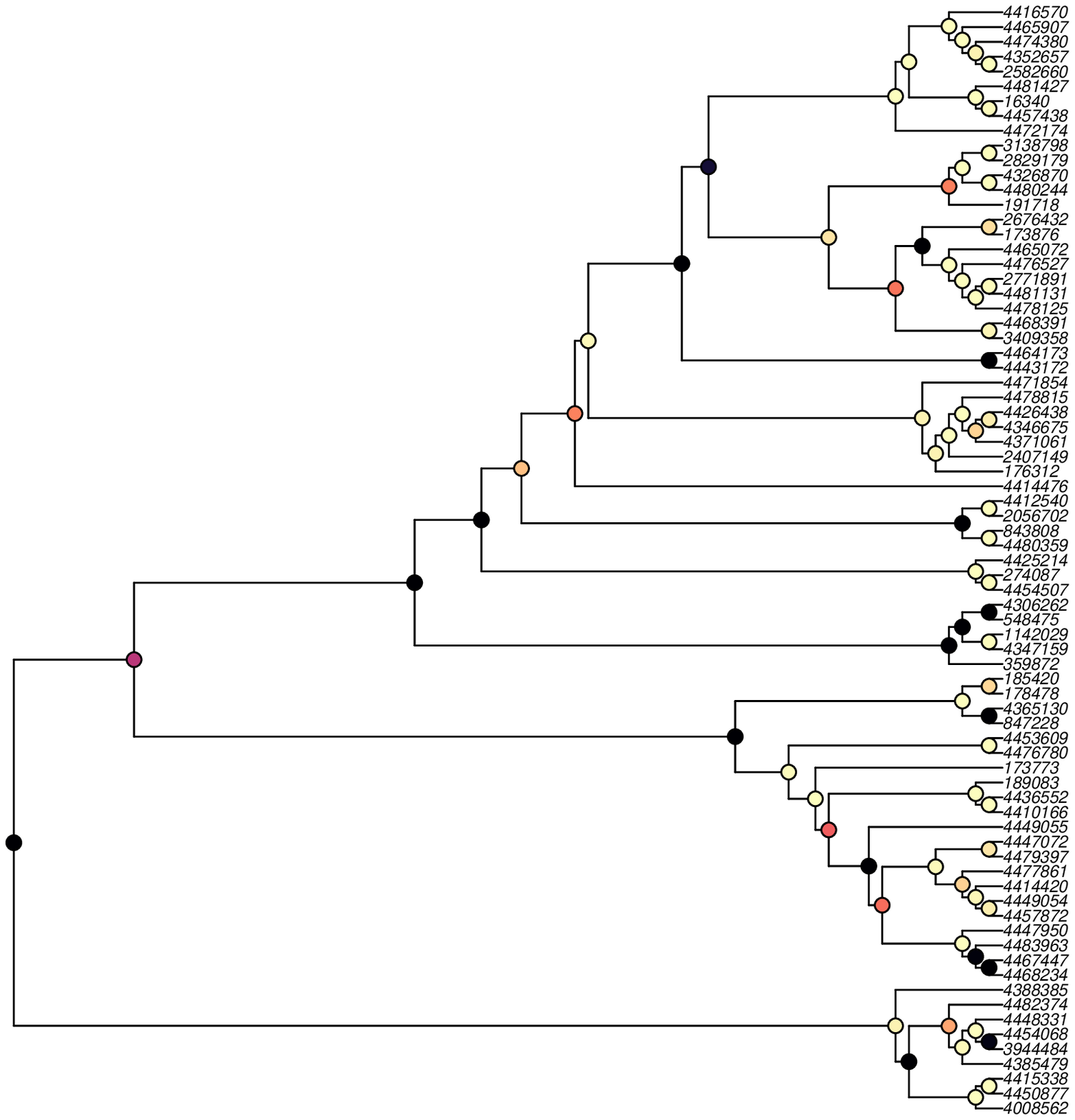} 
   \vspace{0.0cm}\vspace{-0.22cm}
    \end{minipage}\hspace{-0.9cm}
\begin{minipage}[b]{0.4\textwidth}
   \includegraphics[width = 4.7cm]{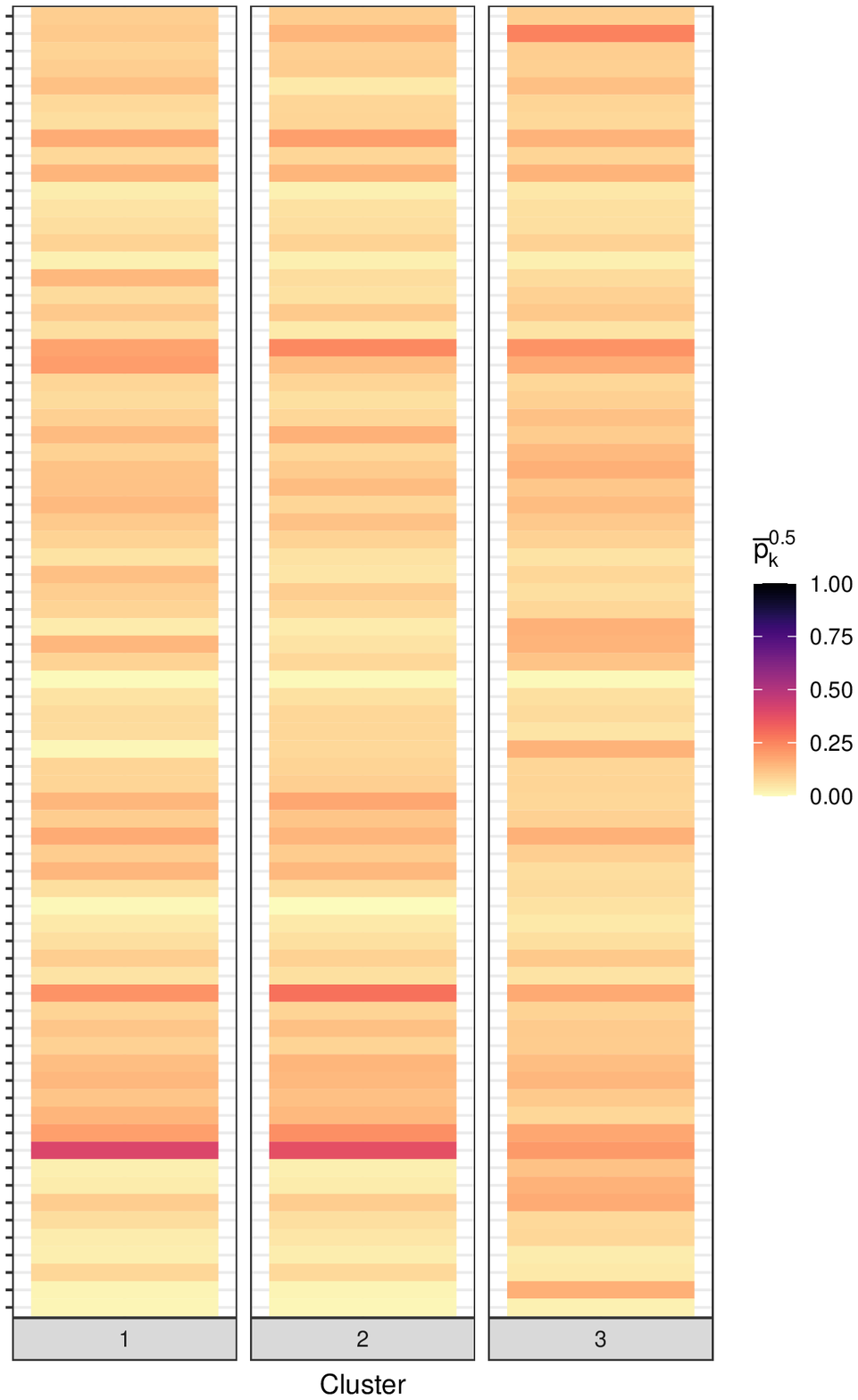}
  \end{minipage}  }
  \caption{(Diabetes application.) Left: Estimated posterior means of the coupling indicators at each node of the phylogenetic tree. Right: The estimated centroids of the three clusters in $\bm{C}_{LS}$ (after the square-root transform). }
  \label{fig:sec_4_centroid_dia}
\end{center}
\end{figure}

\vfill

\clearpage
\null
\vfill
\begin{figure}[!ht]
\begin{center} \resizebox {0.85\textwidth} {!}{
 \begin{minipage}[b]{0.19\textwidth}
    \centering
   \includegraphics[width = 3cm]{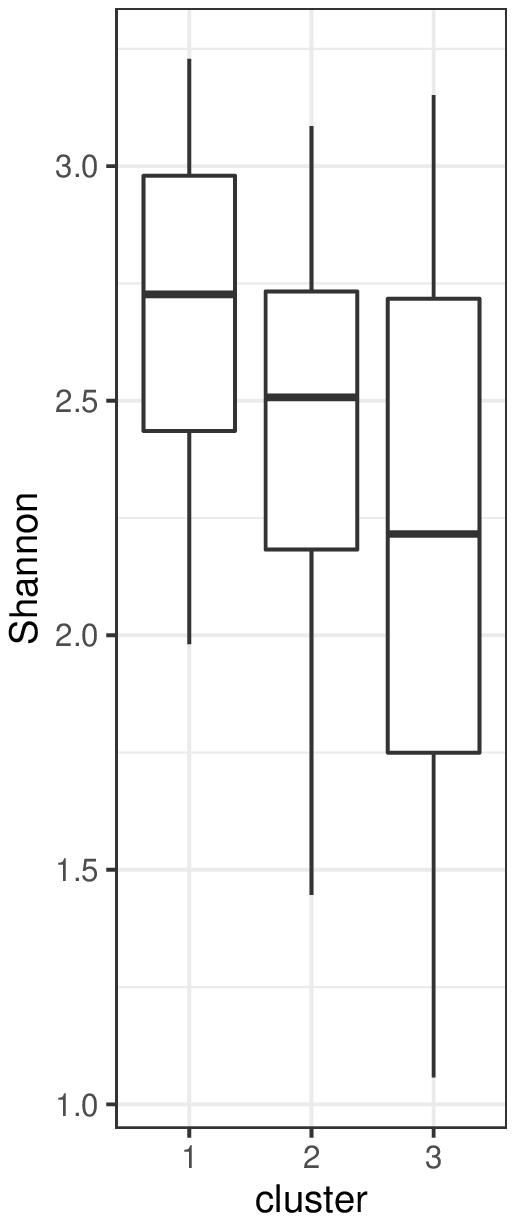} 
    \end{minipage}\hspace{0pt}
    \begin{minipage}[b]{0.8\textwidth}
   \includegraphics[width = 14cm]{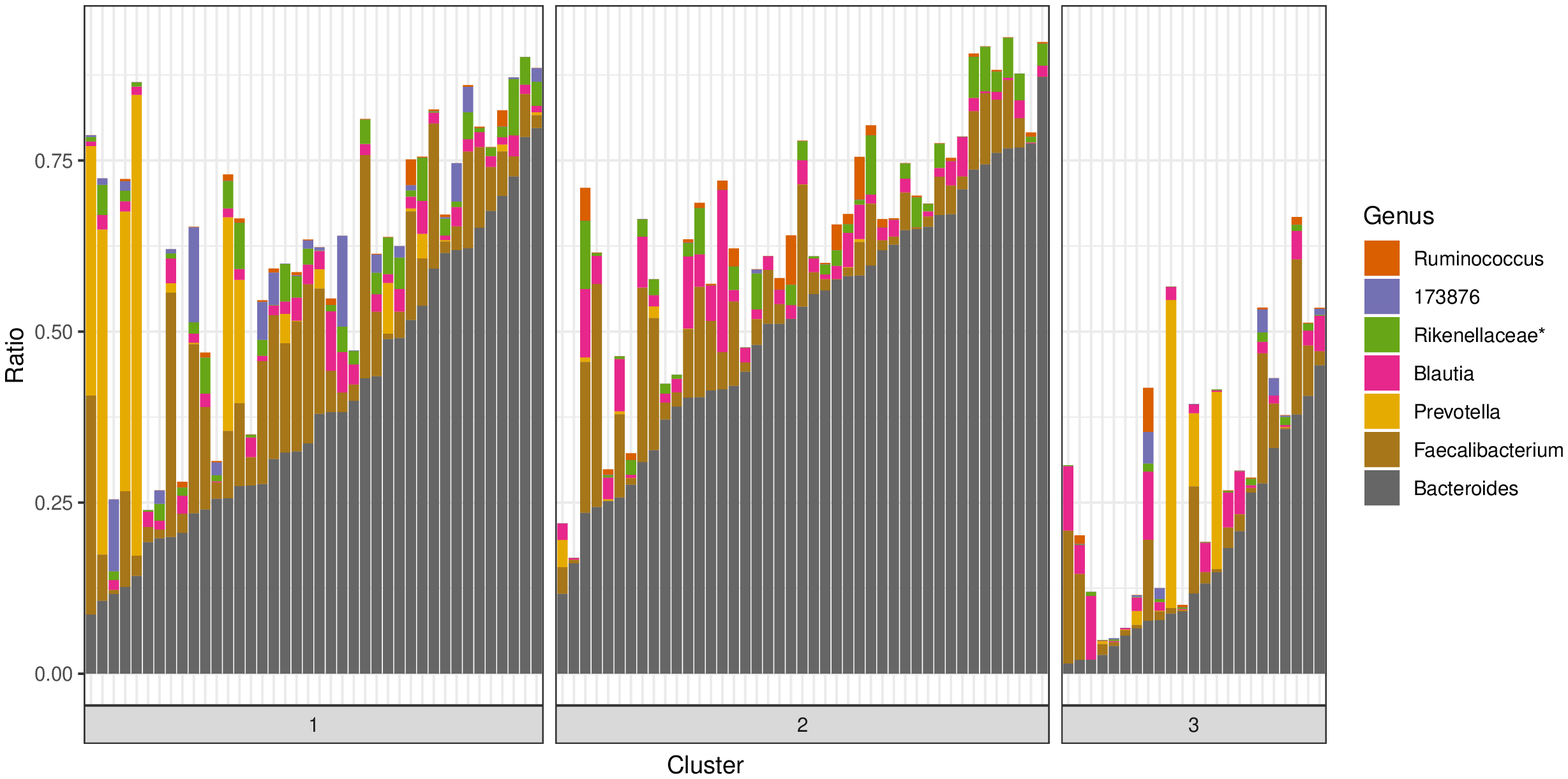}
  \end{minipage}  }
  \caption{(Diabetes application.) Left: Boxplot of the Shannon diversity of samples in each cluster. Right: Relative abundance of 6 genera for each sample. A genus is chosen if its descendant OTUs have large $\vartheta^c_j$ for some $c$. For the 3 OTUs with unavailable genera information, their family is shown instead (indicated by \textit{Rikenellaceae*}). OTU-173876 is important in identifying cluster 1 and is illustrated directly since no genera or family information is available for this OTU. }
  \label{fig:sec_4_cluster_sum_dia}
\end{center}
\end{figure}
\vfill

\clearpage
\null
\vfill
\begin{figure}[!ht]
\begin{center}
\includegraphics[width = 7cm]{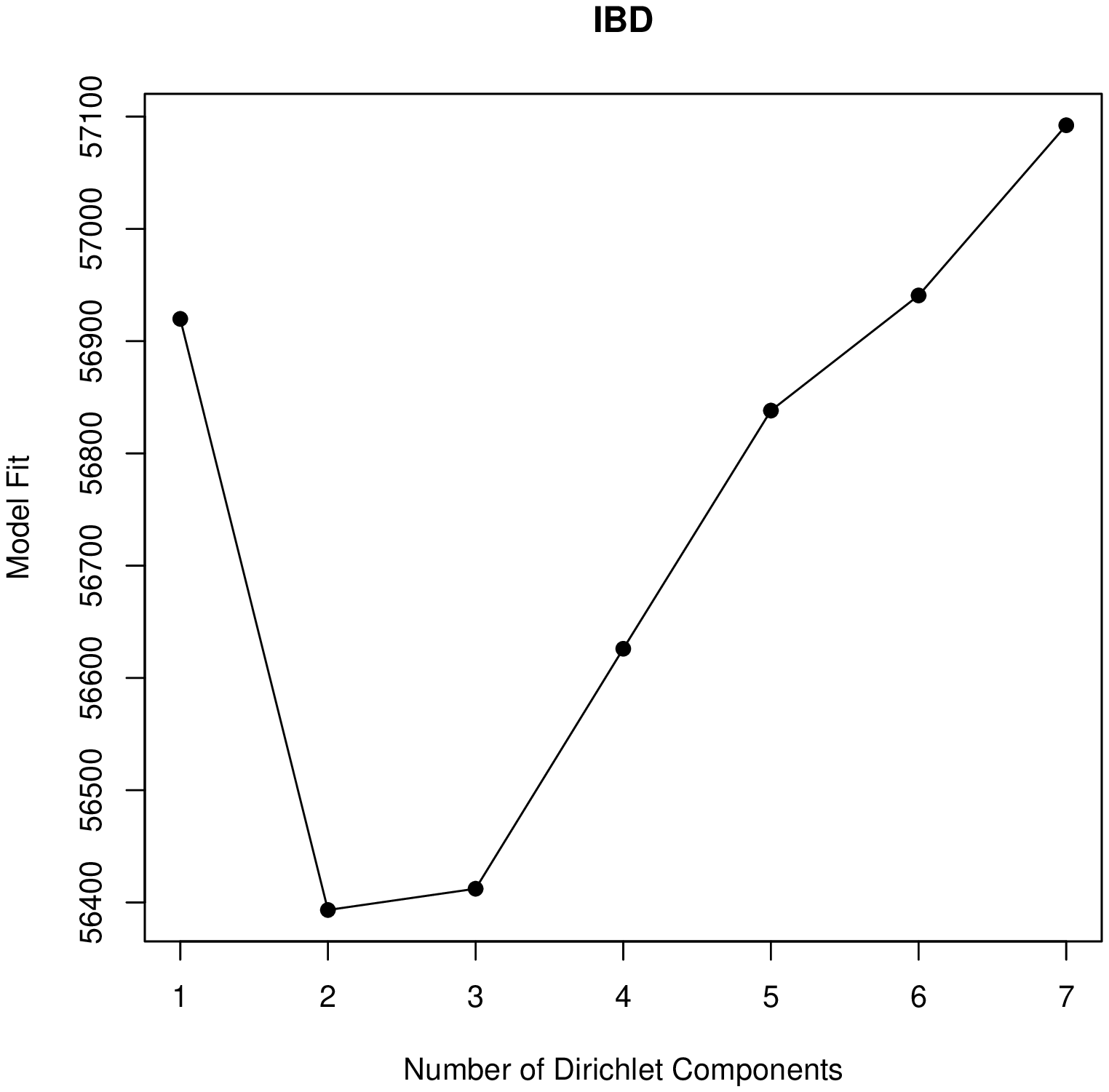}\includegraphics[width = 7cm]{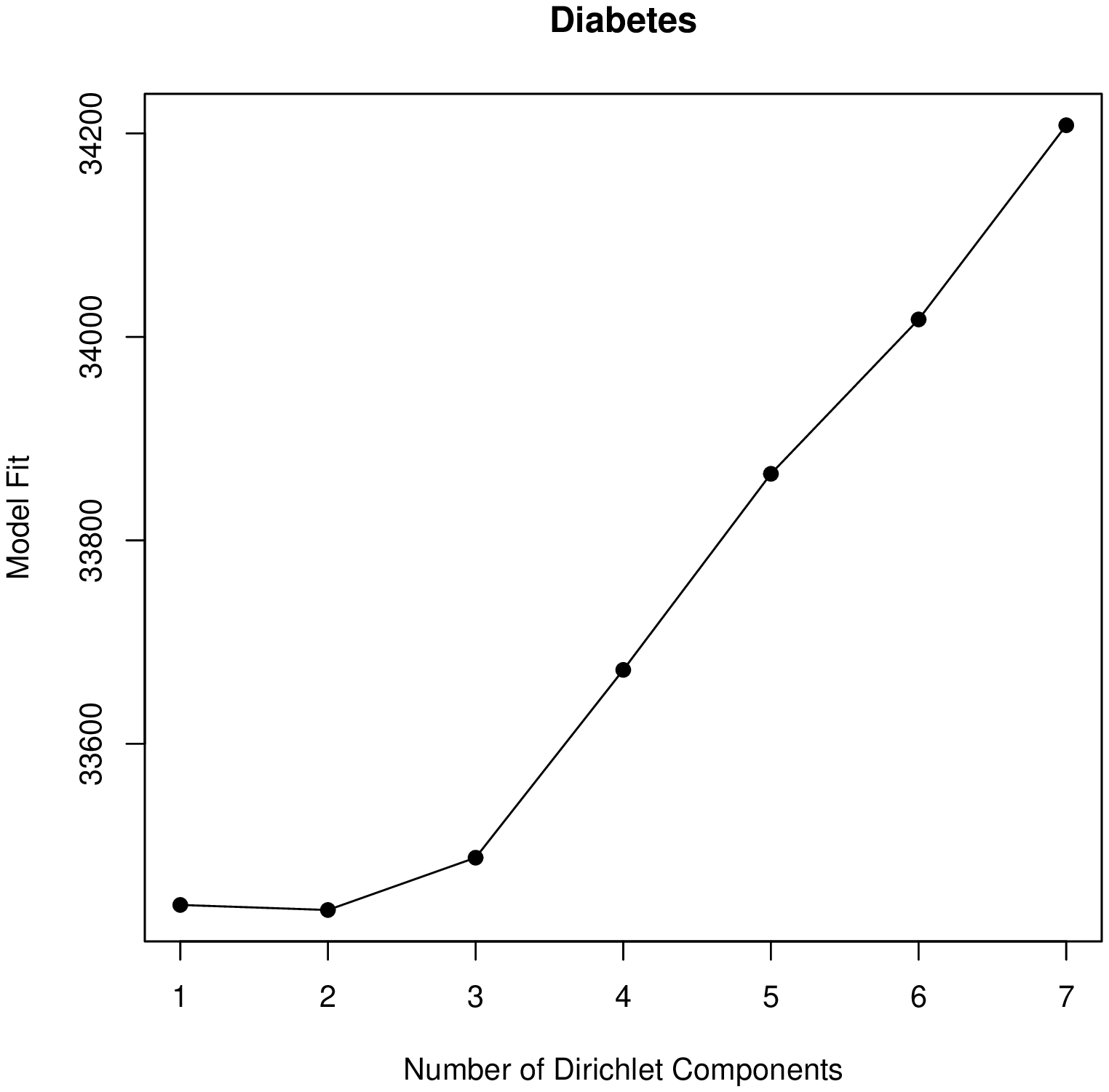}
\caption{Model fit for DMM with different number of clusters. }
\label{fig:sec_spp_dmm}
\end{center}
\end{figure}
\vfill

\clearpage

\null
\vfill
\begin{figure}[!ht]
\begin{center}
\includegraphics[width = 7cm]{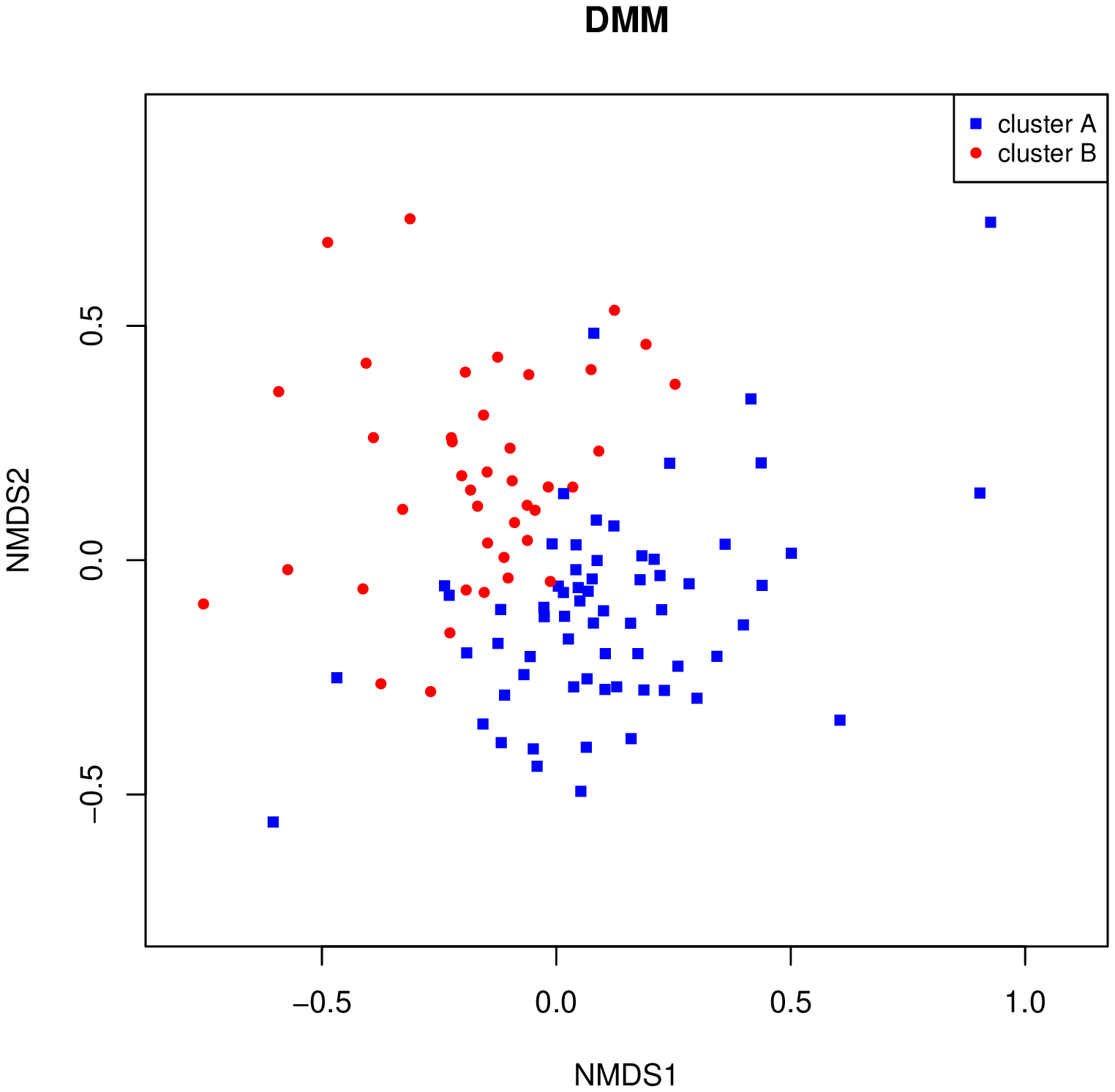}\includegraphics[width = 7cm]{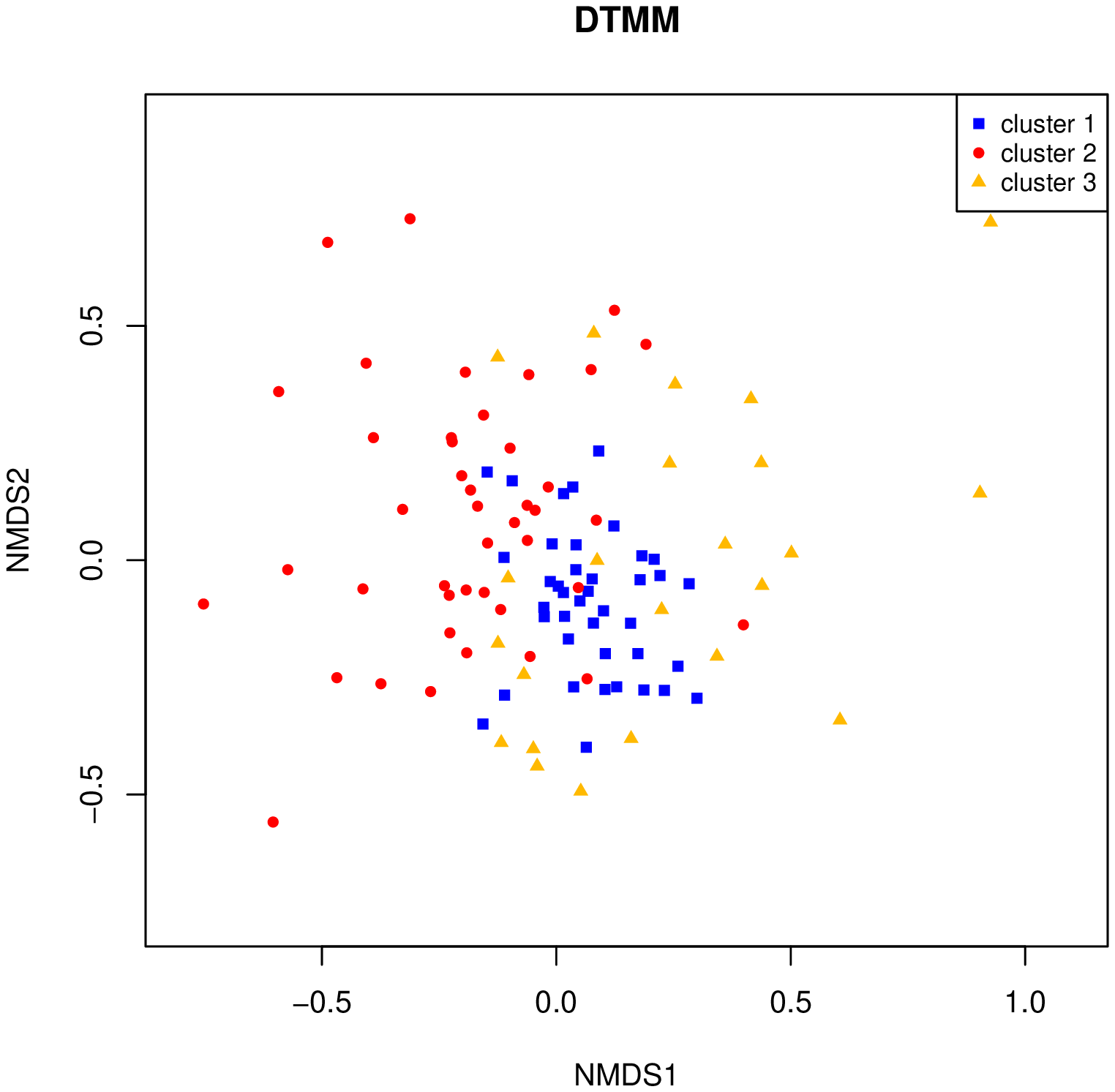} 
\caption{Two-dimensional NMDS plots for the AG diabetes dataset. Points are colored and shaped by the clustering given by DMM (left) or DTMM (right).}
\label{fig:sec_4_dmm_dia}
\end{center}
\end{figure}
\vfill

\end{document}